\documentclass[final]{elsarticle}

\usepackage{changes}
\definechangesauthor[color=red]{R1C1}
\definechangesauthor[color=red]{R1C2}
\definechangesauthor[color=red]{R1C3}
\definechangesauthor[color=blue]{R2C1}
\definechangesauthor[color=blue]{R2C2}
\definechangesauthor[color=blue]{R2C3}
\definechangesauthor[color=blue]{R2C4}
\definechangesauthor[color=blue]{R2C5}
\definechangesauthor[color=blue]{R2C6}
\definechangesauthor[color=magenta]{EC1}
\definechangesauthor[color=magenta]{EC2}
\definechangesauthor[color=magenta]{EC3}
\definechangesauthor[color=magenta]{EC4}
\setremarkmarkup{(#2)}
\usepackage{lineno}
\modulolinenumbers[5]
\usepackage{amsmath}
\usepackage{mathtools}
\usepackage{multirow}
\usepackage{booktabs}
\usepackage{bm}
\usepackage{graphicx}
\usepackage{subfigure}
\usepackage{epstopdf}
\biboptions{numbers,sort&compress}
\usepackage{amssymb}
\usepackage{threeparttable}
\usepackage{algorithm}
\usepackage{algorithmic}
\usepackage{enumerate}

\usepackage{nomencl}
\makenomenclature

\usepackage{etoolbox}
\renewcommand\nomgroup[1]{%
  \item[\bfseries
  \ifstrequal{#1}{C}{Chemical Thermodynamics and Kinetics}{%
  \ifstrequal{#1}{O}{Operation Variables}{%
  \ifstrequal{#1}{N}{Notations}{%
  \ifstrequal{#1}{S}{Superscript and Subscripts}{}}}}%
]}

\journal{Applied Energy}

\begin{document}

\begin{frontmatter}

\title{Optimization of Hydrogen Yield of a High-Temperature Electrolysis System with Coordinated Temperature and Feed Factors at Various Loading Conditions: \\A Model-Based Study\tnoteref{FundingNote}}
\tnotetext[FundingNote]{This work was supported by the Key Program for International S\&T Cooperation Projects of China (2016YFE0102600) and National Natural Science Foundation of China (51577096, 51761135015).}

\author[TsinghuaAddress]{Xuetao Xing}

\author[TsinghuaAddress,TsinghuaInstituteAddress]{Jin Lin\corref{mycorrespondingauthor}}
\cortext[mycorrespondingauthor]{Corresponding author}
\ead{linjin@tsinghua.edu.cn}
\author[TsinghuaAddress,MacauAddress,TsinghuaInstituteAddress]{Yonghua Song}
\author[TsinghuaInstituteAddress,ZhentaiAddress]{Qiang Hu}
\author[NICEAddress]{You Zhou}
\author[NICEAddress]{Shujun Mu}

\address[TsinghuaAddress]{State Key Laboratory of Control and Simulation of Power Systems and Generation Equipment, Department of Electrical Engineering, Tsinghua University, Beijing 100084, China}
\address[MacauAddress]{Department of Electrical and Computer Engineering, University of Macau, Macau, China}
\address[TsinghuaInstituteAddress]{Tsinghua-Sichuan Energy Internet Research Institute, Chengdu 610213, China}
\address[NICEAddress]{National Institute of Clean and Low Carbon Energy, Beijing 102211, China}
\address[ZhentaiAddress]{Zhejiang Zhentai Energy Tech. Co., Ltd., Lishui 321400, Zhejiang Province, China}

\begin{abstract}
  High-temperature electrolysis (HTE) is a promising technology for achieving high-efficiency power-to-gas\deleted[id=EC2]{ (P2G)}, which mitigates the renewable curtailment while promoting decarbonization by transforming wind or solar energy into fuels.
  Different from \replaced[id=R1C1]{low}{room}-temperature electrolysis, a considerable amount of the input energy is consumed by auxiliary equipment in an HTE system for maintaining the temperature, so the studies on systematic energy description and parameter optimization of HTE are essentially required.
  A few published studies investigated HTE's systematic optimization based on simulation, yet there is not a commonly used analytical optimization model which is more suitable for integration with power grid.
  To fill in this blank, a concise analytical operation model is proposed in this paper to coordinate the necessary power consumptions of auxiliaries under various loading conditions of an HTE system.
  First, this paper develops a comprehensive energy flow model for an HTE system based on the fundamentals extracted from the existing work, providing a quantitative description of the impacts of condition parameters, including the temperature and the feedstock flow rates.
  A concise operation model is then analytically proposed to search for the optimal operating states that maximize the hydrogen yield while meeting the desired system loading power by coordinating the temperature, the feedstock flows and the electrolysis current.
  The evaluation of system performance and the consideration of constraints caused by energy balances and necessary stack requirements are both included.
  In addition, analytical optimality conditions are obtained to locate the optimal operating states without performing nonlinear programming by further investigating the optimization method.
  In the case study, a numerical case of an HTE system is employed to validate the proposed operation model and optimization method, which proves that the proposed operation strategies not only improve the overall conversion efficiency but also significantly enlarge the \replaced[id=EC2]{load range}{P2G capacity} of the HTE system.
  \added[id=EC4]{A 24-hour case with fluctuant power input is also simulated to validate the beneficial effects of the proposed operation strategies on producing more hydrogen from a specific profile of surplus electricity.}
  \added[id=EC3]{At last, some results comparisons with existed papers and possible research extensions are discussed briefly.}
\end{abstract}

\begin{keyword}
High-temperature electrolysis\sep system optimization\sep auxiliaries coordination\sep optimal operating state
\end{keyword}

\end{frontmatter}


\nomenclature[C]{$T$}{Temperature}
\nomenclature[C]{$p$}{Pressure}
\nomenclature[C]{$H_T$}{Molar enthalpy at temperature $T$}
\nomenclature[C]{$G_{T,\tilde{\bm{p}}}$}{Molar Gibbs free energy at condition $(T,\tilde{\bm{p}})$}
\nomenclature[C]{$S_{T,\tilde{\bm{p}}}$}{Molar entropy at condition $(T,\tilde{\bm{p}})$}
\nomenclature[C]{$M$}{Molar mass}
\nomenclature[C]{$C_T$}{Isobaric molar heat capacity at temperature $T$}
\nomenclature[C]{$F$}{Faraday constant}
\nomenclature[C]{$R$}{Molar gas constant}
\nomenclature[C]{$n_{\rm e}$}{Number of electron exchanges. For \eqref{eq:overallReact}, $n_{\rm e}= 2$}
\nomenclature[C]{$T_{\rm H}$}{Reference point around operating temperature}
\nomenclature[C]{$T_{\rm L}$}{Reference point around ambient temperature}
\nomenclature[C]{$h$}{Specific enthalpy}
\nomenclature[C]{$c$}{Isobaric specific heat capacity}
\nomenclature[C]{$\Gamma$}{Pre-exponential factor}
\nomenclature[C]{$\xi$}{Activation energy}
\nomenclature[C]{$\sigma$}{Electrical conductivity}

\nomenclature[O]{$U$}{(Equivalent) voltage or overvoltage}
\nomenclature[O]{$I$}{Direct current}
\nomenclature[O]{$i$}{Current density}
\nomenclature[O]{$v$}{Flow velocity of gas mixture}
\nomenclature[O]{$w$}{Mass flow rate of gas mixture}
\nomenclature[O]{$\pi$}{Feed factor}
\nomenclature[O]{$\omega$}{Mass fraction}
\nomenclature[O]{$\chi$}{Molar fraction}
\nomenclature[O]{$\rho$}{Mass density}
\nomenclature[O]{$L$}{Geometric length}
\nomenclature[O]{$P$}{Power}
\nomenclature[O]{$\eta$}{Efficiency}
\nomenclature[O]{$\rm OP$}{Operating points}

\nomenclature[N]{$\tilde{\bm{p}}$}{$\begin{bmatrix} p_{\rm H_2O} & p_{\rm H_2} & p_{\rm O_2} & p_{\rm N_2} \end{bmatrix} ^{\rm T}$}
\nomenclature[N]{$\tilde{\bm{w}},\tilde{\bm{\pi}}$}{$\begin{bmatrix} w_{\rm ca} & w_{\rm an} \end{bmatrix} ^{\rm T},\begin{bmatrix} \pi_{\rm ca} & \pi_{\rm an} \end{bmatrix} ^{\rm T}$}
\nomenclature[N]{$\Delta_{\rm r} X$}{$X_{\rm H_2} + X_{\rm O_2} - 0.5 X_{\rm H_2O}$}
\nomenclature[N]{$\widehat{X}$}{$( w_{\rm ca} X_{\rm ca} + w_{\rm an} X_{\rm an} ) / ( w_{\rm ca} + w_{\rm an} )$}
\nomenclature[N]{${\rm mix}_{\rm ca}(X)$}{$ (p_{\rm H_2O} X_{\rm H_2O} \!+\! p_{\rm H_2} X_{\rm H_2}\!)\!/\!(p_{\rm H_2O} M_{\rm H_2O} \!+\! p_{\rm H_2} M_{\rm H_2}\!)$}
\nomenclature[N]{${\rm mix}_{\rm an}(X)$}{$ (p_{\rm O_2} X_{\rm O_2} + p_{\rm N_2} X_{\rm N_2})/(p_{\rm O_2} M_{\rm O_2} + p_{\rm N_2} M_{\rm N_2})$}
\nomenclature[N]{$\left. X \right|^{\rm s2}_{\rm s1}$}{$X_{\rm s2} - X_{\rm s1}$}
\nomenclature[N]{$\bar{X}$}{$\frac{1}{l_{\rm s4}-l_{\rm s3}} \int_{\rm s3}^{\rm s4} X {\rm d}l$}

\nomenclature[S]{$\ominus$}{Standard state, i.e., pressure at $p^{\ominus}$ for gases}
\nomenclature[S]{$x,y,z,l$}{Spatial dimensions}
\nomenclature[S]{$\rm ca,ele,an$}{Cathode, electrolyte, anode}
\nomenclature[S]{$\rm cha,cel,sys$}{Channel, cell, system}
\nomenclature[S]{$\rm rev,th$}{Reversible, thermo-neutral (voltage)}
\nomenclature[S]{$\rm act,con,ohm$}{Activation, concentration, ohmic (overvoltage)}
\nomenclature[S]{$\rm el,ex$}{Electrolysis, exchange (current density)}
\nomenclature[S]{$\rm fur,com$}{Furnace, compressor}
\nomenclature[S]{$\rm rea,war,dis$}{Reaction, warming, heat dissipation}
\nomenclature[S]{$\rm pum,pre$}{Pump group, preheater}
\nomenclature[S]{$\rm vap,liq,amb$}{Vaporization, liquid, ambient}
\nomenclature[S]{$\rm s0,...,s4$}{At cross section 0,...,4 shown in Fig. \ref{fig:PtG} and Fig. \ref{fig:EnergyBalance}}

\printnomenclature[2.5cm]

\section{Introduction}
The integration of renewable generation sources is faced with formidable challenges because of their undesirable intermittence caused by uncontrollable factors \cite{Waite_WindCurtailment_2016}.
By producing fuel gases with fluctuating electricity and storing the renewable energy that may otherwise be curtailed, the power-to-gas (P2G) approach presents an attractive solution for meeting the increasing energy demand in an ecofriendly manner \cite{Bailera_P2GinIndustry_2017,Meier_PEMandSOEC_2014}.
Moreover, P2G plays an important role in the context of multi-energy systems by functioning as the interaction points to couple the gas network and the power system \cite{Robinius_P2GinDN_2018}. 
Numerous P2G pilot systems have been constructed around the world for research or demonstration purposes as seen in \cite{Gahleitner_P2Gplants_2013} and \cite{Schiebahn_P2G_2015}.
\added[id=EC1]{Actually, it is reported that the levelized fuel cost of P2G is potentially competitive to the prices of diesel and natural gas by 2020 \cite{Mcdonagh_P2G_cost_2018}, and it will not be long before a profitable plant is presented.}

Plenty of studies on P2G's operation and dispatch in the energy supply system have been conducted, such as \cite{McKenna_P2G_2018}, \cite{Clegg_P2G_2015_els} and \cite{Khani_SchedulingP2G_2017_els}, which validated the benefits on energy decarbonization and renewable integration from massive implementation of P2G-based energy storage in the future.
However, these studies mostly treated the P2G systems as simple electricity-gas correlation nodes with fixed conversion efficiencies and simple capacity limits, regardless of their inherent model complexities or local control, \added[id=EC1]{such as the P2G process modeling in \cite{Li_P2G_dispatch_2017}}, which in turn limits the accuracy and practicability of these studies.
To improve this situation, we also need to dig inside of the P2G systems to consider their internal constraints and operating strategies.

Inside a P2G system, the energy conversion is based on the electrolysis process, where $\rm H_2O$ splits into $\rm H_2$, which can be further converted into $\rm CH_4$ with an optional methanation reactor \cite{Lehner_P2G_2014,Qadrdan_P2G_2015}.
Conventionally, the electrolysis occurs \replaced[id=R1C1]{below $100^{\circ}{\rm C}$}{at near room temperature} when using liquid water with the help of an alkaline solution or a polymer electrolyte membrane.
Regarding the mass production of hydrogen, however, high-temperature electrolysis (HTE) is a more promising technology because of its advantage of a substantially higher potential efficiency as described in \cite{Lehner_P2G_2014}, \cite{Brisse_HTE_2008} and \cite{Herring_SOECreview_2007}.
With vaporous feedstock and a solid oxide electrolyte, HTE can achieve a much lower electrolysis voltage by providing an elevated temperature that aids the reaction both kinetically and thermodynamically \cite{Udagawa_SOECmodel_2007}.
In contrast to \replaced[id=R1C1]{low}{room}-temperature electrolysis systems based on alkaline or polymer electrolyte membrane electrolyzers,  a significant fraction of the overall energy consumption is attributed to the auxiliaries in an HTE system, particularly for heating purposes \cite{Frank_P2G_efficiencies_2018,Pan_SOECsystem_experiment_2017}.

As mentioned before, a full investigation inside the HTE system is required for its further implementation in existing power systems and its integration with renewable energy facilities.
The fundamentals of this topic are internal characteristics of HTE stack, such as electrolysis theories, cell models and stack configurations.
These points have long been well supported by many published works.
A one-dimensional HTE model considering overvoltages and internal balances was developed by \cite{Udagawa_SOECmodel_2007} for steady-state evaluation at various temperatures and current densities.
Reference \cite{Kazempoor_SOECmodel_2014} presented a dynamic cell model of a planar solid oxide cell in electrolytic mode (i.e., HTE) and then performed model validation using the available experimental results.
Reference \cite{Navasa_SOEC_CFD_2015} employed a computational fluid dynamics approach to model a cross-flow solid oxide electrolysis cell and then obtained the three-dimensional profiles of current density, temperature, and hydrogen's molar fraction via finite volume method.
\added[id=EC1]{Reference \cite{Cinti_SOEC_Ammonia_2017} simply modeled the solid oxide electrolyzer as an electrical-chemical-heat energy equilibrium coupling with ammonia synthesis process for hydrogen utilization and heat integration.}
A general configuration of an HTE stack of tubular or planar type and a typical design of an HTE system with heat recovery were presented in \cite{Ni_SOECreview_2008}.

Based on the fundamental model of HTE stack, some studies further discussed the parameter impacts and optimization on the stack performance.
Reference \cite{Reytier_HTEstack_2015} studied the impacts of cell number, temperature, inlet composition and current density on the stack voltage by carrying out corresponding HTE experiments.
Reference \cite{Udagawa_SOECoperation_2008} proposed a temperature control strategy based on the steady-state impact of air ratio, and discussed the different distribution of cell temperature in endothermic and exothermic regions.
A comprehensive discussion about the cell-level effects of various operating conditions, such as temperature, current density, steam fraction and steam utilization, was conducted in \cite{Cai_optSOEC_2010}.
An experimental study of a pressurized HTE system was carried out in \cite{Jensen_PressurizedSOEC_2010} to evaluate the cell performance and its dependence on the operating pressure.
These studies above do help improve the HTE performance, but they share a common feature that the focused objects were restricted to an HTE cell or stack rather than a complete system.

For \replaced[id=R1C1]{low}{room}-temperature electrolysis, the operation analyses of cells or stacks may also work in terms of system operation.
Unfortunately, this is not true for HTE systems due to the nonignorable power consumptions of auxiliaries.
That is to say, the studies aimed at systematic energy description and parameter optimization beyond stack are also essentially required for HTE systems.
The problem is that there exist trade-offs in selecting operating parameters between the favors of stack and auxiliaries:
Pursuing higher stack performance with higher temperature and larger feed flow rates leads to higher energy consumption in auxiliaries inevitably.

Despite the extensive research conducting modeling, experiments and parametric analyses at the level of HTE cells or stacks, there are only a few studies of systematic modeling and optimization at HTE system level, where the coordination of auxiliaries' power consumptions based on system efficiency must be considered.
Based on an HTE system model calibrated by cell experiments, reference \cite{Pan_SOECsystem_experiment_2017} described an operating state including the distribution of electric power to electrolyzer, steam generator, furnace and other auxiliaries.
Reference \cite{Frank_P2G_efficiencies_2018} conducted a comprehensive steady-state efficiency calculation in an analytical approach for a generic P2G system considering various of modules and sub-systems.
Note that the parametric study or optimization was not further discussed to improve the system performance in \cite{Pan_SOECsystem_experiment_2017} and \cite{Frank_P2G_efficiencies_2018}.
Reference \cite{Frank_rSOC_op_2018} investigated the impacts of recirculation rate and steam utilization on HTE system efficiency based on a previously validated simulation model of a reversible solid oxide cell plant.
Reference \cite{Brien_NuclearHTE_2010} performed a simulation-based steady-state analysis of the overall thermal-to-hydrogen efficiency and the overall syngas production efficiency of an HTE system coupled with a nuclear reactor.
Reference \cite{Luo_Excergy_SOECsystem_2018} evaluated the exergy efficiency of a simulated HTE system, and also discussed its dependence on operating conditions such as feedstock composition ratio, operating pressure, temperature and steam flow rate.
Reference \cite{Wang_P2M_opt_2018} employed a multi-objective optimization framework to maximize the system efficiency of an HTE simulation model, and depicted the correlation between the optimal efficiency and the current density, pressure, anode/cathode feed rates, etc.
The studies in \cite{Frank_rSOC_op_2018,Brien_NuclearHTE_2010,Luo_Excergy_SOECsystem_2018,Wang_P2M_opt_2018} more or less focused on the parameter optimization and auxiliaries' coordination, and some instructive results were obtained to improve HTE system performance.

Nevertheless, the above studies for HTE system optimization were usually based on a complex simulation model with plenty of equations and professional softwares.
Compared with such approaches of simulation model, an analytical optimization model of the HTE system has the following advantages:
1) The physical constraints can be interpreted concisely and explicitly, which is beneficial to an intuitive understanding.
2) The results and conclusions obtained analytically can be more generic than those drawn from a simulation case.
3) It can be easily developed into an integration model for an HTE system to participate the power grid analysis and operation as a P2G node.
To the best knowledge of the authors, however, there is not an analytical model that is commonly used to optimize HTE system performance with comprehensive coordination of auxiliaries in the published literature.

On purpose of realizing the aforementioned advantages, an analytical operation model of the HTE system that includes various auxiliaries is concisely developed in this paper from a generic energy flow model to maximize the hydrogen production.
Moreover, a discussion of the optimal operating states at various system loading powers, including the maximum production point (MPP) and submaximum production point (SMPP), is also presented.
Both the operation model and the optimization principles are validated with a numerical example.
This paper will hopefully offer theoretical instructions for the local control of HTE systems and provide a foundation for the integrated modeling and analysis of P2G nodes in the power grid with high renewable penetration.

In summary, this paper makes the following main contributions:
1) By introducing the feed factors, an HTE system energy flow model is proposed to formulate the system efficiency and the operating constraints, providing a basis for the coordination of temperature and feed factors with electrolysis current.
2) A concise operation model \eqref{eq:OP_PtG} of the HTE system is then presented to solve the trade-offs of temperature and feed factors to obtain the optimal operating states (SMPP and MPP), where the maximum hydrogen production given a desired system loading power can be achieved and the system capacity is simultaneously extended.
3) An optimization method based on optimality conditions is analytically investigated for SMPP and MPP searching to avoid nonlinear programming.

\section{Problem Description}
Based on the fundamentals, such as typical system structures and classical electrochemical theories, this section introduces the challenge of selecting appropriate state parameters to operate an HTE system safely and efficiently.

\subsection{Fundamentals of the HTE system}

\subsubsection{Operating Structure and Principle}\label{sssec:HTEstructure}
\begin{figure}[!t]
  \centering
  \includegraphics[width=0.99\textwidth]{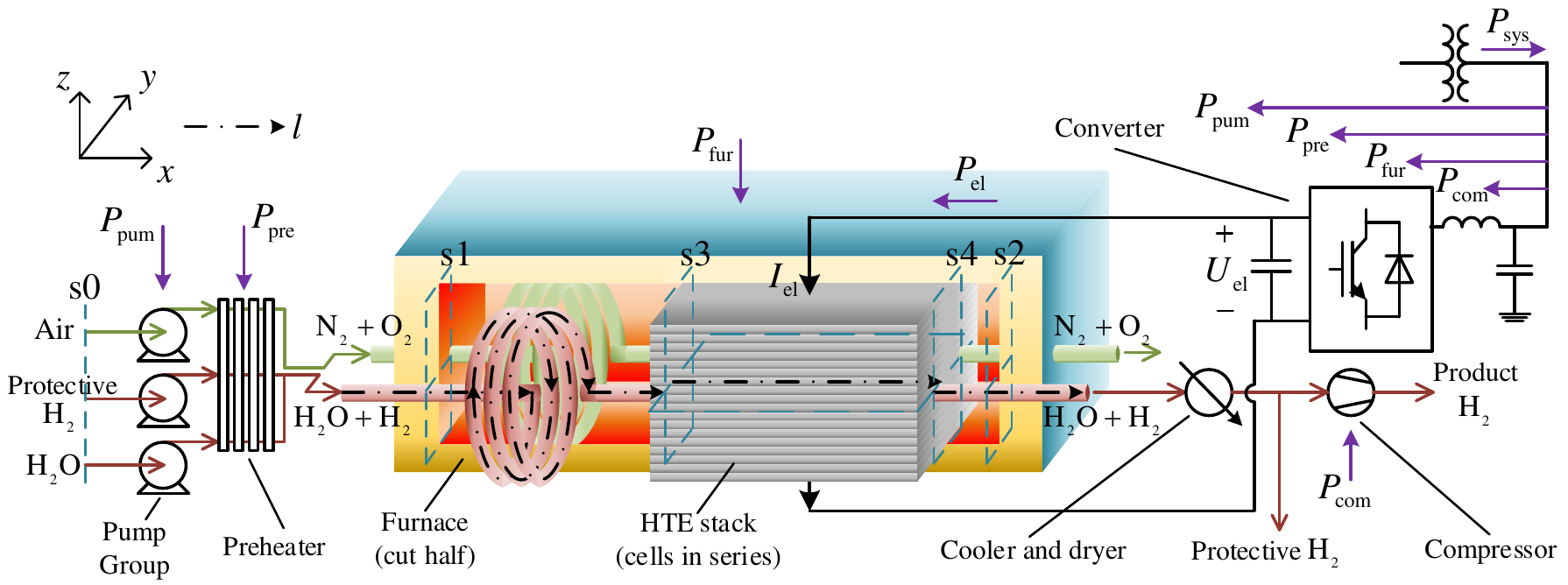}\\
  \caption{Typical structure of an HTE system.}\label{fig:PtG}
\end{figure}

The general structure of an HTE system is illustrated in Fig. \ref{fig:PtG} according to \cite{Zhang_HTEsystem_experiment_2015} and \cite{Frank_rSOC_op_2018}.
The energy conversion mostly occurs in the electric furnace, which provides the high-temperature environment for the electrolysis reaction in the electricity-surplus scenario.
The preheated steam-rich gas and sweep air are pumped in via the cathode pipe and the anode pipe, respectively.
These gases are further heated in coil pipes before being fed into the HTE stack, where $\rm H_2O$ splits into $\rm H_2$ and $\rm O_2$ with the help of direct current from the AC/DC converter.
The $\rm H_2$-rich gas and the $\rm O_2$-rich air are exported out of the furnace; then, the cathode output is cooled, dried and further compressed into the final hydrogen product.
A recirculation of some produced hydrogen  commonly occurs, providing the cathode inlet stream with a specific $\rm H_2$ fraction, e.g., $10\%$ \cite{Frank_rSOC_op_2018,Udagawa_SOECmodel_2007}, which is sufficient to prevent stack oxidation.

\begin{figure}[!t]
  \centering
  \includegraphics[width=0.89\textwidth]{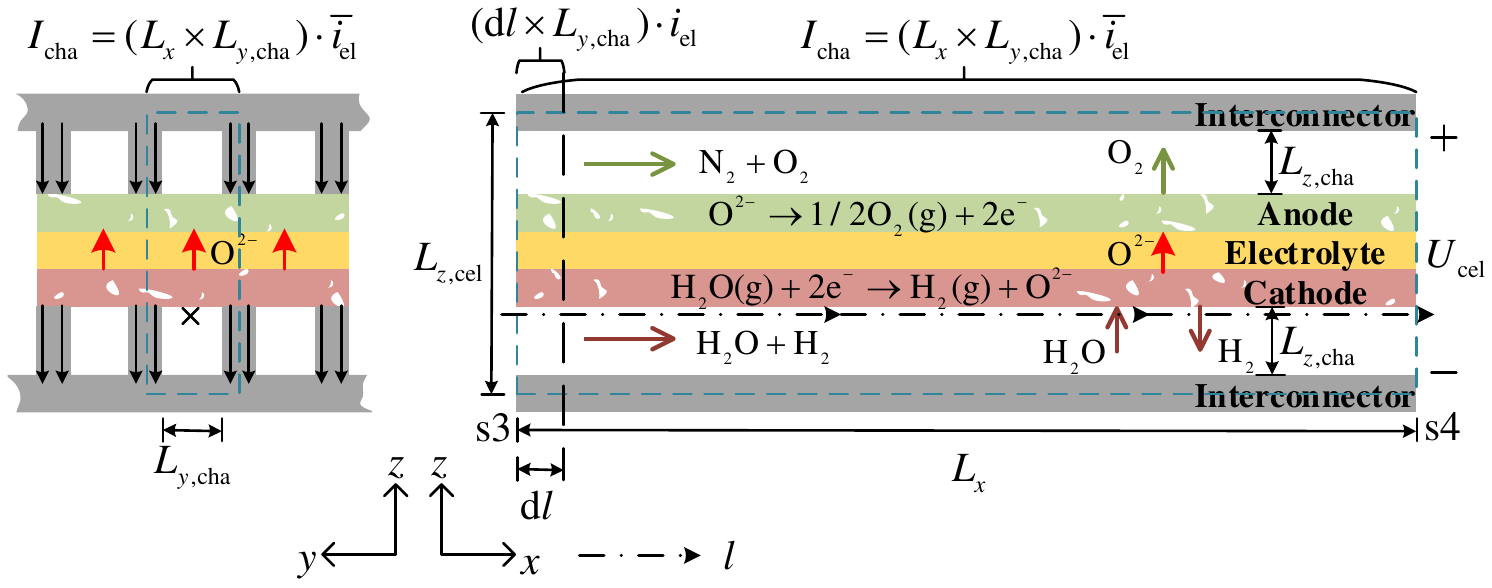}\\
  \caption{Side view and section view of a planar HTE cell.}\label{fig:HTE_cell}
\end{figure}

The HTE stack is typically formed by connecting a number of HTE cells in series, each with tens of channels on the side.
The typical structure of a single planar coflow HTE cell is shown in Fig. \ref{fig:HTE_cell} \cite{Ni_SOECreview_2008,Udagawa_SOECmodel_2007}.
Superheated gases in the two pipes are distributed to every channel of every cell, and then, through porous and catalytic electrodes, the gases diffuse to a triple phase boundary (TPB), where the reactants meet both the electrode and the electrolyte required for reaction.
With the external driving current, the electrolysis reaction occurs at the TPB, generating an amount of $\rm O^{2-}$ and electrons corresponding to the local value of the electrolysis current density.

The overall electrolysis reaction is well known as
\begin{equation}\label{eq:overallReact}
  \rm
  H_2O \rightarrow H_2 + 0.5 O_2
\end{equation}
whose \added[id=R2C1]{molar enthalpy increase $\Delta_{\rm r} H_T$} can be expressed by
\begin{equation}\label{eq:DeltaH}
  \Delta_{\rm r} H_T = \Delta_{\rm r} G_{T,\tilde{\bm{p}}} + T\Delta_{\rm r} S_{T,\tilde{\bm{p}}}
\end{equation}
\added[id=R2C1]{where $\Delta_{\rm r} G_{T,\tilde{\bm{p}}}$ and $\Delta_{\rm r} S_{T,\tilde{\bm{p}}}$ are the molar changes of Gibbs free energy and entropy of reaction \eqref{eq:overallReact}.}
\added[id=R2C1]{Note that $T$ and $\tilde{\bm{p}} = \left[ p_{\rm H_2O} \; p_{\rm H_2} \; p_{\rm O_2} \; p_{\rm N_2} \right] ^{\rm T}$ are temperature and partial pressure conditions of reaction.}
In fact, $\Delta_{\rm r} G_{T,\tilde{\bm{p}}}$ and $T\Delta_{\rm r} S_{T,\tilde{\bm{p}}}$ represent the desired energy in the form of electricity and heat, respectively, and $\Delta_{\rm r} H_T$ denotes the generated chemical energy.
With electrolysis current coupling at the reaction rate of \eqref{eq:overallReact}, the associated voltages of $\Delta_{\rm r} G_{T,\tilde{\bm{p}}}$ and $\Delta_{\rm r} H_T$, namely, reversible voltage $U_{\rm rev}$ and thermo-neutral voltage $U_{\rm th}$, can be introduced:
\begin{equation}\label{eq:UrevUth}
  U_{\rm rev}(T,\tilde{\bm{p}}) \stackrel{\mbox{\tiny def}}{=} \frac{\Delta_{\rm r} G_{T,\tilde{\bm{p}}}}{n_{\rm e}F},\,U_{\rm th}(T) \stackrel{\mbox{\tiny def}}{=} \frac{\Delta_{\rm r} H_T}{n_{\rm e}F}
\end{equation}
where $U_{\rm th}$ is approximately $1.48{\rm V}$ at $25^{\circ}{\rm C}$ and $1.29{\rm V}$ at $800^{\circ}{\rm C}$.
\added[id=R2C1]{Here $n_{\rm e}=2$ is the number of moles of electron exchanges of reaction \eqref{eq:overallReact}, and $F$ is Faraday constant.}
The HTE cell releases excess heat during operation when $U_{\rm cel} > U_{\rm th}$, works only with an external heat source when $U_{\rm rev} < U_{\rm cel} < U_{\rm th}$, and does not perform electrolysis when $U_{\rm cel} < U_{\rm rev}$.

\subsubsection{Classic Voltage Formulations of HTE}\label{sssec:U_cel}
In addition to $U_{\rm rev}$, which provides the necessary Gibbs free energy, additional overvoltages are required for practical electrolysis to overcome kinetic barriers \cite{Kazempoor_SOECmodel_2014,Udagawa_SOECmodel_2007}:
\begin{equation}\label{eq:form_Ucel}
  U_{\rm cel}(i_{\rm el},T,\tilde{\bm{p}}) = U_{\rm rev} + U_{\rm act} + U_{\rm con} + U_{\rm ohm}.
\end{equation}
\added[id=R2C1]{These additional overvoltages $U_{\rm act}$, $U_{\rm con}$ and $U_{\rm ohm}$ are called activation overvoltage, concentration overvoltage and ohmic overvoltage, respectively.}
According to the Nernst equation, $U_{\rm rev}$ can be further formulated as \eqref{eq:form_Urev} \cite{Luo_Excergy_SOECsystem_2018}:
\begin{equation}\label{eq:form_Urev}
  U_{\rm rev}(T,\tilde{\bm{p}}) = U_{\rm rev}^{\ominus}(T) + \frac{RT}{n_{\rm e}F} {\rm ln}\left(\frac{p_{\rm H_2}p_{\rm O_2}^{0.5}}{p_{\rm H_2O}p^{\ominus \, 0.5}}\right)
\end{equation}
\added[id=R2C1]{where $R$ is the molar gas constant, and the expansion of $U_{\rm rev}^{\ominus}(T)$ is included as \eqref{eq:form_Urev0_expansion} in \ref{app:U_rev}.}
The activation overvoltage $U_{\rm act}$ denotes the desired voltage to meet the activation energy at both electrodes, which is generally described by the Butler-Volmer equation \cite{Udagawa_SOECmodel_2007}:
\begin{subequations}
\begin{gather}
  U_{\rm act}(i_{\rm el},T) = \frac{2RT}{n_{\rm e}F} \left[{\rm sinh}^{-1} \left(\frac{i_{\rm el}}{2i_{\rm ex,ca}}\right) + {\rm sinh}^{-1} \left(\frac{i_{\rm el}}{2i_{\rm ex,an}}\right)\right] \label{eq:form_Uact} \\
  i_{\rm ex,ca}(T) = \frac{RT}{n_{\rm e}F} \Gamma_{\rm ex,ca} e^{-\frac{\xi_{\rm ca}}{RT}},
  i_{\rm ex,an}(T) = \frac{RT}{n_{\rm e}F} \Gamma_{\rm ex,an} e^{-\frac{\xi_{\rm an}}{RT}} \label{eq:form_iex}
\end{gather}
\end{subequations}
where $i_{\rm el}$ is the local value of the electrolysis current density, \added[id=R2C1]{and $i_{\rm ex,ca}$ (or $i_{\rm ex,an}$) is the cathode (or anode) exchange current density calculated by \eqref{eq:form_iex} from the corresponding pre-exponential factor $\Gamma_{\rm ex,ca}$ (or $\Gamma_{\rm ex,an}$) and activation energy $\xi_{\rm ca}$ (or $\xi_{\rm an}$)}.
The concentration overvoltage $U_{\rm con}$ reflects the voltage compensation of $U_{\rm rev}$ due to the concentration losses during the component diffusion between the TPB and channel bulk \cite{Udagawa_SOECmodel_2007}:
\begin{equation}\label{eq:form_Uconc}
  U_{\rm con}(i_{\rm el},T,\tilde{\bm{p}}) = \frac{RT}{n_{\rm e}F} {\rm ln} \left( \frac{p_{\rm H_2,TPB}}{p_{\rm H_2}} \frac{p_{\rm H_2O}}{p_{\rm H_2O,TPB}} \frac{p_{\rm O_2,TPB}^{0.5}}{p_{\rm O_2}^{0.5}} \right)
\end{equation}
where the partial pressures at the TPB are related to $i_{\rm el}$ and $\tilde{\bm{p}}$.
In general, $U_{\rm con}$ is negligible unless $p_{\rm H_2}$, $p_{\rm O_2}$ or $p_{\rm H_2O}$ is close to zero.
The ohmic overvoltage $U_{\rm ohm}$ is attributed to the electrical resistances of the cell structure \cite{Udagawa_SOECmodel_2007}:
\begin{equation}\label{eq:form_Uohm}
  U_{\rm ohm}(i_{\rm el},T) = i_{\rm el} \left( \frac{L_{z,{\rm ele}}}{\sigma_{\rm ele}(T)} + \frac{L_{z,{\rm ca}}}{\sigma_{\rm ca}} + \frac{L_{z,{\rm an}}}{\sigma_{\rm an}} \right)
\end{equation}
where \added[id=R2C1]{$\sigma_{\rm ele}$, $\sigma_{\rm ca}$ and $\sigma_{\rm an}$ are electrical conductivities of electrolyte, cathode and anode, while $L_{z,{\rm ele}}$, $L_{z,{\rm ca}}$ and $L_{z,{\rm an}}$ are their geometric thicknesses, as shown in Fig. \ref{fig:HTE_cell}.}
Note that $\sigma_{\rm ele}(T)$ typically decreases as the temperature increases.

\subsection{Coordination Problem}
The steady-state performance of a given HTE cell depends on the electrolysis current density $i_{\rm el}$, the temperature $T$ and the feedstock partial pressures $\tilde{\bm{p}}$ according to \eqref{eq:form_Ucel}-\eqref{eq:form_Uohm}.
The current should adapt to the desired power input; however, the selection of the operating temperature and the feed flow rates at both HTE electrodes needs a strategy, which differs from the polymer electrolyte membrane or alkaline electrolysis where the operating temperature barely varies and the liquid water feed rate has little impact on the reactants' partial pressures \cite{Ursua_DynamicAEC_2012,Carmo_reviewPEMEC_2013}.

\begin{figure}[!t]
  \centering
  \includegraphics[width=0.79\textwidth]{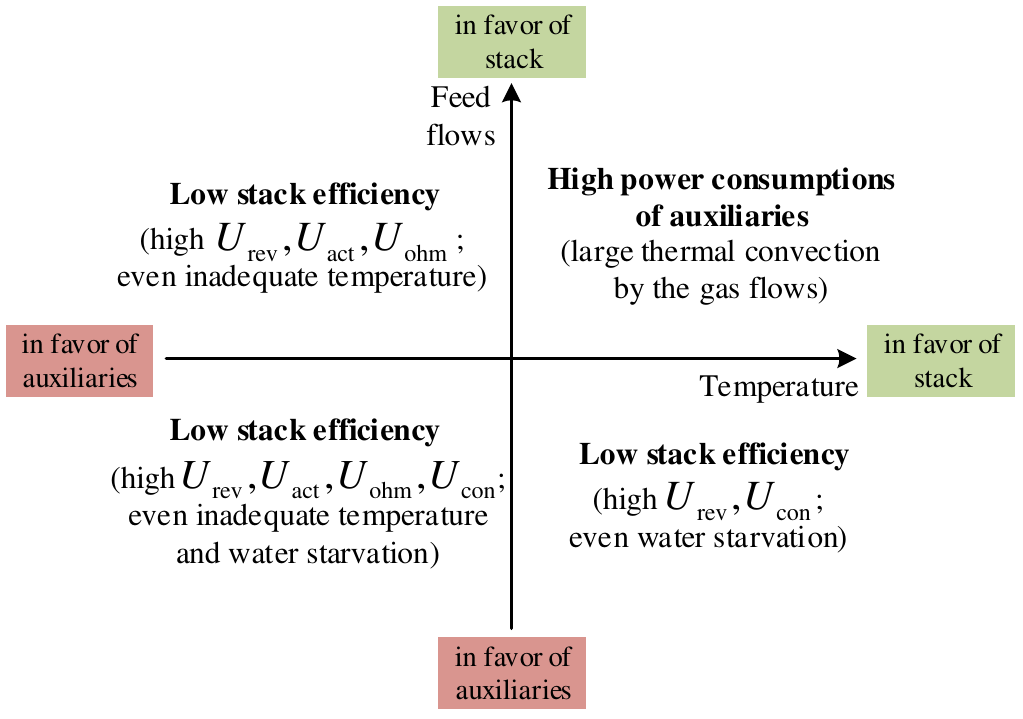}\\
  \caption{Two trade-offs in dimensions of temperature and feed flows.}\label{fig:TradeOff}
\end{figure}

The selection problem of an HTE system's temperature and feed flows is illustrated in Fig. \ref{fig:TradeOff}.
At a higher temperature, the HTE stack will benefit both kinetically (lower $U_{\rm rev}$ as shown in \eqref{eq:form_Urev}) and thermodynamically (lower overvoltages $U_{\rm act}$ and $U_{\rm ohm}$ as shown in \eqref{eq:form_Uact} and \eqref{eq:form_Uohm}), but the power demand for gas heating will also be increased.
This is a trade-off between a better stack performance at high temperature and less power share of auxiliaries at lower temperature.
Meanwhile, larger mass flows of cathode feedstock and anode sweep air will prevent water starvation, lower the oxygen partial pressure and thus improve the stack performance (lower $U_{\rm rev}$ and $U_{\rm con}$ as shown in \eqref{eq:form_Urev} and \eqref{eq:form_Uconc}), while the energy loss due to thermal convection will be simultaneously increased.
Still, it is a trade-off with respect to the feed flows between the stack performance and the auxiliaries' losses.

There has to be an operating state in these trade-offs that optimizes the system hydrogen production.
A comprehensive system model that quantitatively describes the system performance and the operating constraints in terms of temperature and feed flows is required to locate the optimal HTE operating states.

\section{Energy Flow Model of the HTE System}\label{sec:energyFlow}
\begin{figure}[!t]
  \centerline{\includegraphics[width=1.2\textwidth]{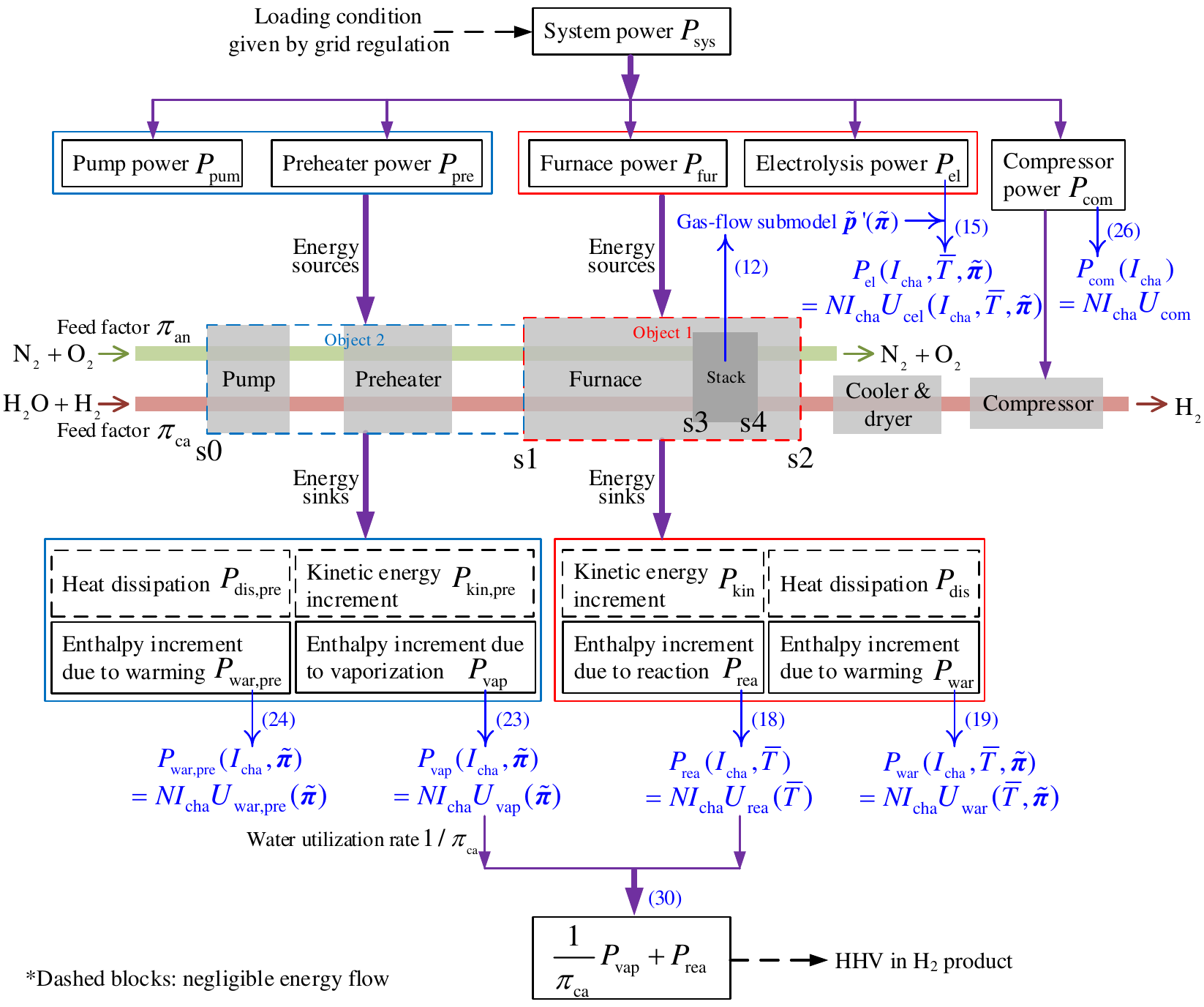}\\}
  \caption{Steady-state energy flow model of the HTE system.}\label{fig:EnergyBalance}
\end{figure}

With the purpose of performing quantitative parameter analysis at the system level, this section proposes an energy flow model that explicitly describes the power sources and sinks of a steady-state operating HTE system and the constraints among them with primary state parameters, including the temperature and the feed flows, as illustrated in Fig. \ref{fig:EnergyBalance}.

\subsection{Methodology}\label{ssec:methodology}
The macroscopic energy balance equation given by \cite{Bird_Transport_2002} is employed as the key principle to develop the energy flow model for the HTE system.

First, regarding the furnace as the object volume (shown as Object 1 in Fig. \ref{fig:EnergyBalance}), the macroscopic energy balance \eqref{eq:energyBalance} can be obtained for an HTE system operating at steady state.
\added[id=R2C1]{Note that $P_{\rm fur}$ and $P_{\rm el}$ on the left side of \eqref{eq:energyBalance} are power inputs of Object 1 via furnace and electrolysis current, respectively, while terms on the right side of \eqref{eq:energyBalance} represent various kinds of energy sinks.}
\begin{equation}\label{eq:energyBalance}
  P_{\rm fur} + P_{\rm el} = \underbrace{( w_{\rm ca,s1} + w_{\rm an,s1} ) \left.\frac{\widehat{v^2}}{2}\right|^{\rm s2}_{\rm s1}}_{\text{$P_{\rm kin}$}} + \underbrace{( w_{\rm ca,s1} + w_{\rm an,s1} ) \left.\widehat{h} ( \tilde{\bm{w}},\tilde{\bm{p}},T )\right|^{\rm s2}_{\rm s1}}_{\text{$P_{\rm rea}+P_{\rm war}$}} + P_{\rm dis}.
\end{equation}
Here, $\widehat{h}(\tilde{\bm{w}},\tilde{\bm{p}},T)$ denotes the specific enthalpy of the mixture flowing in the furnace, which can be calculated with $h_{\rm ca} = {\rm mix_{ca}}(H_T)$ and $h_{\rm an} = {\rm mix_{an}}(H_T)$.
The first term on the right side of \eqref{eq:energyBalance} is $P_{\rm kin}$, the net outflow of kinetic energy.
The second term is the net outflow of enthalpy of the gas mixture, whose enthalpy increase is partially ($P_{\rm rea}$) caused by the chemical reaction of \eqref{eq:overallReact} and partially ($P_{\rm war}$) by gas warming.
$P_{\rm dis}$ denotes the net dissipation of heat by means of conduction and radiation.
The power sources ($P_{\rm fur}$ and $P_{\rm el}$) and sinks ($P_{\rm kin}$, $P_{\rm rea}$, $P_{\rm war}$ and $P_{\rm dis}$) listed above are depicted for Object 1 in \ref{fig:EnergyBalance}.

Second, regarding the preprocessing part as the object volume (shown as Object 2 in Fig. \ref{fig:EnergyBalance}), an energy balance similar to \eqref{eq:energyBalance} can be obtained as \eqref{eq:energyBalance_pre}.
\added[id=R2C1]{Again, $P_{\rm pum}$ and $P_{\rm pre}$ on the left side of \eqref{eq:energyBalance_pre} are power inputs of Object 2 via pump group and preheater, respectively, while terms on the right side of \eqref{eq:energyBalance_pre} represent different energy sinks.}
\begin{equation}\label{eq:energyBalance_pre}
  P_{\rm pum} + P_{\rm pre} = \underbrace{( w_{\rm ca,s1} + w_{\rm an,s1} ) \left.\frac{\widehat{v^2}}{2}\right|^{\rm s1}_{\rm s0}}_{\text{$P_{\rm kin,pre}$}} + \underbrace{( w_{\rm ca,s1} + w_{\rm an,s1} ) \left.\widehat{h}\right|^{\rm s1}_{\rm s0}}_{\text{$P_{\rm vap}+P_{\rm war,pre}$}} + P_{\rm dis,pre}.
\end{equation}
Analogously, \added[id=R2C1]{the energy sinks of Object 2 include kinetic energy increase $P_{\rm kin,pre}$ and non-convective heat dissipation $P_{\rm dis,pre}$}, and the enthalpy increase consists of two parts: the vaporization-induced part ($P_{\rm vap}$) and the warming-induced part ($P_{\rm war,pre}$).
The power sources ($P_{\rm pum}$ and $P_{\rm pre}$) and sinks ($P_{\rm kin,pre}$, $P_{\rm vap}$, $P_{\rm war,pre}$ and $P_{\rm dis,pre}$) listed above are depicted for Object 2 in \ref{fig:EnergyBalance}.

Equations \eqref{eq:energyBalance} and \eqref{eq:energyBalance_pre} present a precise description of the macroscopic energy balances within the HTE system.
Nevertheless, the considered energy flow terms are not explicitly related to the operating parameters, including the temperature and the feed flows.

To present an explicit energy flow model that provides the required efficiency evaluation and operating constraints for later optimization analysis, further formulations and break-downs \eqref{eq:energyBalance} and \eqref{eq:energyBalance_pre} as depicted in Fig. \ref{fig:EnergyBalance} are made in the following subsections.
First, in Section \ref{ssec:gasFlow}, a gas-flow submodel is employed to link the normalized feed flows $\tilde{\bm{\pi}}$ (namely, \emph{feed factors} defined in \eqref{eq:pis}) to the lumped partial pressures $\tilde{\bm{p'}}$, which is required in the subsequent evaluation of $P_{\rm el}$.
Then, in Section \ref{ssec:energyBalance}, based on the aforementioned gas-flow submodel $\tilde{\bm{p'}}(\tilde{\bm{\pi}})$ and necessary approximations, $P_{\rm el}(I_{\rm cha},\bar{T},\tilde{\bm{\pi}})$, $P_{\rm rea}(I_{\rm cha},\bar{T})$ and $P_{\rm war}(I_{\rm cha},\bar{T},\tilde{\bm{\pi}})$ in \eqref{eq:energyBalance} are explicitly formulated.
In Section \ref{ssec:preEnergyBalance}, $P_{\rm vap}(I_{\rm cha},\tilde{\bm{\pi}})$ and $P_{\rm war,pre}(I_{\rm cha},\tilde{\bm{\pi}})$ in \eqref{eq:energyBalance_pre} are analogously formulated.
In addition, the energy flow through the postprocessing compressor $P_{\rm com}(I_{\rm cha})$ is formulated in Section \ref{ssec:compressor}.
Finally, the overall efficiency $\eta_{\rm sys}(I_{\rm cha})$ and the system constraint to meet the desired $P_{\rm sys}$ value are formulated in Section \ref{ssec:efficiency}.

\subsection{Gas-Flow Submodel}\label{ssec:gasFlow}
The reactants' partial pressures $\tilde{\bm{p}}$ have noticeable impacts on $U_{\rm cel}$, as shown in \eqref{eq:form_Ucel}, \eqref{eq:form_Urev} and \eqref{eq:form_Uconc}, especially when a starvation occurs.
Nevertheless, the modeling of $\tilde{\bm{p}}$ is not easy because the components' accumulation rates depend on both the electrolysis current and the inlet gas flows.
Referring to the definition of \emph{air ratio} for high-temperature fuel cells and electrolyzers \cite{Udagawa_SOECoperation_2008}, we introduce \emph{feed factors} $\pi_{\rm ca}$ and $\pi_{\rm an}$ to represent the inlet mass flows of both electrodes of the HTE system:
\begin{subequations}\label{eq:pis}
\begin{align}
\pi_{\rm ca} = \frac{\mbox{\footnotesize ${\rm H_2O}$ provided in cathode stream $w_{\rm ca,s1}$}}{\mbox{\footnotesize ${\rm H_2O}$ consumed by electrolysis at $I_{\rm cha}$}} = \frac{w_{\rm ca,s1} \omega_{\rm H_2O,s1}}{\frac{N I_{\rm cha}}{n_{\rm e}F}M_{\rm H_2O}} \label{eq:pi_ca}\\
\pi_{\rm an} = \frac{\mbox{\footnotesize ${\rm O_2}$ provided in anode stream $w_{\rm an,s1}$}}{\mbox{\footnotesize ${\rm O_2}$ generated by electrolysis at $I_{\rm cha}$}} = \frac{w_{\rm an,s1} \omega_{\rm O_2,s1}}{0.5 \frac{N I_{\rm cha}}{n_{\rm e}F}M_{\rm O_2}} \label{eq:pi_an}
\end{align}
\end{subequations}
\added[id=R2C1]{where $w_{\rm ca,s1}$ (or $w_{\rm an,s1}$) denotes the mass flow rate of cathode (or anode) gas mixture at furnace inlet (position $l_{\rm s1}$), and $\omega_{\rm H_2O,s1}$ (or $\omega_{\rm O_2,s1}$) represents the mass fraction of $\rm H_2O$ in stream $w_{\rm ca,s1}$ (or $\rm O_2$ in stream $w_{\rm an,s1}$).}
$N$ is the total number of channels in the stack and can be calculated by $N = N_{\rm cel} N_{\rm cha}$.
These two feed factors are dimensionless parameters denoting the normalized mass flows.
When $\pi_{\rm ca}=1$ (or $\pi_{\rm an}=1$), the $\rm H_2O$ (or the $\rm O_2$) fed into the system is exactly the amount of $\rm H_2O$ consumption (or $\rm O_2$ production) that occurred in the HTE stack.
Apparently, $\pi_{\rm ca}>1$ and $\pi_{\rm an}\geq0$ must be fulfilled for HTE operation.
$\frac{1}{\pi_{\rm ca}}$ is actually the steam utilization ratio of an operating HTE system.
As mentioned in Section \ref{sssec:HTEstructure}, $\omega_{\rm H_2O, s1}$ and $\omega_{\rm O_2,s1}$ are specific fractions to guarantee a protective gas environment at the cathode; thus, only $w_{\rm ca,s1}$, $w_{\rm an,s1}$ and $I_{\rm cha}$ are variables in \eqref{eq:pis}, and they are correlated by $\pi_{\rm ca}$ and $\pi_{\rm an}$.

\added[id=R2C1]{
Thanks to the information of both current and feed flows included in feed factors $\tilde{\bm{\pi}}$, the detailed gas pressure distribution can be analysed.
In fact, with the inlet composition fixed, the outlet composition ratios of HTE stack are \emph{determined} by $\tilde{\bm{\pi}}$ based on the mass transport continuity of gases (see \eqref{eq:p_H2}-\eqref{eq:p_N2} in \ref{app:p_pi}).
Then the lumped pressure values $\tilde{\bm{p'}}$ at a steady state can be obtained as
}
\begin{subequations}\label{eq:p_p_pi}
\begin{gather}
p_{\rm H_2}'(\pi_{\rm ca}) \simeq p_{\rm ca,s4} \chi_{\rm H_2,s3}^{1-\lambda} \left(\frac{\pi_{\rm ca}\chi_{\rm H_2,s3}+\chi_{\rm H_2O,s3}}{\pi_{\rm ca}}\right)^{\lambda} \label{eq:p_p_H2} \\
p_{\rm H_2O}'(\pi_{\rm ca}) \simeq p_{\rm ca,s4} \chi_{\rm H_2O,s3}^{1-\lambda} \left(\frac{(\pi_{\rm ca}-1)\chi_{\rm H_2O,s3}}{\pi_{\rm ca}}\right)^{\lambda} \label{eq:p_p_H2O} \\
p_{\rm O_2}'(\pi_{\rm an}) \simeq p_{\rm an,s4} \chi_{\rm O_2,s3}^{1-\lambda} \left(\frac{(\pi_{\rm an}+1)\chi_{\rm O_2,s3}}{\pi_{\rm an}+\chi_{\rm O_2,s3}}\right)^{\lambda} \label{eq:p_p_O2} \\
p_{\rm N_2}'(\pi_{\rm an}) \simeq p_{\rm an,s4} \chi_{\rm N_2,s3}^{1-\lambda} \left(\frac{\pi_{\rm an}\chi_{\rm N_2,s3}}{\pi_{\rm an}+\chi_{\rm O_2,s3}}\right)^{\lambda}. \label{eq:p_p_N2}
\end{gather}
\end{subequations}
Note that \added[id=R2C1]{$\chi$ represents the molar fraction of a specific gas,} and $\lambda \in (0,1)$ is a constant selected through a parametric study to calculate the lumped pressure $p'$ from $p_{\rm s3}$ and $p_{\rm s4}$ in the manner of \eqref{eq:p_p_H2_1}.
The above equations \eqref{eq:p_p_pi} present a gas-flow submodel $\tilde{\bm{p'}}(\tilde{\bm{\pi}})$, which provides a quantitative description of the impacts of feedstock flows on the reactants' partial pressures, providing a basis for the evaluation of $U_{\rm cel}$ and thus $P_{\rm el}$ with $\tilde{\bm{\pi}}$ in Section \ref{ssec:energyBalance}.
\added[id=R2C1]{
A detailed derivation of \eqref{eq:p_p_pi} is shown in \ref{app:p_pi}.
}

\subsection{Energy Flows in Furnace (Object 1)}\label{ssec:energyBalance}
$P_{\rm rea}$ and $P_{\rm war}$ in the enthalpy-increment term of \eqref{eq:energyBalance} can be evaluated separately.
According to the definition of $U_{\rm th}$ in \eqref{eq:UrevUth}, the amount of generated chemical energy $P_{\rm rea}$ equals the electric power caused by $U_{\rm th}$, as shown in \eqref{eq:P_rea_1}, which is actually the enthalpy increase due to reaction \eqref{eq:overallReact}:
\begin{subequations}
\begin{align}
P_{\rm rea}&{}=N L_{y,{\rm cha}} \int_{\rm s1}^{\rm s2} i_{\rm el} U_{\rm th}(T) {\rm d}l \label{eq:P_rea_1}\\
&{}\simeq{}N L_{y,{\rm cha}} U_{\rm th}(T_{\rm s2}) \int_{\rm s3}^{\rm s4} i_{\rm el} {\rm d}l \label{eq:P_rea_2}\\
&{}={}N I_{\rm cha} U_{\rm th}(T_{\rm s2}) \label{eq:P_rea_3}
\end{align}
\end{subequations}
\added[id=R2C1]{where $L_{y,{\rm cha}}$ is the channel width as shown in Fig. \ref{fig:HTE_cell}.}
Here, $U_{\rm th}(T) \simeq U_{\rm th}(T_{\rm s2})$ within $[l_{\rm s3},l_{\rm s4}]$ because $\Delta_{\rm r} H_T$ of the vapor electrolysis reaction \eqref{eq:overallReact} varies slightly with temperature (approximately $0.2\%$ every $100^{\circ}{\rm C}$).
\added[id=R1C2]{On the other hand, the enthalpy increase due to warming $P_{\rm war}$ can be formulated with heat capacities and temperature increment according to \cite{Frank_P2G_efficiencies_2018}:}
\begin{equation}
P_{\rm war}=(w_{\rm ca,s1} c_{{\rm ca},T_{\rm H},{\rm s1}} + w_{\rm an,s1} c_{{\rm an},T_{\rm H},{\rm s1}}) (T_{\rm s2} - T_{\rm s1}). \label{eq:P_war_2}
\end{equation}
Note that $c_{{\rm ca},T}={\rm mix_{ca}}(C_T)$ and $c_{{\rm ca},T}={\rm mix_{an}}(C_T)$ \added[id=R2C1]{are isobaric specific heat capacities of cathode and anode mixtures, which can be calculated from gas properties listed in Table \ref{tab:properties}}.
\eqref{eq:P_war_2} can also be obtained by subtracting \eqref{eq:P_rea_3} from the enthalpy-increment term of \eqref{eq:energyBalance}.

For the electrical energy input from the converter, the power $P_{\rm el}$ can be calculated by:
\setlength{\arraycolsep}{0.0em}
\begin{subequations} \label{eq:P_el_sub}
\begin{align}
P_{\rm el}(I_{\rm cha},\bar{T},\tilde{\bm{\pi}})&=N I_{\rm cha} U_{\rm cel}(\bar{i}_{\rm el},\bar{T},\tilde{\bm{p'}}) \label{eq:P_el_1}\\
&= N I_{\rm cha} \cdot U_{\rm cel}(I_{\rm cha},\bar{T},\tilde{\bm{\pi}}). \label{eq:P_el_2}
\end{align}
\end{subequations}
Here, a lumped model based on \eqref{eq:form_Ucel}-\eqref{eq:form_Uohm} is employed to evaluate the cell voltage, which is also used in \cite{Frank_rSOC_op_2018}, \cite{Ni_SOECmodel_2006} and \cite{Xing_HTE_2017} to avoid unnecessary computational costs when the detailed distribution of internal variables is not the concern.
Note that $\bar{i}_{\rm el}$ and $\bar{T}$ are simply spatial average values, while the lumped pressures $\tilde{\bm{p'}}$ are calculated in a geometric mean manner as shown in \eqref{eq:p_p_H2_1}.
\eqref{eq:P_el_2} can be further derived with the help of $\tilde{\bm{p'}}(\tilde{\bm{\pi}})$ formulated in \eqref{eq:p_p_pi} and the proportional relation between $I_{\rm cha}$ and $\bar{i}_{\rm el}$.

On purpose of formulating $P_{\rm rea}$ in \eqref{eq:P_rea_3} and $P_{\rm war}$ in \eqref{eq:P_war_2} with $I_{\rm cha}$, $\bar{T}$ and $\tilde{\bm{\pi}}$ explicitly like $P_{\rm el}$ in \eqref{eq:P_el_2}, the mass flows $w_{\rm ca,s1}$ and $w_{\rm an,s1}$ can be rewritten with $I_{\rm cha}$ and $\tilde{\bm{\pi}}$ based on definition \eqref{eq:pis}:
\begin{subequations}\label{eq:ws_pis}
\begin{gather}
  w_{\rm ca,s1} = \frac{M_{\rm H_2O}}{n_{\rm e} F \omega_{\rm H_2O,s1}} \cdot \pi_{\rm ca} \cdot N I_{\rm cha} \label{eq:w_ca}\\
  w_{\rm an,s1} = \frac{0.5 M_{\rm O_2}}{n_{\rm e} F \omega_{\rm O_2,s1}} \cdot \pi_{\rm an} \cdot N I_{\rm cha}. \label{eq:w_an}
\end{gather}
\end{subequations}
At the same time, an experimental relationship $T_{\rm s2}(\bar{T})$ is employed to substitute $T_{\rm s2}$ in \eqref{eq:P_rea_3} and \eqref{eq:P_war_2} with $\bar{T}$.
In fact, $T_{\rm s2}$ is strongly correlated with $\bar{T}$ in a practical condition with constrained stack temperatures and temperature gradients such that $T_{\rm s2}(\bar{T})$ can be approximately evaluated by
\begin{equation}\label{eq:T_eval}
T_{\rm s2}(\bar{T}) \simeq k \bar{T}
\end{equation}
where $k$ is a constant that is slightly greater than $1$, as estimated by measurements.
Now, we can substitute \eqref{eq:ws_pis} and \eqref{eq:T_eval} into \eqref{eq:P_rea_3} and \eqref{eq:P_war_2} to obtain $P_{\rm rea}(I_{\rm cha},\bar{T})$ and $P_{\rm war}(I_{\rm cha},\bar{T},\tilde{\bm{\pi}})$, respectively:
\begin{subequations} \label{eq:P_rea_final_sub}
\begin{align}
P_{\rm rea}(I_{\rm cha},\bar{T})&=N I_{\rm cha} \cdot U_{\rm th}(k \bar{T}) \label{eq:P_rea_final_1}\\
&{}\stackrel{\mbox{\tiny def}}{=}N I_{\rm cha} \cdot U_{\rm rea}(\bar{T}). \label{eq:P_rea_final}
\end{align}
\end{subequations}
\begin{subequations} \label{eq:P_war_final_sub}
\begin{align}
P_{\rm war}(I_{\rm cha},\bar{T},\tilde{\bm{\pi}})&{}\simeq N I_{\rm cha}(\!\frac{M_{\rm H_2O} c_{{\rm ca},T_{\rm H},{\rm s1}}}{n_{\rm e} F \omega_{\rm H_2O,s1}}\pi_{\rm ca} \!\!+\!\! \frac{0.5 M_{\rm O_2}c_{{\rm an},T_{\rm H},{\rm s1}}}{n_{\rm e} F \omega_{\rm O_2,s1}}\!) (k \bar{T} - T_{\rm s1}) \label{eq:P_war_final_1}\\
&{}\stackrel{\mbox{\tiny def}}{=}N I_{\rm cha} \cdot U_{\rm war}(\bar{T},\tilde{\bm{\pi}}). \label{eq:P_war_final}
\end{align}
\end{subequations}
Note that $U_{\rm rea}(\bar{T})$ and $U_{\rm war}(\bar{T},\tilde{\bm{\pi}})$ are equivalent voltages defined to represent $P_{\rm rea}$ and $P_{\rm war}$, respectively, in a manner of electric power like \eqref{eq:P_el_2}.
\added[id=R2C1]{The explicit formulations of functions $U_{\rm rea}(\bar{T})$ and $U_{\rm war}(\bar{T},\tilde{\bm{\pi}})$ are already included in \eqref{eq:P_rea_final_sub} and \eqref{eq:P_war_final_sub}, respectively.}
The furnace inlet temperature $T_{\rm s1}$ is treated as a fixed command set for the preheater.
The composition ratios at both cathode and anode inlets, i.e., $\omega_{\rm H_2O,s1}$ and $\omega_{\rm O_2,s1}$, are also constant parameters in this paper.

In practical HTE operation, the mixture velocities $v_{\rm ca},v_{\rm an}$ barely change within the furnace.
$P_{\rm dis}$ is also relatively negligible with good lagging casing, a polished material (emissivity lower than 0.1), and an operating temperature under $1000^{\circ}{\rm C}$.
Thus, it can be obtained from \eqref{eq:energyBalance} that
\begin{equation}\label{eq:P_war_balance}
  P_{\rm fur} + \underbrace{N I_{\rm cha} U_{\rm cel}(I_{\rm cha},\bar{T},\tilde{\bm{\pi}})}_{\text{$P_{\rm el}$}} - \underbrace{N I_{\rm cha} U_{\rm rea}(\bar{T})}_{\text{$P_{\rm rea}$}} \simeq \underbrace{N I_{\rm cha} U_{\rm war}(\bar{T},\tilde{\bm{\pi}})}_{\text{$P_{\rm war}$}}
\end{equation}
This balance equation \eqref{eq:P_war_balance} provides us a close look at the energy flow within Object 1, where the furnace power $P_{\rm fur}$ and joule heating $(P_{\rm el}-P_{\rm rea})$ of the HTE stack together contribute to the enthalpy increase of the inlet mixture due to gas warming.

\subsection{Energy Flows in Preprocess (Object 2)}\label{ssec:preEnergyBalance}
Similar to Section \ref{ssec:energyBalance}, the combined enthalpy increase caused by vaporization and prewarming in \eqref{eq:energyBalance_pre} of Object 2 can be decomposed into $P_{\rm vap}$ and $P_{\rm war,pre}$.
\added[id=R1C2]{They can be formulated by \eqref{eq:P_vap} and \eqref{eq:P_war_pre} according to \cite{Frank_P2G_efficiencies_2018}:}
\begin{equation}\label{eq:P_vap}
  P_{\rm vap} = w_{\rm ca,s1} \omega_{\rm H_2O,s1} h_{\rm vap}
\end{equation}
\begin{equation}\label{eq:P_war_pre}
  P_{\rm war,pre} = w_{\rm ca,s1} \Big( c_{{\rm ca(liq)},T_{\rm L},{\rm s1}} \left. T \right|^{\rm vap}_{\rm s0} + c_{{\rm ca},T_{\rm L},{\rm s1}} \left. T \right|^{\rm s1}_{\rm vap} \Big)+ w_{\rm an,s1} \!c_{{\rm an},T_{\rm L},{\rm s1}} \left. T \right|^{\rm s1}_{\rm s0}
\end{equation}
\added[id=R2C1]{where $h_{\rm vap}$ is the vaporization enthalpy of $\rm H_2O$, and the specific heat capacity of liquid water listed in Table \ref{tab:properties} is included in $c_{{\rm ca(liq)},T_{\rm L}}$.}
Again, $P_{\rm vap}(I_{\rm cha},\tilde{\bm{\pi}})$ and $P_{\rm war,pre}(I_{\rm cha},\tilde{\bm{\pi}})$ can be further formulated based on \eqref{eq:ws_pis}:
\begin{subequations}\label{eq:P_vap_final_sub}
\begin{align}
P_{\rm vap}(I_{\rm cha},\tilde{\bm{\pi}})&=N I_{\rm cha} \frac{M_{\rm H_2O} \omega_{\rm H_2O,s1} h_{\rm vap}}{n_{\rm e} F \omega_{\rm H_2O,s1}}\pi_{\rm ca} \label{eq:P_vap_final_1}\\
&{}\stackrel{\mbox{\tiny def}}{=}N I_{\rm cha} \cdot U_{\rm vap}(\tilde{\bm{\pi}}). \label{eq:P_vap_final}
\end{align}
\end{subequations}
\begin{subequations}\label{eq:P_war_pre_final_sub}
\begin{align}
&P_{\rm war,pre}(I_{\rm cha},\!\tilde{\bm{\pi}}) \nonumber\\
&=N I_{\rm cha}\!\! \left[\!\frac{M_{\rm H_2O} (\! c_{{\rm ca(liq)}\!,T_{\rm L}\!,{\rm s1}} \left. \!\!T \right|^{\rm vap}_{\rm s0} \!+\! c_{{\rm ca}\!,T_{\rm L}\!,{\rm s1}} \left.\!\! T \right|^{\rm s1}_{\rm vap} \!)}{n_{\rm e} F \omega_{\rm H_2O,s1}}\pi_{\rm ca} \!\!+\!\! \frac{M_{\rm O_2} c_{{\rm an}\!,T_{\rm L}\!,{\rm s1}} \left.\!\! T \right|^{\rm s1}_{\rm s0}}{2 n_{\rm e} F \omega_{\rm O_2,s1}}\pi_{\rm an}\!\right] \label{eq:P_war_pre_final_1}\\
&{}\stackrel{\mbox{\tiny def}}{=}N I_{\rm cha} \cdot U_{\rm war,pre}(\tilde{\bm{\pi}}). \label{eq:P_war_pre_final}
\end{align}
\end{subequations}
Note that $U_{\rm vap}(\tilde{\bm{\pi}})$ and $ U_{\rm war,pre}(\tilde{\bm{\pi}})$ are equivalent voltages defined to represent $P_{\rm vap}$ and $P_{\rm war,pre}$, respectively.
\added[id=R2C1]{The explicit formulations of these two functions are given by \eqref{eq:P_vap_final_sub} and \eqref{eq:P_war_pre_final_sub}.}

Again, $P_{\rm kin,pre}$ and $P_{\rm dis,pre}$ are generally negligible for such a substantial warming process in a substantial flow rate.
In fact, in \eqref{eq:energyBalance_pre}, $\left.\frac{\widehat{v^2}}{2}\right|^{\rm s1}_{\rm s0}$ can only reach the magnitude of $\left.\widehat{h}\right|^{\rm s1}_{\rm s0}$ with gas velocities of several hundreds of meters per second, which are considerably higher than the practical velocities within an operating stack.
Therefore, the energy flow within the preprocess can be written as
\begin{equation}\label{eq:P_pre_balance}
  P_{\rm pum} + P_{\rm pre} \simeq \underbrace{N I_{\rm cha} U_{\rm vap}(\tilde{\bm{\pi}})}_{\text{$P_{\rm vap}$}} + \underbrace{N I_{\rm cha} U_{\rm war,pre}(\tilde{\bm{\pi}})}_{\text{$P_{\rm war,pre}$}}.
\end{equation}
In other words, the power input of the pump group and the preheater mostly contributes to the feedstock warming, including the vaporization of water.

\subsection{Energy Flow of Compressor}\label{ssec:compressor}
According to \cite{Meier_PEMandSOEC_2014}, the power consumption of a compressor is generally evaluated with an isothermal compression calculation considering its isothermal efficiency $\eta_{\rm com}$ as shown in \eqref{eq:P_com_1}:
\begin{subequations}\label{eq:P_com}
\begin{align}
  P_{\rm com}(I_{\rm cha}) &= \frac{N I_{\rm cha}}{n_{\rm e} F} \frac{R T_{\rm com} \ln(\frac{p_{\rm com}}{p_{\rm ca,s2}})}{\eta_{\rm com}} \label{eq:P_com_1}\\
  &{}\stackrel{\mbox{\tiny def}}{=}{}N I_{\rm cha} \cdot U_{\rm com} \label{eq:P_com_2}
\end{align}
\end{subequations}
where $T_{\rm com}$ is the compressor's inlet temperature and $p_{\rm com}$ is the outlet pressure of hydrogen product.
Apparently, $P_{\rm com}$ is proportional to the logarithm of the pressure ratio $\frac{p_{\rm com}}{p_{\rm ca,s2}}$ and the flow rate of the gas being compressed, which is exactly the net production rate of hydrogen by electrolysis $\frac{N I_{\rm cha}}{n_{\rm e} F}$.
Again, an equivalent voltage $U_{\rm com}$ can be introduced to represent $P_{\rm com}$ as shown in \eqref{eq:P_com_2}.

For the compressor sector, the energy sinks such as enthalpy increase and heat dissipation are not discussed because neither of them contribute to the effective product (HHV of $\rm H_2$ as mentioned later in Section \ref{ssec:efficiency}) of the HTE system.
These sinks are also not included in Fig. \ref{fig:EnergyBalance}.

\subsection{System Efficiency}\label{ssec:efficiency}
In total, the total power consumption \added[id=R2C1]{$P_{\rm sys}$} of such an HTE system is
\begin{subequations}
\begin{align}
P_{\rm sys}&{}={}P_{\rm pum} + P_{\rm pre} + P_{\rm fur} + P_{\rm el} + P_{\rm com} \label{eq:P_sys_source}\\
&{}\simeq{}P_{\rm war,pre} + P_{\rm vap} + P_{\rm rea} + P_{\rm war} + P_{\rm com} \label{eq:P_sys_sink}
\end{align}
\end{subequations}
based on the energy flows described in \eqref{eq:P_war_balance} and \eqref{eq:P_pre_balance}.

Regarding the system efficiency for energy conversion, the higher heating value (HHV) of $\rm H_2$ combustion, 286{\rm kJ/mol}, is generally employed for evaluating yield \added[id=R1C2]{according to \cite{Frank_P2G_efficiencies_2018} and \cite{Petipas_HTE_VariousLoads_2013}}.
Actually HHV is defined as the enthalpy decrease of the hydrogen combustion reaction at the ambient temperature (\added[id=R2C1]{$T_{\rm amb} = 25{\rm ^{\circ}C}$}, to ensure the liquid state of the resulting water), which equals the enthalpy increase $\Delta_{r} H_{T_{\rm amb}}$ of its reverse reaction \eqref{eq:overallReact}.
Considering the definition of $U_{\rm th}$ in \eqref{eq:UrevUth} simultaneously, the system efficiency $\eta_{\rm sys}$ can be formulated as
\begin{equation}\label{eq:efficiency_HHV}
  \eta_{\rm sys} = \frac{\left.w_{\rm H_2}\right|^{\rm s2}_{\rm s0} \cdot \mbox{HHV}_{\rm H_2}}{M_{\rm H_2} P_{\rm sys}} = \frac{\left.w_{\rm H_2}\right|^{\rm s2}_{\rm s0} \cdot \Delta_{r} H_{T_{\rm amb}}}{M_{\rm H_2} P_{\rm sys}} = \frac{N I_{\rm cha} U_{\rm th}(T_{\rm amb})}{P_{\rm sys}}
\end{equation}
\added[id=R2C1]{where $\left.w_{\rm H_2}\right|^{\rm s2}_{\rm s0}=w_{\rm H_2,s2}-w_{\rm H_2,s0}$ represents the net $\rm H_2$ production (in mass rate) of the HTE system.}

$U_{\rm th}(T_{\rm amb})$ (approximately $1.48{\rm V}$) in \eqref{eq:efficiency_HHV} is precise enough to evaluate the system efficiency.
However, \eqref{eq:efficiency_HHV} does not explain the further disposition of various energy flows.
We would have a better understanding of the energy conversion of an HTE system if we can determine which of the energy flows mentioned in \eqref{eq:P_sys_sink} ultimately transform into the hydrogen product.

In fact, as  mentioned for \eqref{eq:P_rea_2}, the temperature dependence of $\Delta_{r} H_{T}$ is quite weak, except that the vaporization enthalpy $h_{\rm vap}$ must be compensated:
\begin{equation}\label{eq:deltaH_decom}
  \stackrel{286}{\Delta_{r} H_{T_{\rm amb}}} \simeq \stackrel{[248.1,248.6]}{\Delta_{r} H_{T_{\rm s2}}} + \stackrel{40.66\quad({\rm unit: \,\, kJ/mol})}{h_{\rm vap}\quad\quad\quad\quad\quad}
\end{equation}
Note that the value of $\Delta_{r} H_{T_{\rm s2}}$ is labeled within the temperature interval of $T_{\rm s2}\in[750^{\circ}{\rm C},850^{\circ}{\rm C}]$. Therefore, the system efficiency can be reformulated as
\begin{equation}\label{eq:efficiency_sink}
  \eta_{\rm sys} \simeq \frac{\left.w_{\rm H_2}\right|^{\rm s2}_{\rm s0} \cdot (\Delta_{r} H_{T_{\rm s2}} + h_{\rm vap})}{M_{\rm H_2} P_{\rm sys}} = \frac{ P_{\rm rea} + \frac{1}{\pi_{\rm ca}}P_{\rm vap}}{P_{\rm sys}}
\end{equation}
with the help of \eqref{eq:pi_ca}, \eqref{eq:P_rea_3} and \eqref{eq:P_vap}.

By comparing \eqref{eq:efficiency_sink} and \eqref{eq:P_sys_sink}, it can be observed that $P_{\rm rea}$ and most of $P_{\rm vap}$ together make up to the energy product among the various types of energy flow sinks.
In other words, the power loss during system operation is caused mainly by the warming from $T_{\rm s0}$ to $T_{\rm s2}$ ($P_{\rm war,pre}+P_{\rm war}$) of feedstock, whose heat will dissipate soon (in the cooler) after departing the furnace; the vaporization of excess water ($\frac{\pi_{\rm ca}-1}{\pi_{\rm ca}}P_{\rm vap}$); and the electric power required by the compressor ($P_{\rm com}$).
For generalization, heat recovery is not considered in this paper.

\section{Optimization of Hydrogen Yield of the HTE system}\label{sec:optimization}
Based on the above energy flow model, this section proposes a steady-state operation model of the HTE system to discuss locating the optimal operating states that maximize the hydrogen production with coordinated parameters.
This optimization is supposed to play an important part in the practical system control strategy.

\subsection{Operation Model of the HTE system}\label{ssec:OPmodel}
\added[id=R2C1]{According to the energy flow model Fig. \ref{fig:EnergyBalance} presented in Section \ref{sec:energyFlow}, the energy flows $P_{\rm el}$, $P_{\rm rea}$, $P_{\rm war}$, $P_{\rm vap}$ and $P_{\rm war,pre}$ of an operating HTE system can be represented by corresponding (equivalent) voltages $U_{\rm el}$, $U_{\rm rea}$, $U_{\rm war}$, $U_{\rm vap}$ and $U_{\rm war,pre}$, respectively, which can in turn be formulated as explicit functions of operating parameters $(I_{\rm cha},\bar{T},\tilde{\bm{\pi}})$ as illustrated in \eqref{eq:P_el_sub}, \eqref{eq:P_rea_final_sub}, \eqref{eq:P_war_final_sub}, \eqref{eq:P_vap_final_sub} and \eqref{eq:P_war_pre_final_sub}.}
In fact, the following two operation constraints \eqref{eq:Ueq_fur} and \eqref{eq:Ueq_sys} for Object 1 and the whole system (including Object 1, Object 2 and the compressor), respectively, can be obtained if we divide both sides of \eqref{eq:P_war_balance} and \eqref{eq:P_sys_sink} by $N I_{\rm cha}$:
\begin{equation}\label{eq:Ueq_fur}
U_{\rm fur} + U_{\rm cel}(I_{\rm cha},\bar{T},\tilde{\bm{\pi}}) - U_{\rm rea}(\bar{T}) = U_{\rm war}(\bar{T},\tilde{\bm{\pi}})
\end{equation}
\begin{equation}\label{eq:Ueq_sys}
U_{\rm rea}(\bar{T}) + U_{\rm war}(\bar{T},\tilde{\bm{\pi}}) + U_{\rm war,pre}(\tilde{\bm{\pi}}) + U_{\rm vap}(\tilde{\bm{\pi}}) + U_{\rm com} = \frac{P_{\rm sys}}{NI_{\rm cha}}
\end{equation}
where $U_{\rm fur}$ is an equivalent voltage defined as:
\begin{equation}\label{eq:U_fur}
U_{\rm fur} \stackrel{\mbox{\tiny def}}{=} \frac{P_{\rm fur}}{N I_{\rm cha}}
\end{equation}
The equivalent voltages above are more intuitive than the power rates because they are normalized values regardless of the electrolysis load and capacity.

To ensure that no stack damage will be caused by inadequate temperature or an excessive temperature gradient, the following temperature constraints are necessarily set:
\begin{equation}\label{eq:constr_Ts3}
T_{\rm s3} \geq T_{\rm min}
\end{equation}
\begin{equation}\label{eq:constr_Tbar}
T_{\rm min} \leq \bar{T} \leq T_{\rm min} + \frac{L_x}{2} ({\rm d}T/{\rm d}l)_{\rm max}
\end{equation}
where $T_{\rm min}$ is the minimum local temperature for HTE operation and $({\rm d}T/{\rm d}l)_{\rm max}$ is the maximum acceptable temperature gradient \cite{Cai_SOECoperation_2014}.
In fact, given specific feed flows $\tilde{\bm{\pi}}$, the stack inlet temperature $T_{\rm s3}$ is barely impacted by the electrolysis reaction occurring behind but largely determined by the furnace output $U_{\rm fur}$.
Therefore, a minimum furnace output $U_{\rm fur,min}$ can be measured according to \eqref{eq:constr_Ts3} at various values of $\tilde{\bm{\pi}}$:
\begin{equation}\label{eq:constr_Ufur}
U_{\rm fur} \geq U_{\rm fur,min}(\tilde{\bm{\pi}})
\end{equation}
where the measured $U_{\rm fur,min}(\tilde{\bm{\pi}})$ can also be theoretically derived from $T_{\rm min}$ in a similar manner as $U_{\rm war,pre}(\tilde{\bm{\pi}})$ in \eqref{eq:P_war_pre_final_sub}.
In other words, the presence of a furnace is necessary for superheating the streams (to at least $T_{\rm min}$) before stack inlets.
Without the extra electrical heating provided by furnace, $\bar{T}$ can be high enough if HTE operates at significantly exothermic states, but $T_{\rm s3}$ will still be inadequate and the temperature gradient will be too large.
Then, we can use \eqref{eq:Ueq_fur} and \eqref{eq:constr_Ufur} together to replace \eqref{eq:constr_Ts3}.

According to \eqref{eq:efficiency_HHV}, \eqref{eq:Ueq_fur}, \eqref{eq:Ueq_sys}, \eqref{eq:constr_Tbar} and \eqref{eq:constr_Ufur}, the following operation model \eqref{eq:OP_PtG} can be obtained:
\begin{subequations}\label{eq:OP_PtG}
\begin{gather}
\max_{I_{\rm cha},\bar{T},\tilde{\bm{\pi}}} \frac{N I_{\rm cha} U_{\rm th}(T_{\rm amb})}{P_{\rm sys}} \label{eq:OP_PtG_J}\\
{\rm s.t.} \quad U_{\rm rea}(\bar{T}) \!\!+\! U_{\rm war}(\bar{T},\!\tilde{\bm{\pi}}) \!\!+\! U_{\rm war,pre}(\tilde{\bm{\pi}}) \!\!+\! U_{\rm vap}(\tilde{\bm{\pi}}) \!\!+\! U_{\rm com} \!\!=\!\! \frac{P_{\rm sys}}{NI_{\rm cha}} \label{eq:OP_PtG_Ueq}\\
U_{\rm rea}(\bar{T}) + U_{\rm war}(\bar{T},\tilde{\bm{\pi}}) \geq U_{\rm cel}(I_{\rm cha},\bar{T},\tilde{\bm{\pi}}) + U_{\rm fur,min}(\tilde{\bm{\pi}}) \label{eq:OP_PtG_U}\\
I_{\rm cha,min} \leq I_{\rm cha} \leq I_{\rm cha,max} \label{eq:OP_PtG_I}\\
\bar{T}_{\rm min} \leq \bar{T} \leq \bar{T}_{\rm max} \label{eq:OP_PtG_T}\\
\pi_{\rm ca} > 1, \pi_{\rm an} \geq 0. \label{eq:OP_PtG_pi}
\end{gather}
\end{subequations}
Note that $\bar{T}_{\rm min}$ and $\bar{T}_{\rm max}$ in \eqref{eq:OP_PtG_T} are defined as \eqref{eq:T_limits} according to \eqref{eq:constr_Tbar}:
\begin{subequations}\label{eq:T_limits}
\begin{gather}
\bar{T}_{\rm min} \stackrel{\mbox{\tiny def}}{=} T_{\rm min} \label{eq:def_T_min}\\
\bar{T}_{\rm max} \stackrel{\mbox{\tiny def}}{=} T_{\rm min} + \frac{L_x}{2} ({\rm d}T/{\rm d}l)_{\rm max}. \label{eq:def_T_max}
\end{gather}
\end{subequations}
This nonlinear programming model \eqref{eq:OP_PtG} can be solved numerically using the interior-point algorithm, \added[id=R2C6]{which proves to be accurate and efficient for nonconvex nonlinear programming with arbitrary explicit objective and constraint formulations as long as they are twice continuously differentiable according to \cite{Vanderbei_interior_1999}.}
The solutions $I_{\rm cha}^*$, $\bar{T}^*$, and $\tilde{\bm{\pi}}^*$ represent the suggested setpoints of the power converter, the electric furnace, and the pump groups (or mass flow controllers), respectively.
By implementing these solutions in practical operation to coordinate the condition parameters $\bar{T}$, $\tilde{\bm{\pi}}$ and $I_{\rm cha}$, the maximum hydrogen yield of the HTE system can be achieved.

\subsection{Submaximum Production Point (SMPP): Solving the Temperature's Trade-Off}\label{ssec:SMPP}
Although the aforementioned operation model \eqref{eq:OP_PtG} is adequate for solving the trade-offs in Fig. \ref{fig:TradeOff} numerically, some discussions to understand the optimization intuitively from a physical perspective are worth conducting.

For this subsection, let us consider only the optimization of $I_{\rm cha}$ and $\bar{T}$, assuming fixed $\tilde{\bm{\pi}}$, which leads to the submaximum production point (SMPP).
Based on the energy balance principle, we intersect a source curve and a sink curve to study the steady-state operating point and its parameter sensitivity in the following analyses.

\begin{figure}[!t]
  \centerline{\includegraphics[width=1.49\textwidth]{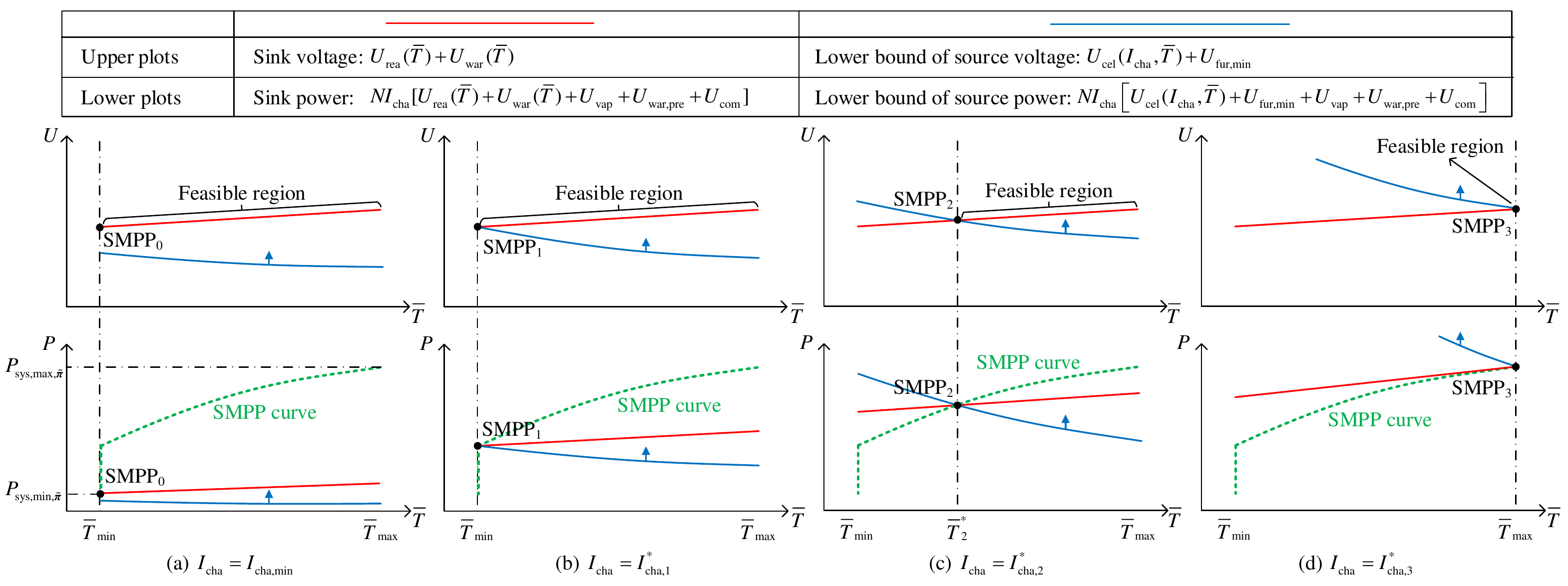}\\}
  \caption{The SMPP curve based on $P$-$\bar{T}$ and $U$-$\bar{T}$ plots with $\tilde{\bm{\pi}}$ fixed.}\label{fig:IT_opt}
\end{figure}

The constraint \eqref{eq:OP_PtG_U} can be illustrated in $U$-$\bar{T}$ plots, as shown in the upper plots of Fig. \ref{fig:IT_opt}.
Apparently, with $\tilde{\bm{\pi}}$ fixed, the sink voltage $[U_{\rm rea}(\bar{T})+U_{\rm war}(\bar{T})]$ is a fixed single curve (depicted as red lines in the upper half of Fig. \ref{fig:IT_opt}) in a $U$-$\bar{T}$ plot, which is upward sloping because a higher $\bar{T}$ barely influences $U_{\rm rea}(\bar{T})$ (the enthalpy increase of reaction \eqref{eq:overallReact}) but increases the value of $U_{\rm war}(\bar{T})$ (the conventional heat transfer rate).
Meanwhile, the source voltage $[U_{\rm cel}(I_{\rm cha},\bar{T})+U_{\rm fur}]$ is a family of curves with a lower bound $[U_{\rm cel}(I_{\rm cha},\bar{T})+U_{\rm fur,min}]$ (depicted as blue lines in the upper half of Fig. \ref{fig:IT_opt}) determined by $I_{\rm cha}$, which has a downward trend because $U_{\rm cel}(I_{\rm cha},\bar{T})$ decreases under the improved electrolysis condition as $\bar{T}$ increases.
In the feasible regions labeled in Fig. \ref{fig:IT_opt}, practical steady-state operating points can be obtained by intersecting the curves of $[U_{\rm cel}(I_{\rm cha},\bar{T})+U_{\rm fur}]$ and $[U_{\rm th}(\bar{T})+U_{\rm war}(\bar{T})]$.
Outside the feasible regions, these two curves clearly cannot intersect with the other one without violating \eqref{eq:OP_PtG_U}.

Based on \eqref{eq:OP_PtG_Ueq}, we can calculate the value of $P_{\rm sys}$ using the intersection point obtained with the aforementioned $U$-$\bar{T}$ curves to locate a corresponding intersection point in $P$-$\bar{T}$ plots, as shown in the lower plots of Fig. \ref{fig:IT_opt}.
Note that the curve trends of the source power (depicted as blue lines in the lower half of Fig. \ref{fig:IT_opt}) and sink power (depicted as red lines in the lower half of Fig. \ref{fig:IT_opt}) are inherited from the corresponding voltage curves.
However, the absolute values of the $P$-$\bar{T}$ curves vary almost proportionally with $I_{\rm cha}$, which is different from the normalized voltages in the $U$-$\bar{T}$ plots.
The vertical coordinates of crossing points of the $P$-$\bar{T}$ curves provide the actual loading $P_{\rm sys}$ of the HTE system.

A combined analysis to locate the SMPP with optimized $\bar{T}$ at 4 different electrolysis current values ($I_{\rm cha,min}$, $I_{\rm cha,1}^*$, $I_{\rm cha,2}^*$ and $I_{\rm cha,3}^*$, in increasing order) is depicted in Fig. \ref{fig:IT_opt} (a), (b), (c) and (d), respectively.
Apparently, the desired SMPP (e.g., SMPP$_{\rm 0}$, SMPP$_{\rm 1}$, SMPP$_{\rm 2}$ and SMPP$_{\rm 3}$ in Fig. \ref{fig:IT_opt}) is always the left end point of the current feasible region, where $P_{\rm sys}$ reaches its minimum while the hydrogen yield rate (determined by $I_{\rm cha}$) remains unchanged.
This result makes sense in practice: With the energy sink of hydrogen production given by fixed $I_{\rm cha}$, a higher temperature within the feasible interval will only increase the convectional heat loss and thus the overall energy sink.

The SMPP curve in Fig. \ref{fig:IT_opt} is obtained by tracking the SMPPs at various system loading conditions.
It is clearly a two-piece curve divided by SMPP$_{\rm 1}$, namely, the head piece and the middle piece.
From this SMPP curve it can be seen that a lower temperature is preferred at low loading conditions -- similar results were obtained by simulation in \cite{Wang_P2M_opt_2018}.
We can conclude that $U_{\rm fur}^*$ at an MPP is always kept at $U_{\rm fur,min}$ to ensure the optimized $\bar{T}^*$ at the middle-piece loading conditions.
For the head piece (with $I_{\rm cha}^* \in [I_{\rm cha,min},I_{\rm cha,1}^*)$), however, additional furnace output (with $U_{\rm fur}^*>U_{\rm fur,min}$) is required to maintain $\bar{T}^*$ at $\bar{T}_{\rm min}$.
The optimality conditions used to locate SMPPs are summarized in Table \ref{tab:conditions}.
\begin{table}[!t]
  \renewcommand{\arraystretch}{1.3}
  \caption{Optimality conditions to locate $(I_{\rm cha}^*, \bar{T}^*, \tilde{\bm{\pi}}^*)$ at SMPPs and MPPs}\label{tab:conditions}
  \centering
  \begin{threeparttable}
  \begin{tabular}{ccc}\toprule
  Piece &SMPP curve &MPP curve \\ \midrule
  Head &$\bar{T}=\bar{T}_{\rm min},\pi_{\rm ca} = \text{given constant}$ &$\bar{T}=\bar{T}_{\rm min},U_{\rm fur}=U_{\rm fur,min}$ \\
  \!\!Middle\!\! &\!\!\!\!$U_{\rm fur}\!\!=\!\!U_{\rm fur,min},\!\pi_{\rm ca} \!\!=\!\! \text{given constant}$\!\!\!\! &\eqref{eq:pi_ca_opt}, $U_{\rm fur}=U_{\rm fur,min}$ \\
  Tail &(No tails) &$\bar{T}=\bar{T}_{\rm max},U_{\rm fur}=U_{\rm fur,min}$ \\ \bottomrule
  \end{tabular}
  \begin{tablenotes}
  \footnotesize
  \item{*} In most cases, $\pi_{\rm an}$ is fixed at 0 because \eqref{eq:pi_an_opt} has no solution.
  \end{tablenotes}
  \end{threeparttable}
\end{table}

The projection of the entire SMPP curve on the $P$ axis provides the acceptable range of system loading $P_{\rm sys}$, which can be calculated by
\begin{subequations}
\begin{align}
 \frac{P_{{\rm sys,min},\tilde{\bm{\pi}}}}{ N I_{\rm cha,min}} \!&=\! U_{\rm rea}(\bar{T}_{\rm min}) + U_{\rm war}(\bar{T}_{\rm min},\tilde{\bm{\pi}}) + U_{\rm war,pre}(\tilde{\bm{\pi}}) + U_{\rm vap}(\tilde{\bm{\pi}}) + U_{\rm com} \label{eq:P_min_pi}\\
 \frac{P_{{\rm sys,max},\tilde{\bm{\pi}}}}{N I_{\rm cha,3}^*} \!&=\! U_{\rm rea}(\bar{T}_{\rm max}) + U_{\rm war}(\bar{T}_{\rm max},\tilde{\bm{\pi}}) + U_{\rm war,pre}(\tilde{\bm{\pi}}) + U_{\rm vap}(\tilde{\bm{\pi}})+U_{\rm com} \label{eq:P_max_pi}
\end{align}
\end{subequations}
as depicted in Fig. \ref{fig:IT_opt}.
If $P_{\rm sys}<P_{{\rm sys,min},\tilde{\bm{\pi}}}$, then the power input will be insufficient even for electrolysis at $I_{\rm cha,min}$ and $\bar{T}_{\rm min}$.
If $P_{\rm sys}>P_{{\rm sys,max},\tilde{\bm{\pi}}}$, then the power input will be too high to prevent the violation of $\bar{T}_{\rm max}$ at steady state.
Within $\left[P_{{\rm sys,min},\tilde{\bm{\pi}}},P_{{\rm sys,max},\tilde{\bm{\pi}}}\right]$, however, if a desired loading $P_{\rm sys}$ is given by grid regulation, which may match the renewable generation to avoid curtailments, then an SMPP with a coordinated $\bar{T}^*$ can be obtained via the SMPP curve in Fig. \ref{fig:IT_opt}, solving the trade-off problem of temperature in Fig. \ref{fig:TradeOff}.

\subsection{Maximum Production Point (MPP): Solving the Feed Factors' Trade-Off}\label{ssec:MPP}
\begin{figure}[!t]
  \centering
  \includegraphics[width=0.59\textwidth]{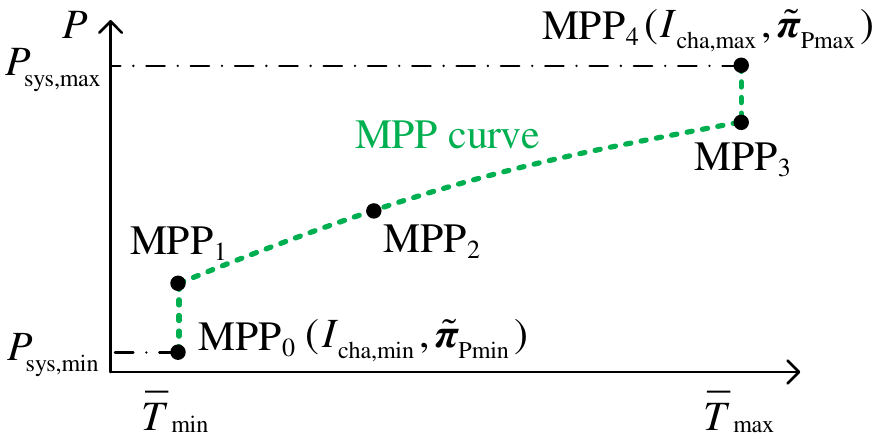}\\
  \caption{The MPP curve that optimizes the hydrogen yield.}\label{fig:pi_opt}
\end{figure}

In this subsection, let us include the optimization of $\tilde{\bm{\pi}}$ as well to locate the maximum production point (MPP).
Karush-Kuhn-Tucker (KKT) conditions are employed in combination with the results of the aforementioned SMPP analysis to select the trade-off decision $\tilde{\bm{\pi}}^*$ of feed factors $\tilde{\bm{\pi}}$.
At the end, an MPP curve as shown in Fig. \ref{fig:pi_opt} will be obtained as a comparison with the SMPP curve in Fig. \ref{fig:IT_opt}.

For an MPP with $\bar{T}^*>\bar{T}_{\rm min}$, such as MPP$_{\rm 2}$, MPP$_{\rm 3}$ and MPP$_{\rm 4}$ in Fig. \ref{fig:pi_opt}, its $U_{\rm fur}^*$ must equal $U_{\rm fur,min}$:
If not, $U_{\rm fur}^*$ can always be lowered to achieve the same hydrogen yield with less power $P_{\rm sys}$ according to the analysis of Fig. \ref{fig:IT_opt}, which contradicts the MPP's criteria (the maximum \eqref{eq:OP_PtG_J}).
Therefore, we obtain a useful conclusion that the constraint \eqref{eq:OP_PtG_U} is always active at MPPs with $\bar{T}^*>\bar{T}_{\rm min}$.
Specifically, the following two subsituations can be further discussed:

1) On the middle-piece MPP curve ($\bar{T}^* \in (\bar{T}_{\rm min},\bar{T}_{\rm max})$, such as MPP$_{\rm 2}$ in Fig. \ref{fig:pi_opt}) with \eqref{eq:OP_PtG_U} being the only active inequality constraint, the following KKT condition related to $\pi_{\rm ca}$ can be obtained from optimization problem \eqref{eq:OP_PtG}:
\begin{equation}\label{eq:pi_ca_opt}
 \frac{\frac{\partial U_{\rm cel}}{\partial \bar{T}}}{\frac{\partial (U_{\rm th}+U_{\rm war})}{\partial \bar{T}}} = \frac{\frac{\partial (U_{\rm cel}+U_{\rm fur,min}+U_{\rm war,pre}+U_{\rm vap})}{\partial \pi_{\rm ca}}}{\frac{\partial (U_{\rm war}+U_{\rm war,pre}+U_{\rm vap})}{\partial \pi_{\rm ca}}}.
\end{equation}
In fact, \eqref{eq:pi_ca_opt} evaluates the impacts of $\pi_{\rm ca}$ on both stack performance and auxiliaries' power losses simultaneously, and it provides the trade-off solution of $\pi_{\rm ca}$ together with \eqref{eq:OP_PtG_Ueq} and active \eqref{eq:OP_PtG_U}.
Note that $\frac{\partial U_{\rm cel}}{\partial \pi_{\rm ca}}$ is much more sensitive to $\pi_{\rm ca}$ compared with the other terms in \eqref{eq:pi_ca_opt} when $p'_{\rm H_2O,TPB}$ is close to zero (i.e., near water starvation); hence, the solution $\pi_{\rm ca}^*$ of \eqref{eq:pi_ca_opt} varies little with $\bar{T}^*$.
In fact, $\pi_{\rm ca}^*$ is usually a bit larger than $1$, because excess steam feed rate (i.e., small water utilization) decreases the system efficiency, which is commonly agreed in the literature such as \cite{Frank_rSOC_op_2018}, \cite{Luo_Excergy_SOECsystem_2018} and \cite{Wang_P2M_opt_2018}.
Meanwhile, $\pi_{\rm an}$ can be analogously analyzed, providing the following KKT condition:
\begin{equation}\label{eq:pi_an_opt}
 \frac{\frac{\partial (U_{\rm cel}+U_{\rm fur,min}+U_{\rm war,pre}+U_{\rm vap})}{\partial \pi_{\rm an}}}{\frac{\partial (U_{\rm war}+U_{\rm war,pre}+U_{\rm vap})}{\partial \pi_{\rm an}}} = \frac{\frac{\partial (U_{\rm cel}+U_{\rm fur,min}+U_{\rm war,pre}+U_{\rm vap})}{\partial \pi_{\rm ca}}}{\frac{\partial (U_{\rm war}+U_{\rm war,pre}+U_{\rm vap})}{\partial \pi_{\rm ca}}}.
\end{equation}
However, \eqref{eq:pi_an_opt} can rarely be fulfilled because $p'_{\rm O_2,TPB}$ never approaches zero according to \eqref{eq:p_p_O2} and $\pi_{\rm an}$ has little impact on $U_{\rm cel}$ compared with $\pi_{\rm ca}$.
In this case, $\pi_{\rm an}^*$ is preferably set at its lower limit regardless of \eqref{eq:pi_an_opt}, i.e., $\pi_{\rm an}^*=0$.
Because of the slight variation in $\tilde{\bm{\pi}}^*$, the middle piece shape of the MPP curve in Fig. \ref{fig:pi_opt} is basically inherited from the SMPP curve with fixed $\tilde{\bm{\pi}}$ in Fig. \ref{fig:IT_opt}.

2) When $\bar{T}^*$ reaches its upper limit of $\bar{T}_{\rm max}$, such as MPP$_{\rm 3}$ and MPP$_{\rm 4}$ in Fig. \ref{fig:pi_opt}, $\pi_{\rm ca}^*$ can be increased accordingly to prevent overheating by enhanced convection cooling.
A tail-piece MPP curve is thus distinctively developed as shown in Fig. \ref{fig:pi_opt} in comparison with the SMPP curve.
Note that the above discussion of $\pi_{\rm an}^*$ still applies.
In other words, unless $\pi_{\rm ca}^*$ near the tail end MPP$_{\rm 4}$ becomes large enough to ensure the solvability of $\pi_{\rm an}^*$ in \eqref{eq:pi_an_opt}, $\pi_{\rm an}^*=0$ still holds.
Then, the MPP can be simply located by solving $\bar{T}=\bar{T}_{\rm max}$, $\eqref{eq:OP_PtG_Ueq}$ and active \eqref{eq:OP_PtG_U} together.

For an MPP with $\bar{T}^* $ held at $\bar{T}_{\rm min}$ because of insufficient $P_{\rm sys}$, such as MPP$_{\rm 0}$ and MPP$_{\rm 1}$ in Fig. \ref{fig:pi_opt}, additional $U_{\rm fur}^*$ is necessarily required by an SMPP to maintain temperature, as mentioned in the head-piece SMPP analysis.
In the MPP case, however, $\pi_{\rm ca}^*$ can be turned down properly to slow  the convection cooling and meet the energy balance while keeping $U_{\rm fur}^* = U_{\rm fur,min}$.
In other words, constraint \eqref{eq:OP_PtG_U} is still active on the head-piece MPP curve of Fig. \ref{fig:pi_opt} in contrast to the head-piece SMPP curve.
The above discussion of $\pi_{\rm an}^*$ still applies, and the MPP can be simply located by solving $\bar{T}=\bar{T}_{\rm min}$, $\eqref{eq:OP_PtG_Ueq}$ and active \eqref{eq:OP_PtG_U} together.

The above optimality conditions to locate MPPs are summarized in Table \ref{tab:conditions}.
The projection of the entire MPP curve on the P axis provides the acceptable range of system loading $P_{\rm sys}$, which can be calculated at MPP$_{\rm 0}$ and MPP$_{\rm 4}$ in Fig. \ref{fig:pi_opt}:
\begin{subequations}
\begin{gather}
 \frac{P_{\rm sys,min}}{ \!N I_{\rm cha,min}\!} \!=\! U_{\rm rea}\!(\bar{T}_{\rm min}\!) + U_{\rm war}\!(\bar{T}_{\rm min},\!\tilde{\bm{\pi}}_{\rm Pmin}\!) + U_{\rm war,pre}\!(\tilde{\bm{\pi}}_{\rm Pmin}\!) + U_{\rm vap}\!(\tilde{\bm{\pi}}_{\rm Pmin}\!) + U_{\rm com} \label{eq:P_min}\\
 \frac{P_{\rm sys,max}}{\!N I_{\rm cha,max}\!} \!=\! U_{\rm rea}\!(\bar{T}_{\rm max}\!) + U_{\rm war}\!(\bar{T}_{\rm max},\!\tilde{\bm{\pi}}_{\rm Pmax}\!) + U_{\rm war,pre}\!(\tilde{\bm{\pi}}_{\rm Pmax}\!) + U_{\rm vap}\!(\tilde{\bm{\pi}}_{\rm Pmax}\!) + U_{\rm com} \label{eq:P_max}
\end{gather}
\end{subequations}
If $P_{\rm sys} < P_{\rm sys,min}$, then electrolysis at $I_{\rm cha,min}$ and $\bar{T}_{\rm min}$ cannot be maintained as a steady state, even with controllable gas flows.
If $P_{\rm sys} > P_{\rm sys,max}$, then the system power consumption can still be increased by increasing $\tilde{\bm{\pi}}$; however, it is meaningless because the maximum yield of $\rm H_2$ has been achieved.
If a desired loading $P_{\rm sys}$ within $\left[P_{{\rm sys,min}},P_{{\rm sys,max}}\right]$ is given by grid regulation considering the renewables, then an MPP with coordinated $\bar{T}^*$ and $\tilde{\bm{\pi}}^*$ can be obtained based on the above analysis, as shown in Table \ref{tab:conditions}, solving the trade-off problems of temperature and feed flows in Fig. \ref{fig:TradeOff}.

To conclude, the system optimization of hydrogen yield at various loading conditions requires proper coordination of the temperature and the feed flows; this process can be described as a three-piece MPP curve in Fig. \ref{fig:pi_opt}.
Nevertheless, the SMPP curve in Fig. \ref{fig:IT_opt} with only head and middle pieces with $\tilde{\bm{\pi}}$ assumed constant, which appears more practical for easy chart reading implementation, is also effective because the slight fluctuation of $\tilde{\bm{\pi}}^*$ has limited impact on the system within its primary working area.
In addition to the conversion efficiency, the operation range $\left[P_{{\rm sys,min}},P_{{\rm sys,max},}\right]$ is also maximized by the MPP approach of Fig. \ref{fig:pi_opt} because the system adaptability is enhanced by the coordination of auxiliaries.

\section{Numerical Study}\label{sec:numericalStudy}
In this section, an operation model of an HTE system is constructed as a numerical case study.
The model is first verified with a benchmark model in COMSOL and then analyzed in terms of SMPP and MPP strategies using the results of the numerical optimization, based on which the optimality conditions in Table \ref{tab:conditions} are validated.
\added[id=R2C3]{At last, a 24-hour operation case with dynamic power input is presented to show the effects of the proposed SMPP and MPP strategies on improving hydrogen yield from surplus electricity.}

\subsection{Model Verification}\label{ssec:COMSOL}
By employing the cell parameters in \cite{Udagawa_SOECmodel_2007} and the necessary gas properties, a lumped operation model in the form of \eqref{eq:OP_PtG} can be derived from the energy flow model as mentioned in Section \ref{sec:energyFlow} for a 200-cell-stack-based HTE system.
The primary parameters are listed in Table \ref{tab:parameters}.
\begin{table}[!t]
  \footnotesize
  \renewcommand{\arraystretch}{1.3}
  \caption{Primary parameters of the HTE system for numerical study}\label{tab:parameters}
  \centerline{
  \begin{threeparttable}
  \begin{tabular}{cccc}\toprule
  Parameters &Values &Units &Annotations\\ \midrule
  Stack config.& & &\\
  $N_{\rm cel}$ &$200$ &$1$ &\footnotesize Num. of cells within the HTE stack.\\
  $N_{\rm cha}$ &$100$ &$1$ &\footnotesize Num. of channel pairs within an HTE cell.\\
  Geometry& & &\\
  $L_x$ &$20$ &$\rm cm$ &\footnotesize Length of an HTE cell.\\
  $L_{y,{\rm cha}}$ &$2.67$ &$\rm mm$ &\footnotesize Width of an HTE channel.\\
  $L_{z,{\rm ele}}$ &$0.14$ &$\rm mm$ &\footnotesize Thickness of electrolyte layer.\\
  $L_{z,{\rm ca}}$ &$0.023$ &$\rm mm$ &\footnotesize Thickness of cathode layer.\\
  $L_{z,{\rm an}}$ &$0.023$ &$\rm mm$ &\footnotesize Thickness of anode layer.\\
  Electrochemistry & & &\\
  $\sigma_{\rm ele}$ &$9.35{\rm E}3 \cdot e^{-\frac{7237{\rm K}}{T}}$ &$\rm S/m$ &\footnotesize Electrical conductivity of electrolyte.\\
  $\sigma_{\rm ca}$ &$1.13{\rm E}5$ &$\rm S/m$ &\footnotesize Electrical conductivity of cathode.\\
  $\sigma_{\rm an}$ &$7.02{\rm E}3$ &$\rm S/m$ &\footnotesize Electrical conductivity of anode.\\
  $\Gamma_{\rm ex,ca}$ &$6.54{\rm E}11$ &$\rm S/m^2$ &\footnotesize Pre-exponential factor of $i_{\rm ex,ca}$.\\
  $\Gamma_{\rm ex,an}$ &$2.35{\rm E}11$ &$\rm S/m^2$ &\footnotesize Pre-exponential factor of $i_{\rm ex,an}$.\\
  $\xi_{\rm ca}$ &$1.40{\rm E}5$ &$\rm J/mol$ &\footnotesize Activation energy of cathode reaction.\\
  $\xi_{\rm an}$ &$1.37{\rm E}5$ &$\rm J/mol$ &\footnotesize Activation energy of anode reaction.\\
  $i_{\rm el,min}$ &$0.1$ &$\rm A/cm^2$ &\footnotesize Min. electrolysis current density.\\
  $i_{\rm el,max}$ &$0.9$ &$\rm A/cm^2$ &\footnotesize Max. electrolysis current density.\\
  $I_{\rm cha,min}$ &$0.533$ &$\rm A$ &\footnotesize Min. channel current, calculated from $i_{\rm el,min}$.\\
  $I_{\rm cha,max}$ &$4.80$ &$\rm A$ &\footnotesize Max. channel current, calculated from $i_{\rm el,max}$.\\ 
  Feed streams & & &\\
  \multicolumn{3}{c}{\footnotesize Cathode inlet compositions} &\footnotesize $\rm H_2O$ of $\rm 90vol.\%$, and $\rm H_2$ of $10 \rm vol.\%$ to prevent oxidation.\\
  \multicolumn{3}{c}{\footnotesize Anode inlet compositions} &\footnotesize $\rm O_2$ of $\rm 21vol.\%$, and $\rm N_2$ of $79 \rm vol.\%$, as in the air.\\
  $\lambda$ &$0.65$ &$1$ &\footnotesize Selected ratio to evaluate $\tilde{\bm{p'}}$ with \eqref{eq:p_H2}-\eqref{eq:p_N2}.\\
  $p_{\rm s2}$ &$1$ &$\rm bar$ &\footnotesize HTE operating pressure.\\
  $p_{\rm com}$ &$100$ &$\rm bar$ &\footnotesize $\rm H_2$ product pressure (behind compressor).\\
  $\eta_{\rm com}$ &$70\%$ &$1$ &\footnotesize Isothermal efficiency of the compressor.\\
  Temperatures & & &\\
  $T_{\rm s0}$ &$25$ &$\rm ^{\circ}{\rm C}$ &\footnotesize System input temperature.\\
  $T_{\rm s1}$ &$110$ &$\rm ^{\circ}{\rm C}$ &\footnotesize Furnace input temperature, ensured by preheater.\\
  $T_{\rm min}$ &$750$ &$\rm ^{\circ}{\rm C}$ &\footnotesize Min. HTE temperature.\\
  $({\rm d}T/{\rm d}l)_{\rm max}$ &$10$ &$\rm ^{\circ}{\rm C}/cm$ &\footnotesize Max. acceptable temperature gradient.\\
  $\bar{T}_{\rm min}$ &$750$ &$\rm ^{\circ}{\rm C}$ &\footnotesize Min. stack average temperature, calculated from $T_{\rm min}$.\\
  $\bar{T}_{\rm max}$ &$850$ &$\rm ^{\circ}{\rm C}$ &\footnotesize Max. stack average temperature, calculated from $T_{\rm min}$ and $({\rm d}T/{\rm d}l)_{\rm max}$.\\
  $T_{\rm com}$ &$75$ &$\rm ^{\circ}{\rm C}$ &\footnotesize Compressor temperature (behind cooler).\\
  $k$ &$1.085$ &$1$ &\footnotesize Selected ratio to evaluate $\bar{T}$ with \eqref{eq:T_eval}.\\ \bottomrule
  \end{tabular}
  \end{threeparttable}}
\end{table}

\begin{figure}[!t]
  \centering
  \subfigure[Validation of pressure formulations \eqref{eq:p_H2O} and \eqref{eq:p_O2}]{
  \includegraphics[width=0.89\textwidth]{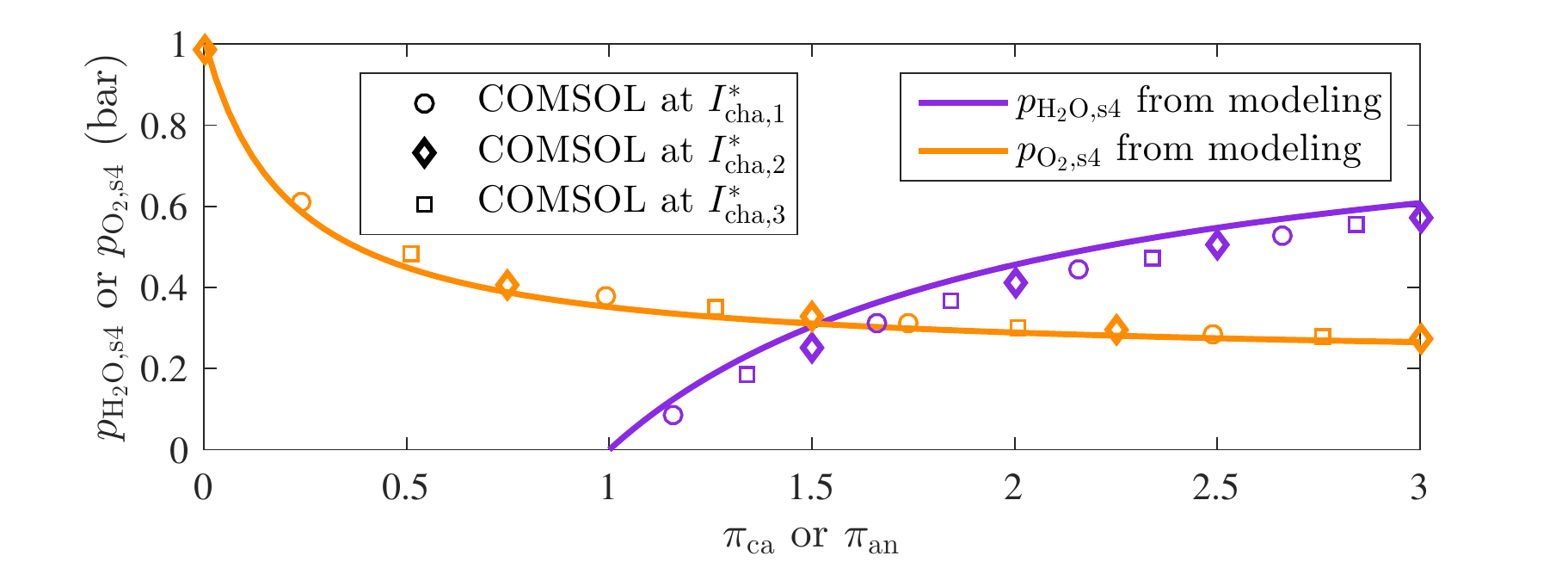}
  }
  \subfigure[Validation of $U$-$T$ curves in Fig. \ref{fig:IT_opt}]{
  \includegraphics[width=0.89\textwidth]{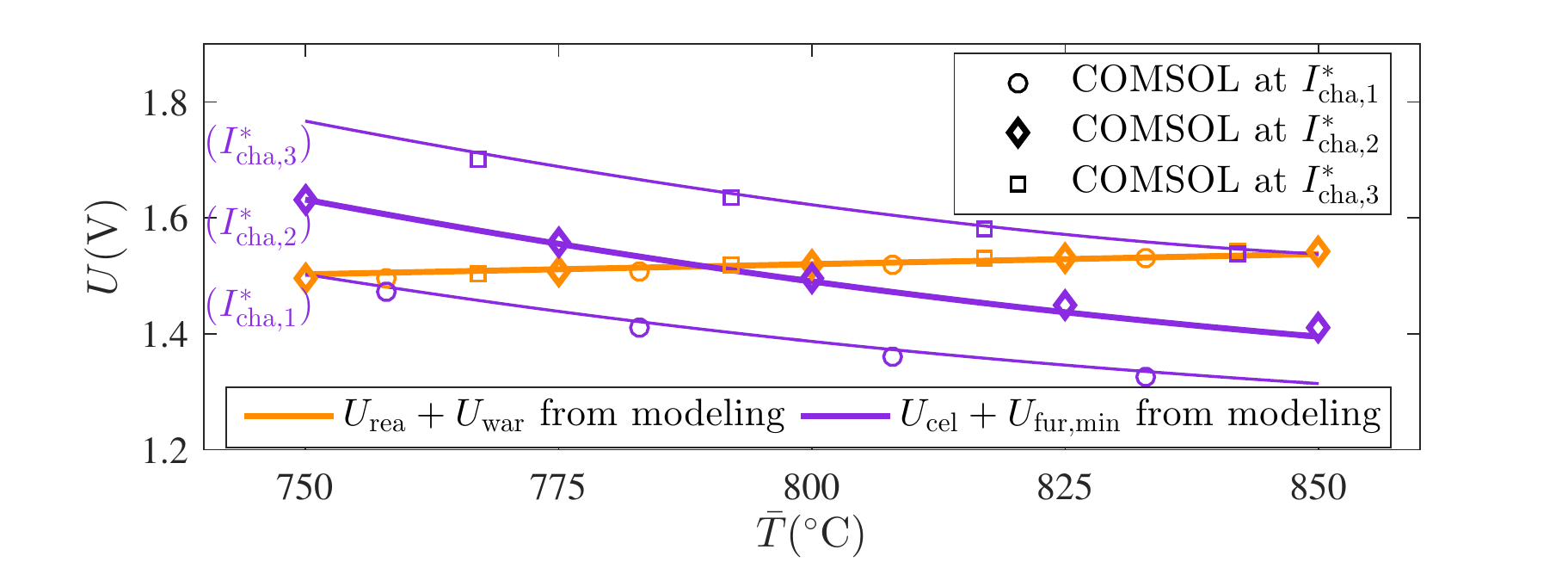}
  }
  \caption{Model verification based on the COMSOL benchmark.}\label{fig:model_validation}
\end{figure}

Simultaneously, we construct a refined HTE 3D-model in COMSOL with the same parameters as shown in Table \ref{tab:parameters}, which describes the multiphysics coupling of the electrochemical process, gas flows and thermal transmission.
This model in COMSOL is employed as a benchmark to validate the accuracy of the aforementioned operation model.
Fig. \ref{fig:model_validation}(a) depicts the feed-factor dependence of the outlet compositions given by \eqref{eq:p_H2}-\eqref{eq:p_N2} of the operation model, which is well confirmed by the sampling results of the COMSOL benchmark at three different electrolysis currents $I_{\rm cha,1}^*$, $I_{\rm cha,2}^*$ and $I_{\rm cha,3}^*$.
With $\tilde{\bm{\pi}}$ fixed at $[1.22,0]^{\rm T}$, Fig. \ref{fig:model_validation}(b) presents the $U$-$T$ plot similar to that of Fig. \ref{fig:IT_opt} calculated from the operation model.
The results of the proposed model coincide with the results of COMSOL.

In addition, 4 equilibrium points of the furnace calculated by COMSOL are listed in Table \ref{tab:balances}, demonstrating the energy flows in Fig. \ref{fig:EnergyBalance} in terms of the equivalent voltages.
Apparently, the assumption of neglectable $P_{\rm dis}$ and $P_{\rm kin}$ proves reasonable in various situations.
\begin{table}[!t]
  \renewcommand{\arraystretch}{1.3}
  \caption{Energy balances in the furnace (Object 1) at several operating points}\label{tab:balances}
  \centering
  \begin{threeparttable}
  \begin{tabular}{ccc|cccccc}\toprule
  \multicolumn{3}{c|}{Operating points} &\multicolumn{6}{|c}{Equivalent voltages of energy flows}\\ \hline
  $I_{\rm cha}$ &$\bar{T}$ &$\pi_{\rm ca}$&$U_{\rm cel}$ &$U_{\rm fur}$ &$U_{\rm rea}$ &$U_{\rm war}$ &$U_{\rm dis}$ &$U_{\rm kin}$\\ \hline
  2.50 &800 &1.22 &\!\!1.2074\!\! &\!\!0.3021\!\! &\!\!1.2894\!\! &\!\!0.2121\!\! &\!\!0.0080\!\! &\!\!6.8E-8\!\!\\ 
  2.00 &800 &1.22 &\!\!1.1534\!\! &\!\!0.3501\!\! &\!\!1.2894\!\! &\!\!0.2016\!\! &\!\!0.0125\!\! &\!\!4.0E-8\!\!\\
  2.50 &775 &1.22 &\!\!1.2627\!\! &\!\!0.2486\!\! &\!\!1.2887\!\! &\!\!0.2160\!\! &\!\!0.0066\!\! &\!\!7.0E-8\!\!\\
  2.50 &800 &1.15 &\!\!1.2221\!\! &\!\!0.2773\!\! &\!\!1.2894\!\! &\!\!0.2021\!\! &\!\!0.0079\!\! &\!\!5.8E-8\!\!\\ \bottomrule
  \end{tabular}
  \begin{tablenotes}
  \footnotesize
  \item{*} Units: $I_{\rm cha}$-$\rm A$, $\bar{T}$-$^{\circ}{\rm C}$, $\pi_{\rm ca}$-1, $U$-$\rm V$. Note that $\pi_{\rm an}$ is fixed at 0.
  \end{tablenotes}
  \end{threeparttable}
\end{table}

\subsection{SMPP and MPP Operating Strategies}\label{ssec:SMPPandMPPcase}

\subsubsection{\replaced[id=R2C2]{Locating SMPP and MPP at Specific Load}{Cross-Sectional Analysis under Given Loading Power}}\label{sssec:specificLoad}
\begin{figure}[!t]
  \centerline{\includegraphics[width=1.4\textwidth]{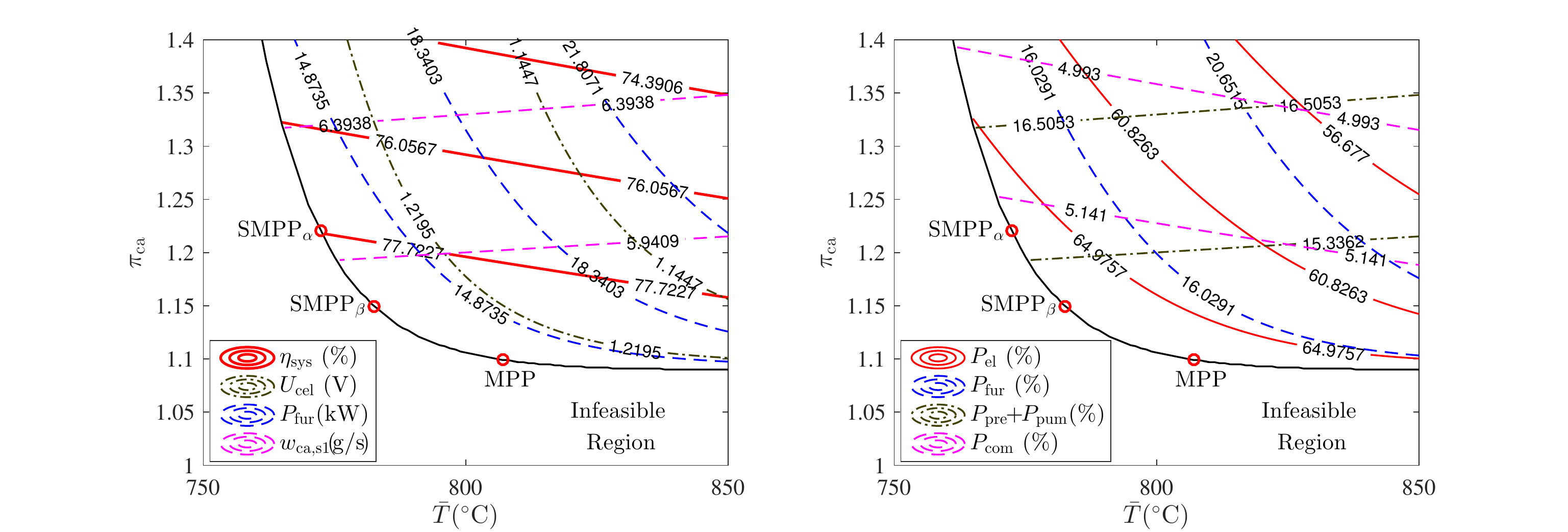}}
  \caption{The trade-off decisions of temperature and feed factors: \added[id=R2C5]{system performances (left) and energy allocations (right) at various operating parameters given load of $P_{\rm sys}=100{\rm kW}$ based on description of model \eqref{eq:OP_PtG}}.}\label{fig:Tpi_contour}
\end{figure}

Fig. \ref{fig:Tpi_contour} presents a detailed cross-sectional analysis of the proposed operation model \eqref{eq:OP_PtG} given $P_{\rm sys}=100{\rm kW}$, which quantitatively illustrates the trade-off decisions of temperature and feed factors mentioned in Fig. \ref{fig:TradeOff}.
The feasible region to ensure the stack safety, the efficiency variation with primary operating parameters, and the locations of MPP and SMPP are depicted with contour plots.
The legends SMPP$_{\rm \alpha}$ and SMPP$_{\rm \beta}$ are obtained from the optimization of \eqref{eq:OP_PtG} with $\tilde{\bm{\pi}}$ fixed at $\tilde{\bm{\pi}}_{\rm \alpha}=[1.22,0]^{\rm T}$ and $\tilde{\bm{\pi}}_{\rm \beta}=[1.15,0]^{\rm T}$, respectively.

The primary infeasibilities include water starvation (with $p'_{\rm H_2O,TPB}$ approaching 0) due to low $\pi_{\rm ca}$ at the bottom of Fig. \ref{fig:Tpi_contour} and inadequate inlet temperature $T_{\rm s3}$ due to insufficiently set $\bar{T}$ (which occurs less at large feed flows with smaller temperature increment along $l$) at the left side of Fig. \ref{fig:Tpi_contour}.
The locations of the infeasible region (bottom-left) and low-efficiency region (top-right) in Fig. \ref{fig:Tpi_contour} coincide with the analysis in Fig. \ref{fig:TradeOff}.

Within the feasible region, each point denotes a system operating state, whose operating conditions (such as cell voltage $U_{\rm cel}$, cathode stream feed rate $w_{\rm ca,s1}$ \added[id=R2C5]{and energy allocations on different units}) and efficiency performance can be depicted by corresponding contours.
For example, we can know from the left plot of Fig. \ref{fig:Tpi_contour} that the system efficiency at SMPP$_{\rm \alpha}$ is approximately $77.72\%$, which is clearly lower than the efficiency at MPP.
\added[id=R2C5]{We can also know from the right plot of Fig. \ref{fig:Tpi_contour} that the energy consumed by electrolyser is above $64.98\%$ of the total load at SMPP$_{\rm \alpha}$, but that consumed by furnace is below $16.03\%$.}
\added[id=R2C2]{The MPP is found at the feasible region bound (FB) with constraint \eqref{eq:OP_PtG_U} being active, which coincides with the obtained conditions in Table \ref{tab:conditions}.}
\added[id=R2C5]{In fact, the contours in Fig. \ref{fig:Tpi_contour} indicate that the optimized parameters in maximizing system efficiency result in higher energy allocations to electrolyser and compressor, but lower energy allocations to furnace, preheater and pump group, which makes sense considering a specific $P_{\rm sys}$ has been given.}

\begin{figure}[!t]
  \centering
  \includegraphics[width=0.89\textwidth]{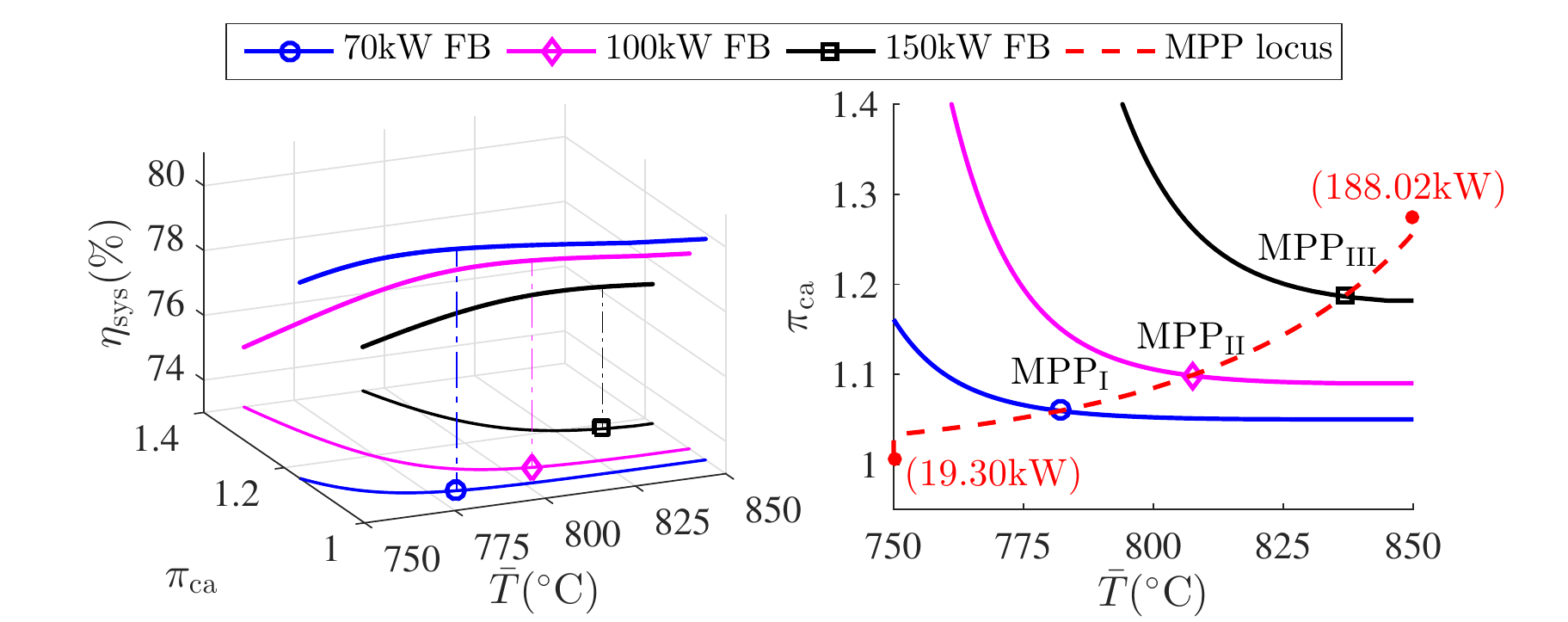}
  \caption{\added[id=R2C2]{Locating MPPs at loads $70{\rm kW}$, $100{\rm kW}$ and $150{\rm kW}$ based on numerical optimization of \eqref{eq:OP_PtG} to get MPP locus (FB: Feasible region bound as shown in Fig. \ref{fig:Tpi_contour}).}}\label{fig:MPP_obtain}
\end{figure}

\added[id=R2C2]{
By extending Fig. \ref{fig:Tpi_contour} to other loading powers like $70{\rm kW}$ and $150{\rm kW}$, we can obtain more MPPs to get the MPP locus at full-range loads, as illustrated in Fig. \ref{fig:MPP_obtain}.
Each MPP is located at the corresponding FB, and performs the maximum system efficiency possible at the corresponding load.
The load range under MPP operation $[19.30,188.02]{\rm kW}$ is calculated from \eqref{eq:P_min} and \eqref{eq:P_max}.
Based on the numerical solutions of MPPs, it can be validated that the electrolysis current density $i_{\rm el}$ reaches its minimum $0.1{\rm A/cm^2}$ and maximum $0.9{\rm A/cm^2}$ at loads of $19.30{\rm kW}$ and $188.02{\rm kW}$, respectively.
}

\subsubsection{\replaced[id=R2C2]{SMPP and MPP Loci at Full-Range Loads}{Operation Analysis under Various Loading Powers}}\label{sssec:fullRange}
\begin{figure}[!t]
  \centering
  \subfigure[Furnace set points (curves in 2 or 3 pieces)]{
  \includegraphics[width=0.89\textwidth]{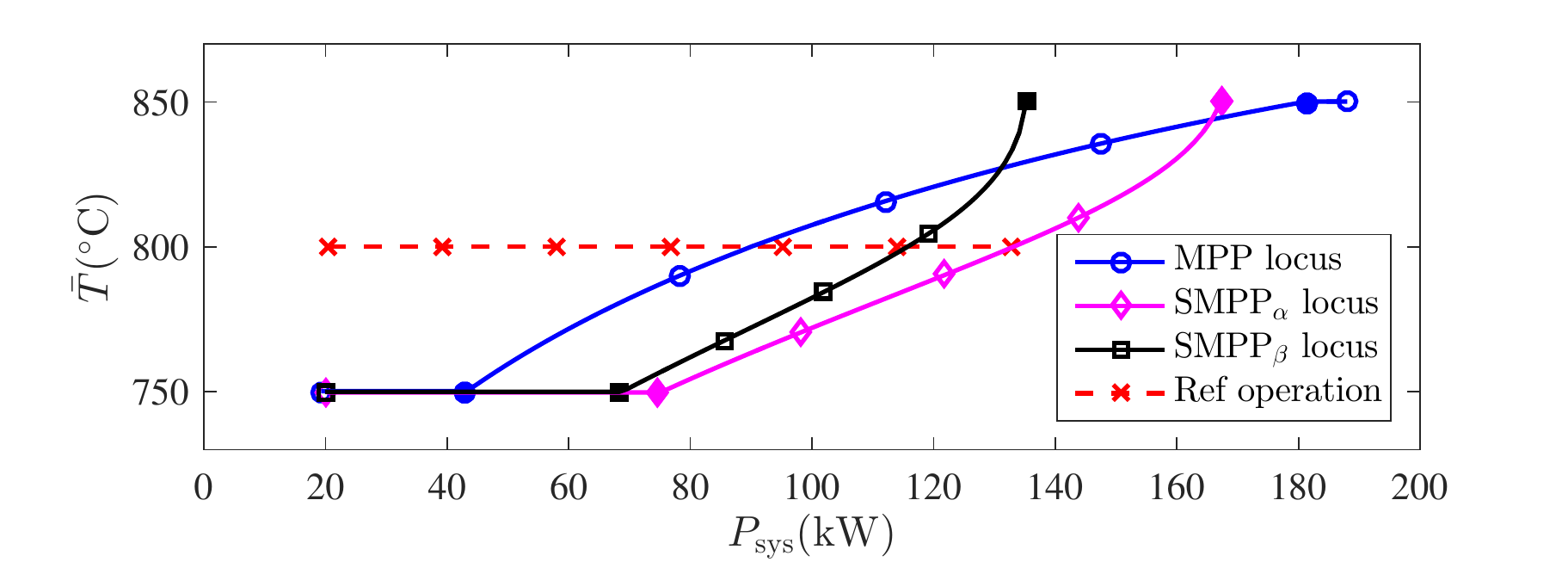}
  }
  \subfigure[Mass flow set points]{
  \includegraphics[width=0.89\textwidth]{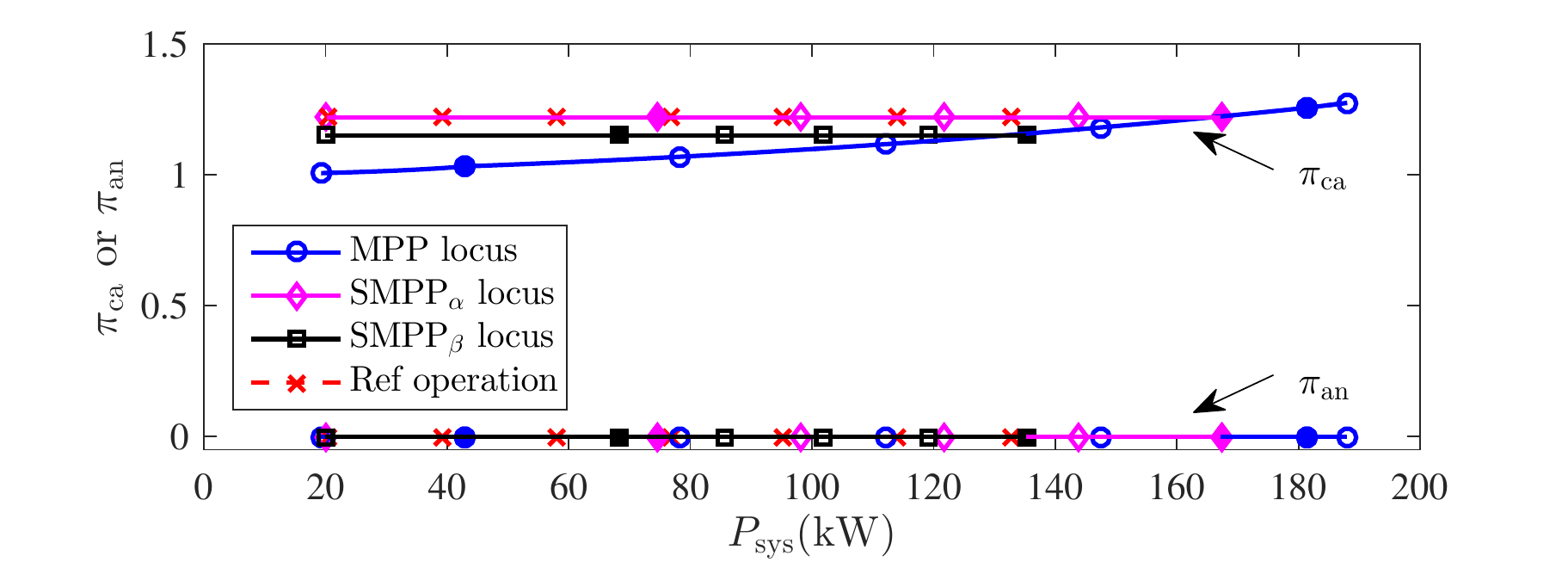}
  }
  \subfigure[Improvement of efficiency and load range]{
  \includegraphics[width=0.89\textwidth]{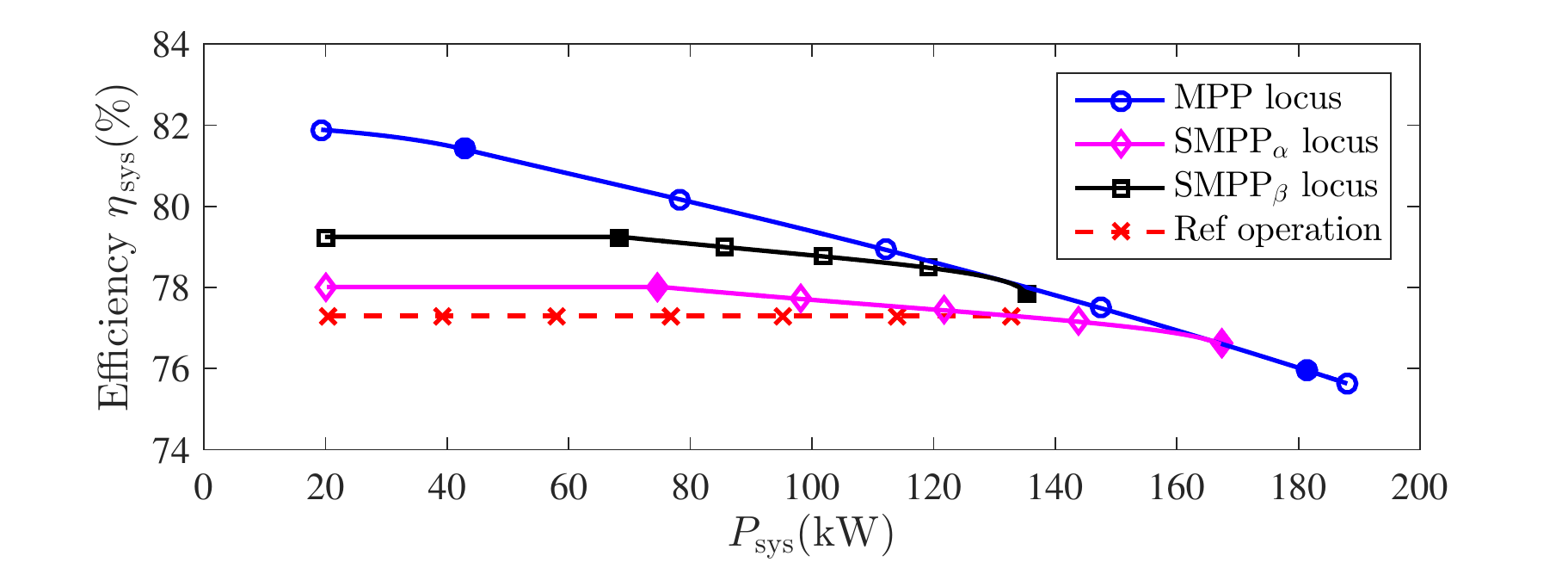}
  }
  \caption{Full-range set points of SMPP and MPP strategies based on numerical optimization of model \eqref{eq:OP_PtG}.}\label{fig:MPPT_operation}
\end{figure}

\added[id=R2C2]{Based on the method described in Fig. \ref{fig:MPP_obtain}, the SMPP$_{\rm \alpha}$ locus and SMPP$_{\rm \beta}$ locus can be analogously obtained from operation model \eqref{eq:OP_PtG}.}
The two-piece shape for the SMPP$_{\rm \alpha}$ and SMPP$_{\rm \beta}$ curves and the three-piece shape for the MPP curve in Fig. \ref{fig:MPPT_operation}(a) coincide with the analyses in Sections \ref{ssec:SMPP} and \ref{ssec:MPP}, as shown in Fig. \ref{fig:IT_opt} and Fig. \ref{fig:pi_opt}.
Note that the segment points are labeled by solid markers in Fig. \ref{fig:MPPT_operation}.
Moreover, the limited variations of $\pi_{\rm ca}^*$ and $\pi_{\rm an}^*$ for MPP in Fig. \ref{fig:MPPT_operation}(b) are consistent with the conclusion of Section \ref{ssec:MPP}, thus confirming the applicability of SMPP$_{\rm \alpha}$ and SMPP$_{\rm \beta}$.

Fig. \ref{fig:MPPT_operation}(c) presents the beneficial effects of the proposed SMPP and MPP strategies compared with the reference operation, where $\bar{T}$ and $\tilde{\bm{\pi}}$ are fixed at $800{\rm ^{\circ}C}$ and $\tilde{\bm{\pi}}_{\rm \alpha}$, respectively, and only $I_{\rm cha}^*$ is altered to meet the desired system loading $P_{\rm sys}$.
The MPP strategy increases the average efficiency from $77.30\%$ to $79.12\%$.
The optimal efficiency decreases as the loading power increases, which is consistent with the results of \cite{Wang_P2M_opt_2018}, \cite{Menon_SOEC_model_2014} and \cite{Petipas_HTE_VariousLoads_2013}.
Moreover, the load range is increased from $[20.44,132.86]{\rm kW}$ to $[19.30,188.02]{\rm kW}$, which significantly improves the system's capacity of P2G operation.
\added[id=R2C2]{The reason of the smaller range of reference operation is that: To maintain $\bar{T}=800{\rm ^{\circ}C}$ at loads larger than $133.86{\rm kW}$, the furnace output $U_{\rm fur}$ has to fall below $U_{\rm fur,min}$ to give place to the increasing $U_{\rm cel}$, causing inadequate inlet temperature $T_{\rm s3}$ but excess temperature gradient ${\rm d}T/{\rm d}l$.}
In addition, the option to achieve part of these effects by the SMPP$_{\rm \alpha}$ strategy with a lower computational cost is more practical.

\begin{figure}[!t]
  \centering
  \includegraphics[width=0.89\textwidth]{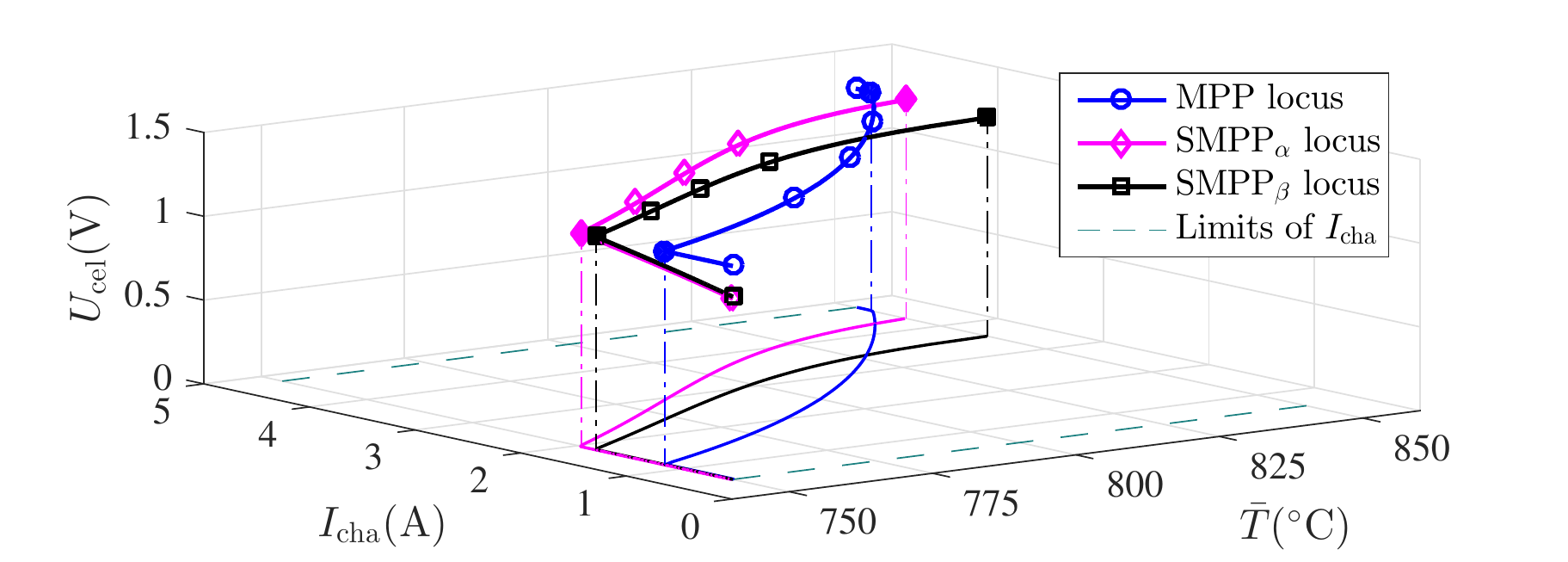}
  \caption{The locus of electrolysis status at SMPP and MPP strategies.}\label{fig:UIT_track}
\end{figure}

The locus of cell status at these operating states are depicted in Fig. \ref{fig:UIT_track}.
\added[id=R2C2]{
Apparently, the MPP strategy makes best use of the cell active area, because the limit of $I_{\rm cha,max}$ can only be achieved by the tail piece of the MPP curve.
Fig. \ref{fig:PP_energy} presents the composition of system energy sinks at several MPPs.
Meanwhile, it can be deduced that steam starvation does not occur during the SMPP or MPP operation due to the normal cell voltages $U_{\rm cel}$ at full-range loads as shown in Fig. \ref{fig:UIT_track}.}

\subsubsection{\added[id=R2C5]{Energy Balances under MPPs at Full-Range Loads}}\label{sssec:energyBalances}
\begin{figure}[!t]
  \centering
  \includegraphics[width=0.89\textwidth]{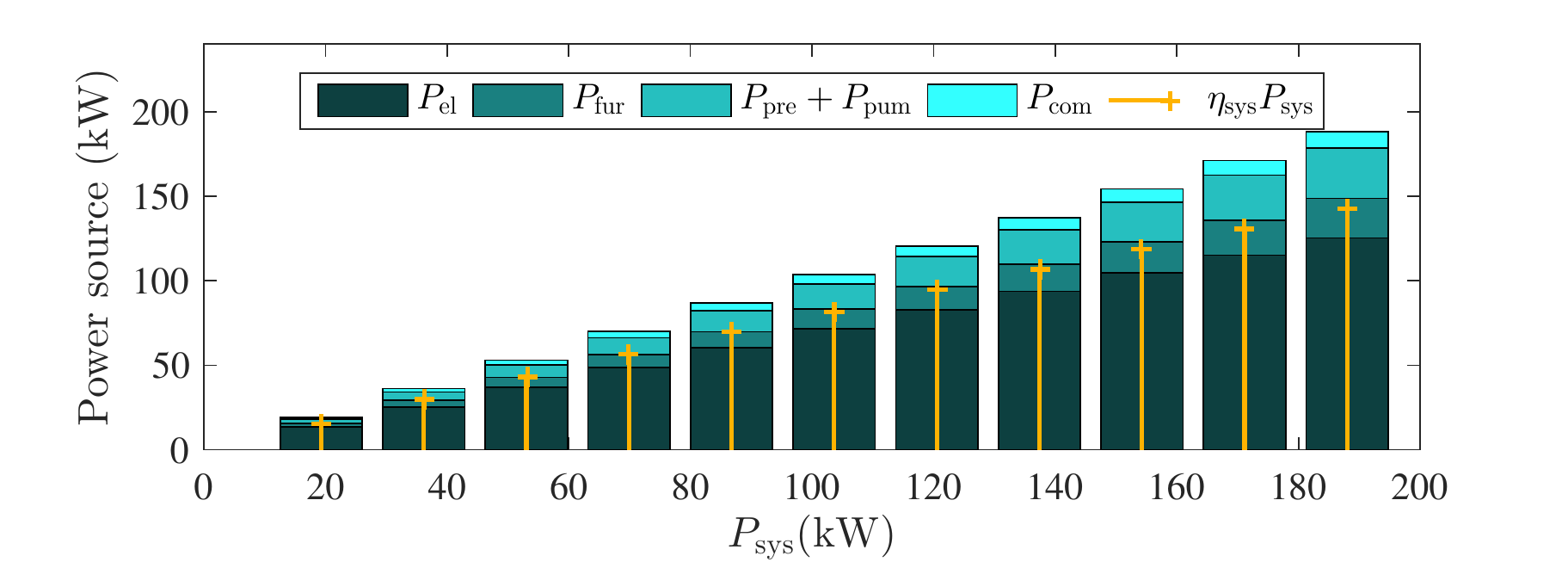}
  \caption{\added[id=R2C5]{The system energy sources at different MPPs.}}\label{fig:PP_energy_source}
\end{figure}

\added[id=R2C5]{
Fig. \ref{fig:PP_energy_source} depicts the contributions of system energy sources under MPPs at full-range loads, including energy consumptions of preheater and pump ($P_{\rm pre}+P_{\rm pum}$), furnace ($P_{\rm fur}$), dc electrolysis current ($P_{\rm el}$), and hydrogen compressor ($P_{\rm com}$).
Apparently the power converter ($P_{\rm el}$) always serves as the largest contributor to supply the energy demands of HTE.
However, the preheater, pump group and furnace also provide a significant fraction ($P_{\rm pre}+P_{\rm pum}+P_{\rm fur}$) to ensure full vaporization and adequate stack temperature.
}

\begin{figure}[!t]
  \centering
  \includegraphics[width=0.89\textwidth]{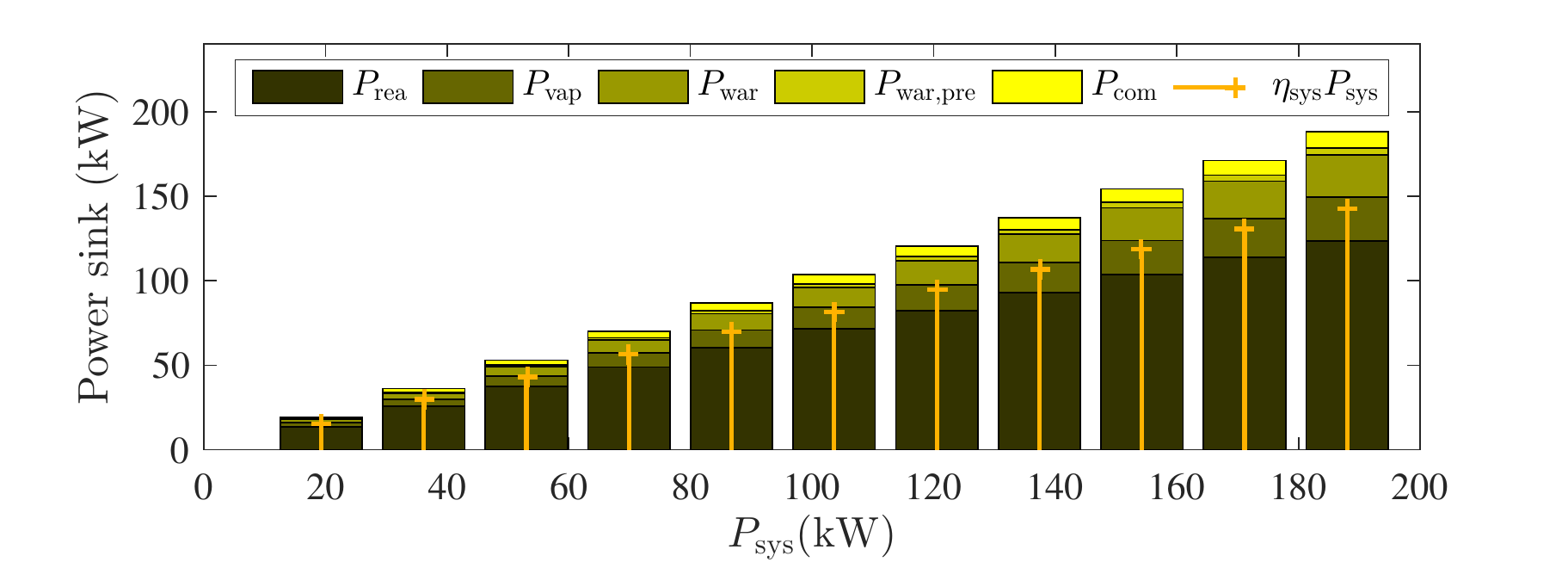}
  \caption{The system energy sinks at different MPPs.}\label{fig:PP_energy}
\end{figure}

\added[id=R2C5]{
Fig. \ref{fig:PP_energy} depicts the allocations of system energy sinks under MPPs at full-range loads, including feedstock enthalpy increments due to prewarming ($P_{\rm war,pre}$), vaporization ($P_{\rm vap}$), electrolysis reaction ($P_{\rm rea}$), furnace warming  ($P_{\rm war}$) and pressurization ($P_{\rm com}$).
Apparently the reaction-induced and vaporization-induced enthalpy changes ($P_{\rm rea}+P_{\rm vap}$) always take the primary fraction, representing that most of the energy input is converted into a chemical form in $\rm H_2$.
}
The energy loss has contributions from the vaporization of excess water (part of $P_{\rm var}$), the feedstock warming including preheating ($P_{\rm war,pre}+P_{\rm war}$) and the compressor power ($P_{\rm com}$), which coincides with the analysis based on \eqref{eq:efficiency_sink}.

\begin{figure}[!t]
  \centering
  \includegraphics[width=0.89\textwidth]{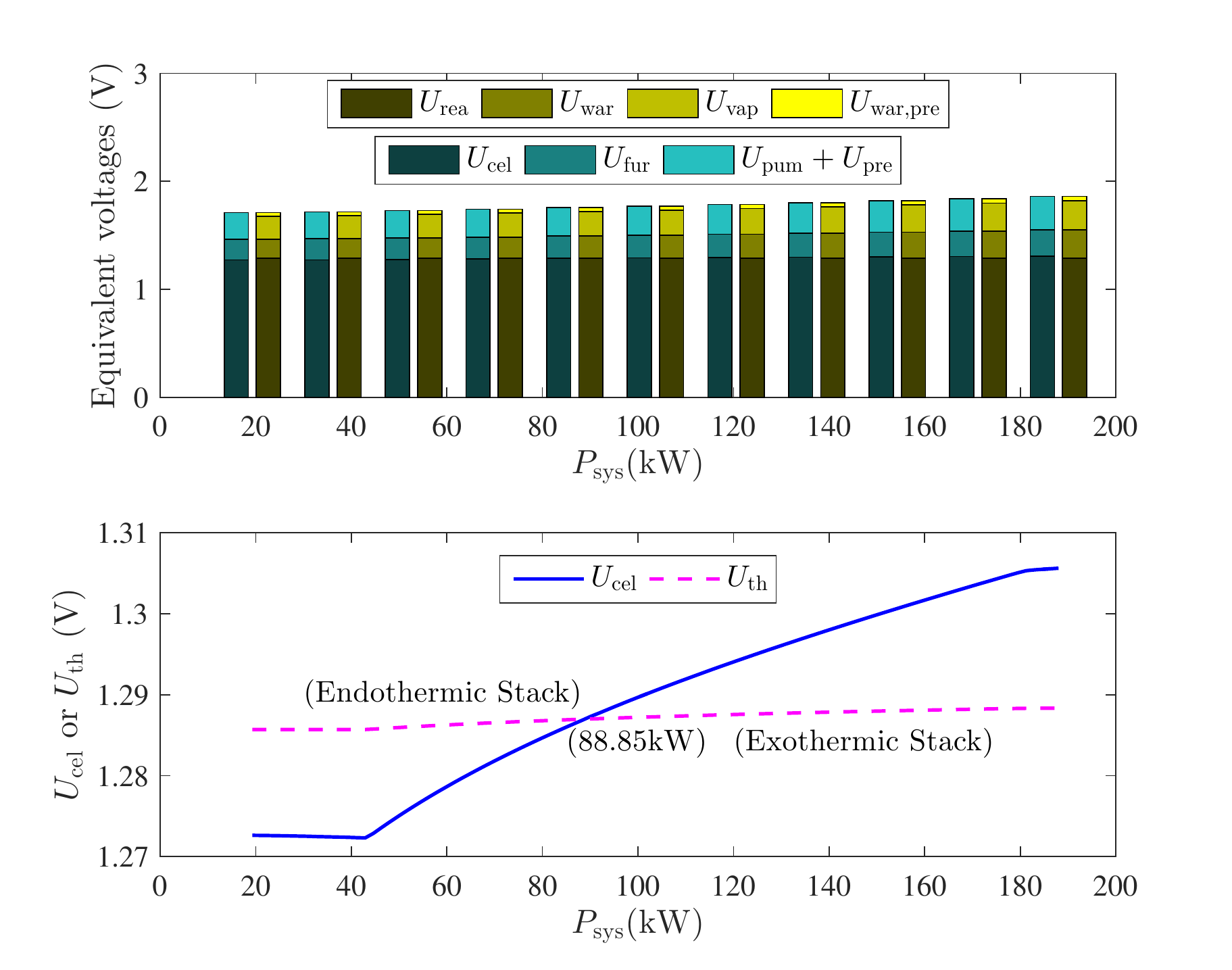}
  \caption{\added[id=R2C5]{The system energy balances at different MPPs.}}\label{fig:UP_energy}
\end{figure}

\added[id=R2C5]{
Based on Fig. \ref{fig:PP_energy_source} and Fig. \ref{fig:PP_energy}, we can simultaneously display energy sources and sinks in Fig. \ref{fig:UP_energy} to observe the energy balances of the HTE system at MPPs.
This time the energy is described with equivalent voltages defined in \eqref{eq:OP_PtG} for normalized comparison among different loading conditions.
Apparently the absolute values of these equivalent voltages barely change with loading power $P_{\rm sys}$.
In other words, the relative proportions of aforementioned energy sources and sinks in Fig. \ref{fig:PP_energy_source} and Fig. \ref{fig:PP_energy} basically stay unvaried at different MPPs.
Moreover, the HTE stack always operates near the thermo-neutral point under the proposed MPP strategy since $U_{\rm cel} \simeq U_{\rm th}$ in Fig. \ref{fig:UP_energy}.
But strictly speaking, the stack is endothermic ($U_{\rm el}$ slightly lower than $U_{\rm th}$) at loads below $88.85{\rm kW}$, while exothermic at higher loads.
}

\subsubsection{Validation of Optimality Conditions}\label{sssec:optimalityConditions}
\begin{figure}[!t]
  \centering
  \subfigure[Validation of $U_{\rm fur}=U_{\rm fur,min}$]{
  \includegraphics[width=0.89\textwidth]{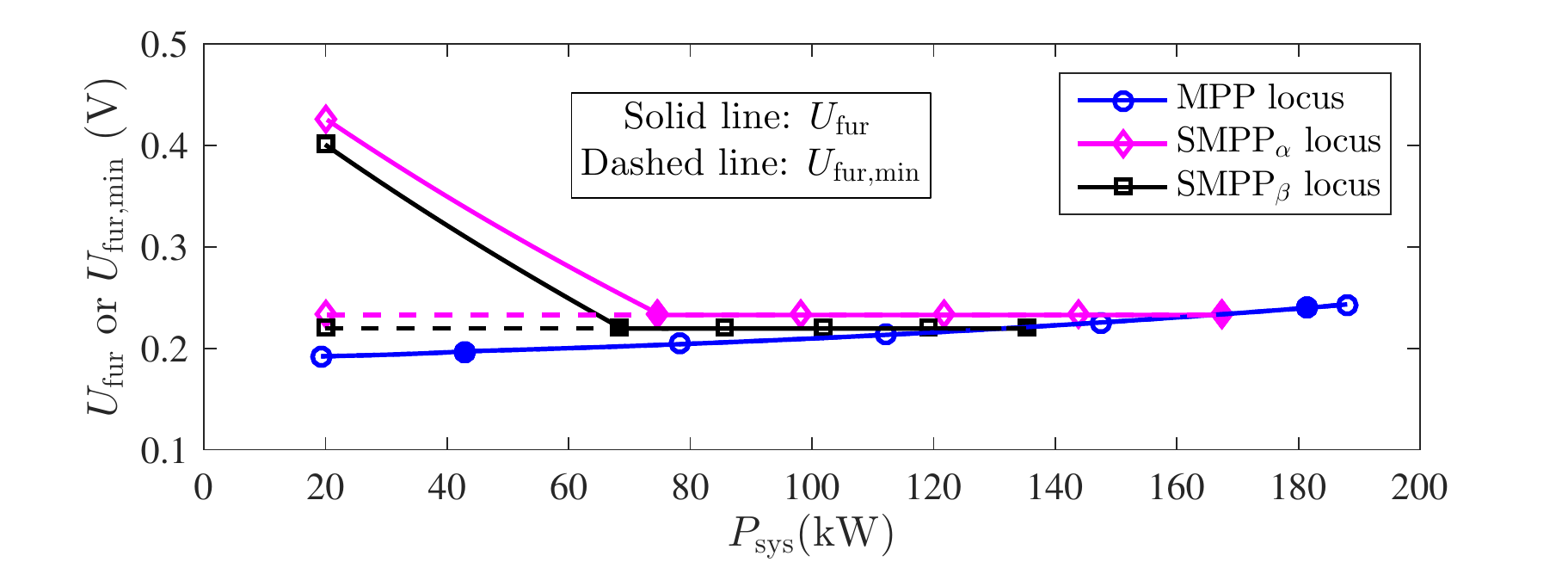}
  }
  \subfigure[Validation of \eqref{eq:pi_ca_opt}]{
  \includegraphics[width=0.89\textwidth]{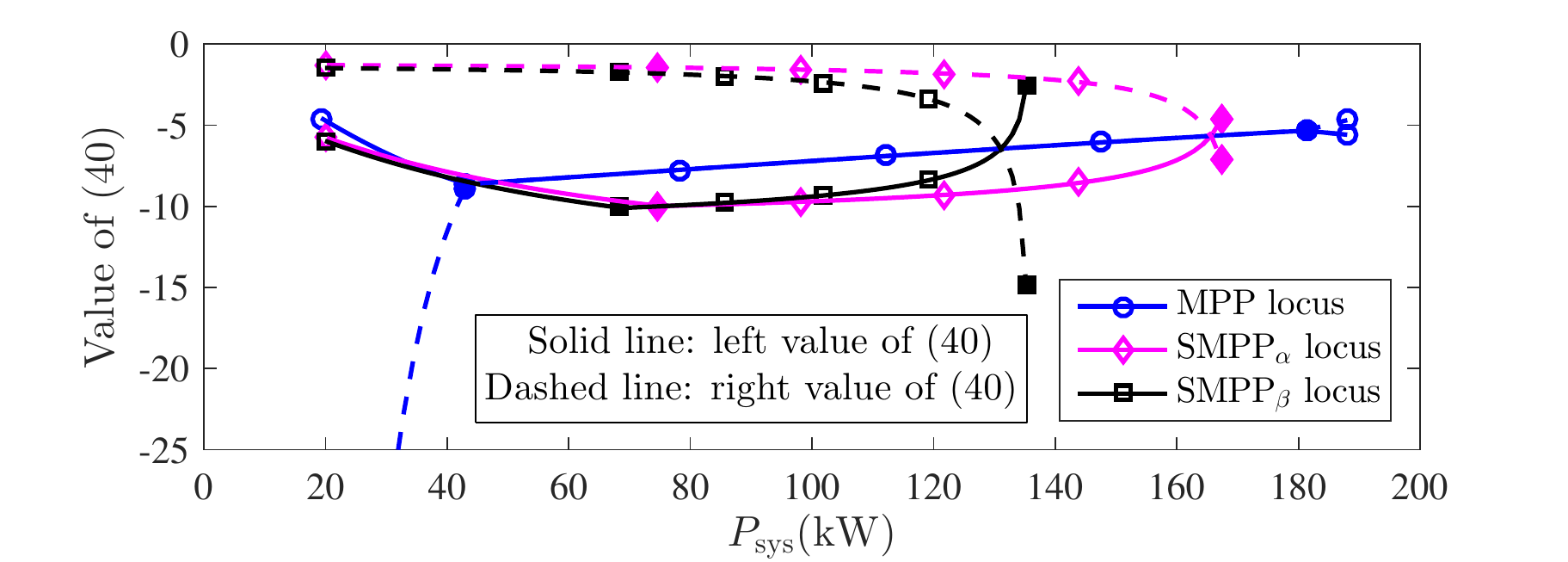}
  }
  \caption{Validation of the optimality conditions in Table \ref{tab:conditions}.}\label{fig:conditions}
\end{figure}

The optimality conditions of Table \ref{tab:conditions} in terms of $\bar{T}$ are easily checked by Fig. \ref{fig:MPPT_operation}(a) and Fig. \ref{fig:UIT_track}, while those in terms of $U_{\rm fur}$ and $\pi_{\rm ca}$ can be validated by the numerical results shown in Fig. \ref{fig:conditions}.

According to Table \ref{tab:conditions}, $U_{\rm fur}=U_{\rm fur,min}$ should be fulfilled on the middle-piece SMPP curve.
In Fig. \ref{fig:conditions}(a), the solid line $U_{\rm fur}$ and the dashed line $U_{\rm fur,min}$ do overlap  on the middle piece for SMPP$_{\alpha}$ and SMPP$_{\beta}$.
Note that the ``middle piece'' in Fig. \ref{fig:conditions}(a) can be located with the solid markers labeled previously in Fig. \ref{fig:MPPT_operation}(a).
Meanwhile, the solid line $U_{\rm fur}$ and the dashed line $U_{\rm fur,min}$ show no numerical differences on the entire MPP curve in Fig. \ref{fig:conditions}(a), which coincides with the analytical optimality condition that $U_{\rm fur}$ is always set at the minimum at MPPs, as shown in Table \ref{tab:conditions}.

Another analytically obtained optimality condition in Table \ref{tab:conditions} is that the KKT equation \eqref{eq:pi_ca_opt} holds on the middle-piece MPP curve.
As shown in Fig. \ref{fig:conditions}(b), there is indeed an overlap between the left value of \eqref{eq:pi_ca_opt} (in solid line) and its right value (in dashed line) on the middle-piece MPP curve.
Still, the ``middle piece'' in Fig. \ref{fig:conditions}(b) can be located with the solid markers labeled previously in Fig. \ref{fig:MPPT_operation}(a).

With these validated conditions, nonlinear programming of \eqref{eq:OP_PtG} can be replaced by solving a set of equations during SMPP or MPP implementation to ensure reduced computational cost for the local controller of an HTE system.

\subsection{\added[id=R2C3]{Implementation Case under Dynamic Power Input}}\label{ssec:dynamicPower}
\begin{figure}[!t]
  \centering
  \subfigure[Operating loads to track surplus electricity]{
  \includegraphics[width=0.89\textwidth]{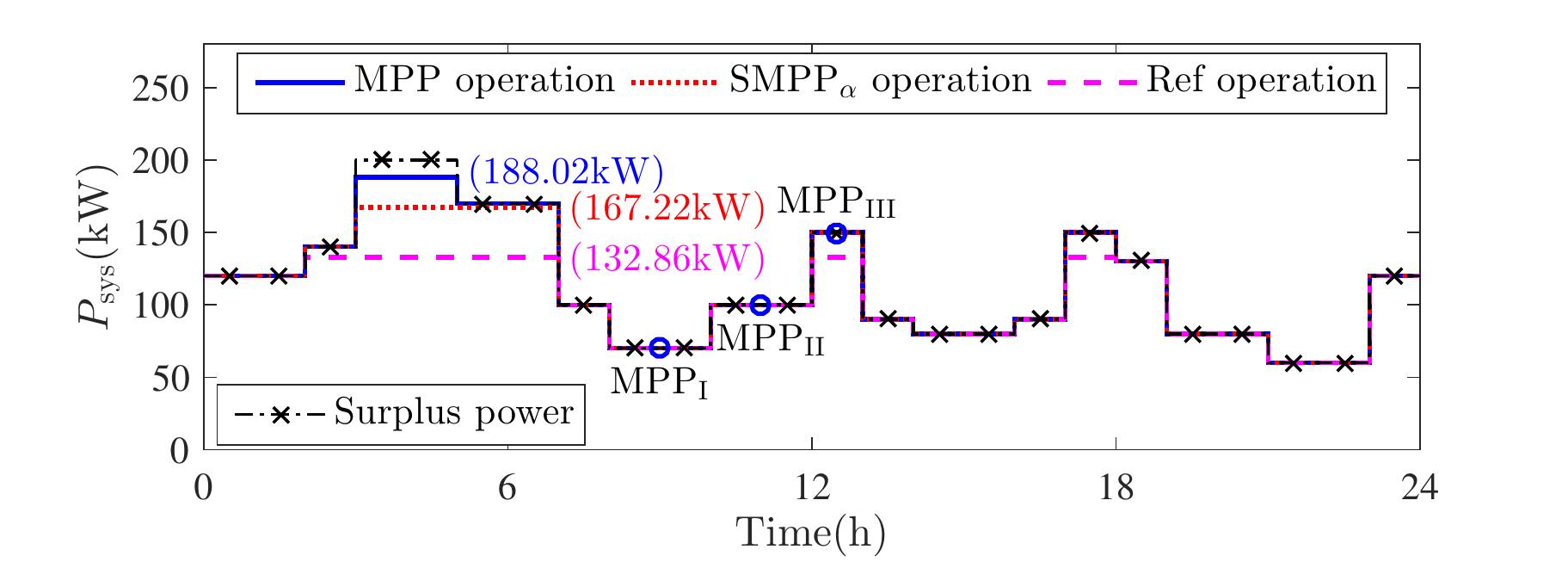}
  }
  \subfigure[Furnace operation]{
  \includegraphics[width=0.89\textwidth]{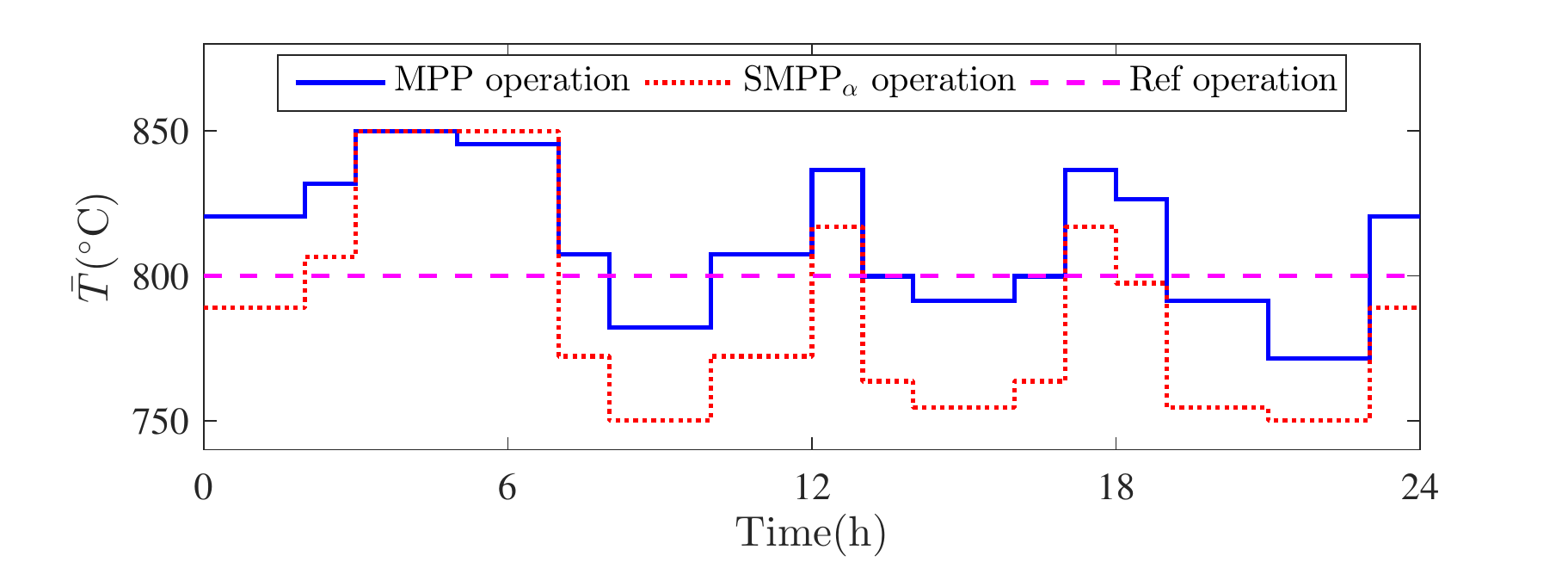}
  }
  \subfigure[Cathode mass flow operation]{
  \includegraphics[width=0.89\textwidth]{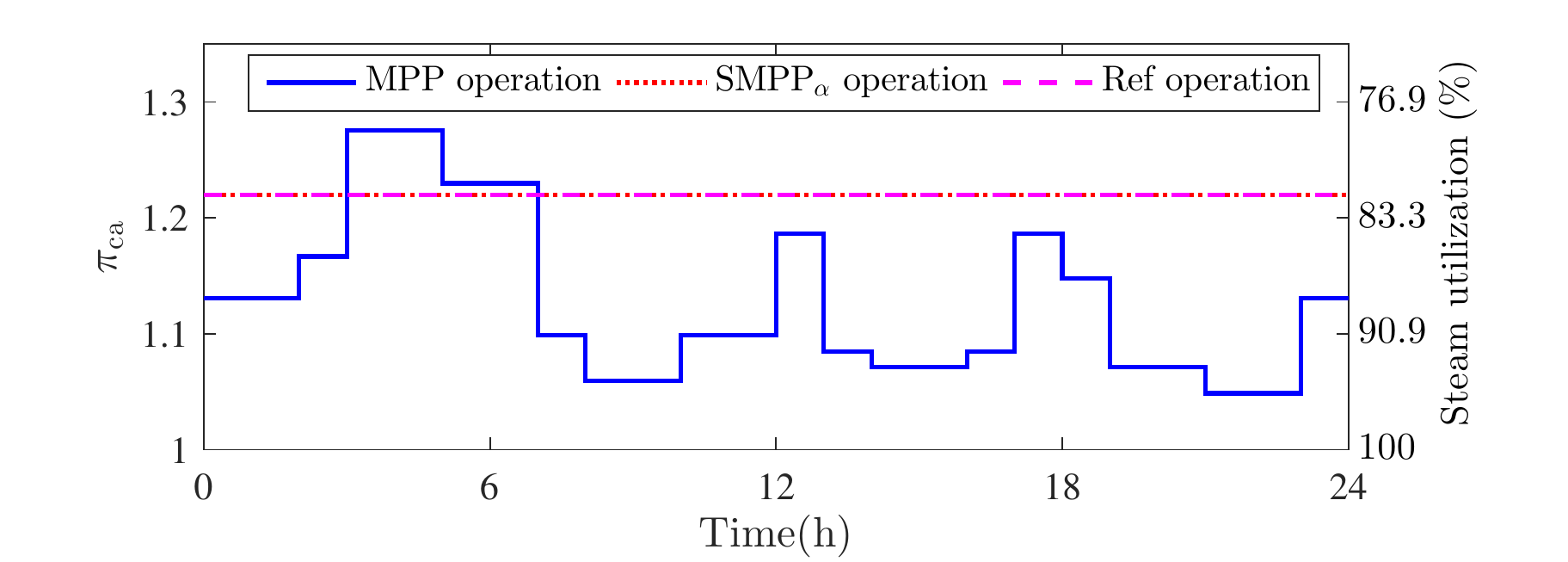}
  }
  \caption{\added[id=R2C3]{24-hour implementation of SMPP and MPP strategies under dynamic power input.}}\label{fig:day_operation}
\end{figure}

\added[id=EC4]{
To demonstrate the effects of proposed method in practical scenarios, we implement the SMPP$_{\rm \alpha}$ and MPP strategies under a 24-hour (from 0:00 to 24:00) dynamic power input as shown in Fig. \ref{fig:day_operation}(a).
The hourly scheduling commands for $P_{\rm sys}$ is generated to ensure the HTE system to track the profile of surplus electricity power so as to utilize it as much as possible.
In a case of hybrid microgrid, for example, the surplus power profile may come from load scheduling to moderate the peak power based on daily power profiles of renewable generation and residential load, which are usually fluctuant.
Given specific $P_{\rm sys}$, the operating parameters $I_{\rm cha}$, $\bar{T}$ and $\tilde{\bm{\pi}}$ for MPP, SMPP$_{\rm \alpha}$ or reference operation can be selected by looking up the set points in Fig. \ref{fig:MPPT_operation}.
The proposed strategies are expected to increase the $\rm H_2$ production from the given surplus energy.
}

\begin{figure}[!t]
  \centering
  \subfigure[Improved electrolysis current and $\rm H_2$ yield]{
  \includegraphics[width=0.89\textwidth]{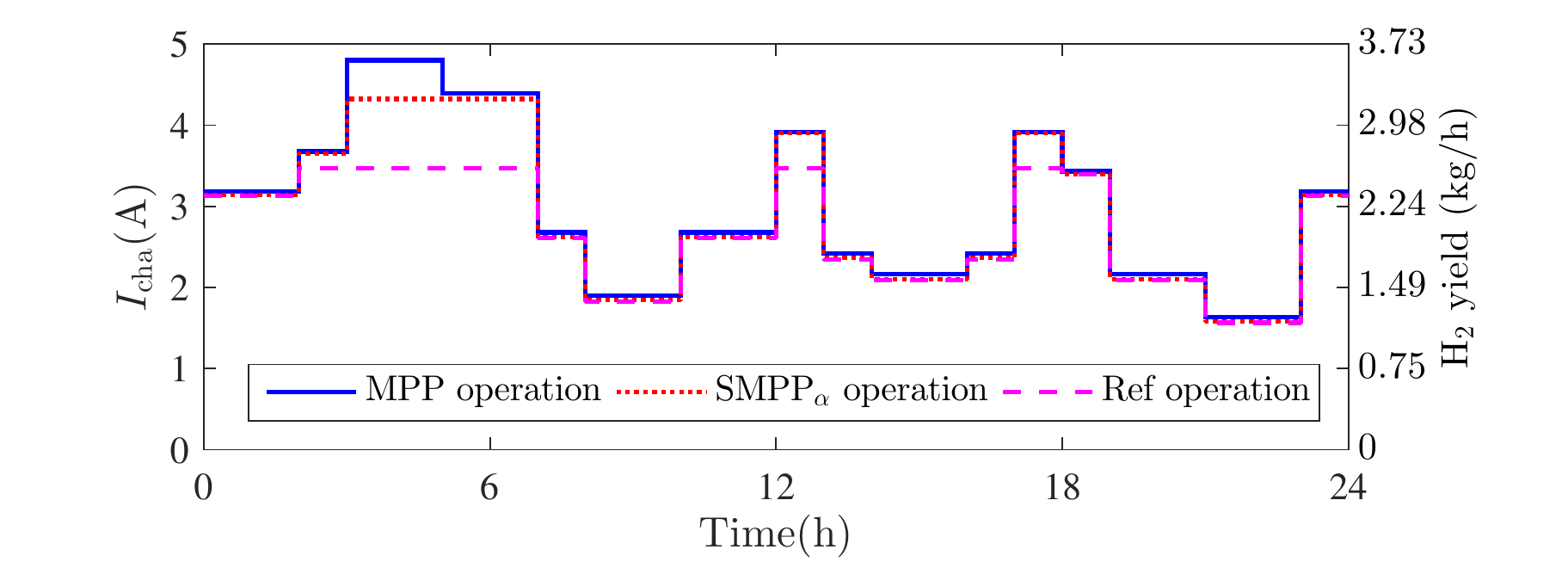}
  }
  \subfigure[Improved system efficiency]{
  \includegraphics[width=0.89\textwidth]{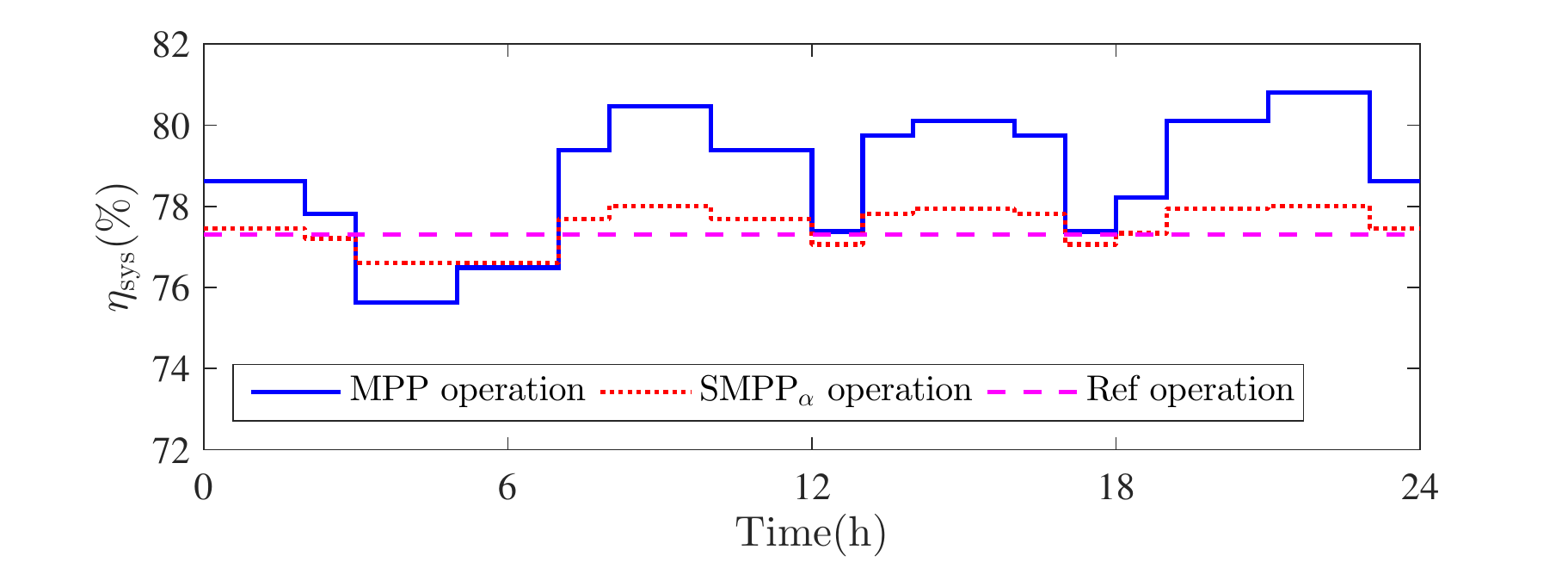}
  }
  \caption{\added[id=R2C3]{Beneficial effects of SMPP and MPP strategies under dynamic power input.}}\label{fig:day_performance}
\end{figure}

\added[id=R2C3]{
The simulated results are depicted in Fig. \ref{fig:day_operation}, Fig. \ref{fig:day_performance} and Table \ref{tab:performance}.
Fig. \ref{fig:day_operation}(a) shows the HTE system's actual load curves under different strategies.
It can be observed that the actual load follows the surplus power very well for most of the time, except for several hours when the surplus power exceeds the maximum system load, which is $188.02{\rm kW}$, $167.22{\rm kW}$ and $132.86{\rm kW}$ for MPP, SMPP$_{\rm \alpha}$ and reference operation, respectively.
Apparently the enlarged load ranges of MPP and SMPP$_{\rm \alpha}$ benefit the HTE system by significantly improving its available capacity to track the fluctuant curve of surplus power.
As summarized in Table \ref{tab:performance}, the utilization ratios of surplus electricity under MPP, SMPP$_{\rm \alpha}$ and reference operation are $99.12\%$, $97.40\%$ and $90.84\%$, respectively.
}

\added[id=R2C3]{
Fig. \ref{fig:day_operation}(b) and Fig. \ref{fig:day_operation}(c) present the operation of stack temperature $\bar{T}$ and cathode feed factor $\pi_{\rm ca}$, respectively.
As summarized in Table \ref{tab:performance}, the reference strategy employs constant $\bar{T}$ ($800{\rm ^{\circ}C}$) and $\pi_{\rm ca}$ ($1.22$) and only alters electrolysis current $I_{\rm cha}$ to meet the desired load power $P_{\rm sys}$.
By comparison, the SMPP$_{\rm \alpha}$ strategy adds $\bar{T}$ into the controllable parameters to improve system efficiency and load range, but $\pi_{\rm ca}$ is still kept as constant ($1.22$).
When it comes to the MPP strategy, both $\bar{T}$ and $\pi_{\rm ca}$ are considered flexible, so that all the primary HTE auxiliaries (power converter, furnace and pump group) can be coordinately utilized to achieve a further optimization.
}
\begin{table}[!t]
  \renewcommand{\arraystretch}{1.3}
  \caption{\added[id=R2C2]{24-hour $\rm H_2$ yields and other performance of proposed MPP and SMPP strategies under dynamic power input.}}\label{tab:performance}
  \centerline{
  \begin{threeparttable}
  \begin{tabular}{cccc}\toprule
  Operation &Ref operation &SMPP$_{\rm \alpha}$ strategy &MPP strategy\\ \midrule
  Total $\rm H_2$ yield ($\rm kg$) &48.29 &51.80 &53.35 \\
  Average system efficiency ($\eta_{\rm sys}$) &77.34$\%$ &77.38$\%$ &78.31$\%$ \\
  Surplus energy utilization &90.84$\%$ &97.40$\%$ &99.12$\%$ \\
  Surplus energy to $\rm H_2$ efficiency &70.26$\%$ &75.37$\%$ &77.62$\%$ \\
  Load range ($\rm kW$) &[20.44,132.86] &[20.25,167.22] &[19.30,188.02] \\
  Electrolysis current &Varies with $P_{\rm sys}$ &Varies with $P_{\rm sys}$ &Varies with $P_{\rm sys}$ \\
  Stack temperature &Fixed at $800{\rm ^{\circ}C}$ &Varies with $P_{\rm sys}$ &Varies with $P_{\rm sys}$ \\
  Cathode feed factor &Fixed at $\pi_{\rm ca}=1.22$ &Fixed at $\pi_{\rm ca}=1.22$ &Varies with $P_{\rm sys}$ \\ \bottomrule
  \end{tabular}
  \end{threeparttable}
  }
\end{table}

\added[id=R2C3]{
Fig. \ref{fig:day_performance}(a) depicts the variation of electrolysis current $I_{\rm cha}$ and $\rm H_2$ yield under MPP, SMPP$_{\rm \alpha}$ and reference operation.
It can be observed that at all hours, the temporal $\rm H_2$ yields under reference, SMPP$_{\rm \alpha}$ and MPP operation increase successively.
The corresponding system efficiency $\eta_{\rm sys}$ is depicted in Fig. \ref{fig:day_performance}(b).
Most of the time,  the temporal values of $\eta_{\rm sys}$ under reference, SMPP$_{\rm \alpha}$ and MPP operation increase successively.
There are also some hours when $\eta_{\rm sys}$ decreases after implementation of SMPP$_{\rm \alpha}$ or MPP strategies, but these situations are actually due to different $P_{\rm sys}$ caused by insufficient capacity under reference operation, and only occur when $P_{\rm sys}$ turns out to be significantly larger under SMPP$_{\rm \alpha}$ or MPP strategies.
After all, MPP $\eta_{\rm sys}$ at a larger load can be lower than the reference $\eta_{\rm sys}$ at a smaller load, as shown in Fig. \ref{fig:MPPT_operation}(c).
Anyway, even though $\eta_{\rm sys}$ can be lower under SMPP$_{\rm \alpha}$ and MPP operation sometimes, the $\rm H_2$ yield ($\eta_{\rm sys} P_{\rm sys}$) will always be improved.
As summarized in Table \ref{tab:performance}, the 24-hour average system efficiencies under MPP, SMPP$_{\rm \alpha}$ and reference operation are $78.31\%$, $77.38\%$ and $77.34\%$, respectively.
The cumulative yields of $\rm H_2$ under MPP, SMPP$_{\rm \alpha}$ and reference operation are $53.35{\rm kg}$, $51.80{\rm kg}$ and $48.29{\rm kg}$, representing surplus energy to $\rm H_2$ efficiencies of $77.62\%$, $75.37\%$ and $70.26\%$, respectively.
}

\added[id=EC4]{
From the above implementation case, it is validated that the proposed SMPP and MPP strategies both have advantageous effects on improving the HTE system's performance on $\rm H_2$ production from fluctuant power input, which will surely benefits the HTE plants by increasing the energy conversion profits during practical operation.
Moreover, the enlarged load ranges due to proposed strategies can increase the dispatchable capacities as shown in Fig. \ref{fig:day_operation}(a), which will in turn benefit the grid operation when the HTE plant participates grid services such as economic dispatch or peak regulation.
}

\section{Discussions}

\subsection{\added[id=EC3]{Comparisons with Similar Work}}
\added[id=EC3]{
As discussed in Section \ref{sec:optimization} and Section \ref{sec:numericalStudy}, some useful principles of controlling operating parameters can be obtained from model \eqref{eq:OP_PtG} to improve the HTE system performance.
Here, these results are compared with those of some published papers that studied HTE operating conditions as well.
}

\added[id=EC3]{
Reference \cite{Wang_P2M_opt_2018} performed a comprehensive multi-objective optimization to improve the system performance of an HTE based P2G plant, indicating that the system HHV efficiency increases with steam utilization, but decreases with gas yield, electrical heating and anode feed rate.
Apparently these results coincided with our our model \eqref{eq:OP_PtG} as shown in Fig. \ref{fig:Tpi_contour} and Fig. \ref{fig:MPPT_operation}(c).
However, reference \cite{Wang_P2M_opt_2018} kept the inlet temperature constant ($750{\rm ^{\circ}C}$) and did not discuss its optimization.
Also, a professional simulator like Aspen was necessarily required as a single step of the optimizing iterations.
Model \eqref{eq:OP_PtG} in our paper is more concise and intuitive for understanding, and easier for practical computation.
}

\added[id=R1C3]{
Reference \cite{Buttler_HTE_heatIntegration_2015} managed to locate an optimal loading point to solve the trade-off between lower specific electrical energy consumption and higher hydrogen production for an isothermal operating HTE system with heat integration.
The authors also took the investment costs into consideration by performing a techno-economic analysis.
However, reference \cite{Buttler_HTE_heatIntegration_2015} focused on the stack efficiency excluding the energetic consumptions of necessary auxiliaries and heat integration, and the temperature optimization was not discussed as the inlet temperature was fixed at $850 {\rm ^{\circ}C}$.
In terms of feed flow rates, the authors indicated that the HTE performance rises significantly at a high steam utilization ratio (low $\pi_{\rm ca}$), but this ratio is limited by the starvation phenomenon.
Apparently this point is consistent with our model \eqref{eq:OP_PtG} as shown in Fig. \ref{fig:Tpi_contour}.
}

\added[id=R1C3]{
Reference \cite{Laurencin_SOEC_conditions_2011} carried out a comprehensive parametric study for HTE's irreversible losses based on the simulation of a refined 2D stack model and thermal model.
Dozens of spacial distribution profiles and sensitive plots of primary HTE characteristics were to discuss the impacts of current densities, gas compositions, geometric designs, etc.
The behaviors of our lumped cell model $U_{\rm cel}(I_{\rm cha},\bar{T},\tilde{\bm{\pi}}$ can be validated in \cite{Laurencin_SOEC_conditions_2011}, including the large proportion of concentration overvoltage at a high steam utilization ratio (low $\pi_{\rm ca}$).
However, reference \cite{Laurencin_SOEC_conditions_2011} focused on the stack polarisation curves and the system performance was not discussed.
}

\added[id=R1C3]{
Reference \cite{Menon_SOEC_model_2014} employed a detailed cell model to study the dependence of system efficiency on hydrogen production rate, operating temperature and gas composition.
It was pointed out that the overall efficiency decreases at higher hydrogen production rate, which agreed with our model \eqref{eq:OP_PtG} as shown in Fig. \ref{fig:MPPT_operation}(c).
It was also indicated that the operating temperature has advantageous effects on overall efficiency, which seemed inconsistent with our Fig. \ref{fig:Tpi_contour}.
However, it should be noted that the thermal utilization ratio was assumed to be constant (0.9) for all cases when calculating the overall efficiency in \cite{Menon_SOEC_model_2014}.
In our paper, this ratio is expected to decrease at higher temperature because the unutilized heat output (in thermal convection) increases with temperature.
}

\added[id=EC3]{
Reference \cite{Petipas_HTE_VariousLoads_2013} simulated steady-state operating points of an HTE system at various loads, where the system efficiency is calculated based on hydrogen production (HHV) and the overall energy consumption including auxiliary equipment like the compressor, the heater and the pump.
The results indicated that the system efficiency decreases as loading power increases, which coincided with our model \eqref{eq:OP_PtG} as shown in Fig. \ref{fig:MPPT_operation}(c).
Meanwhile, the authors pointed out that an improved system efficiency can be reached with no air sweep, or with a steam utilization ratio ``as high as possible'', which agreed with our conclusions about $\pi_{\rm an}$ and $\pi_{\rm ca}$ as shown in Fig. \ref{fig:Tpi_contour}.
However, reference \cite{Petipas_HTE_VariousLoads_2013} focused on presenting the system behaviors at different electrolysis currents while keeping inlet temperature, steam utilization ratio and air ratio constants ($800{\rm ^{\circ}C}$, $75\%$ and $0$, respectively).
No optimization or constraints modeling was performed.
In fact, reference \cite{Petipas_HTE_VariousLoads_2013} concluded that control strategies to regulate operating parameters were needed to improve the system performance, to enlarge the operating power range, and to make the HTE system ``suitable for variable operation''.
This is exactly what we have done in above sections.
}

\added[id=EC3]{
To sum up, a concise operation model \eqref{eq:OP_PtG} is novelly abstracted from a general and complicate HTE system in this paper, providing an intuitional method to optimize HTE operating parameters precisely.
Our mathematically obtained operating principles are consistent with previous experimental or simulative results of relevant papers, but our method shows valuable advantages such as comprehensiveness, quantitativeness, analyticity and easy applicability.
}

\subsection{\added[id=R2C4]{Further Research on Dynamics}}
\added[id=R2C4]{
In Section \ref{ssec:dynamicPower}, the proposed SMPP and MPP strategies are implemented to improve the $\rm H_2$ production from dynamic power input, and their beneficial effects on the HTE system are validated as expected.
However, it should be noted that our paper is aimed to present a optimization of steady-state performance of an HTE system.
Both our energy flow model of Fig. \ref{fig:EnergyBalance} and operation model \eqref{eq:OP_PtG} are steady-state models in essence, because they are developed from steady-state energy balances \eqref{eq:energyBalance} and \eqref{eq:energyBalance_pre} as mentioned in the modeling methodology in Section \ref{ssec:methodology}.
Strictly speaking, the proposed operation model \eqref{eq:OP_PtG} only applies to steady-state operating points, such as MPP$_{\rm I}$, MPP$_{\rm II}$ and MPP$_{\rm III}$ in Fig. \ref{fig:MPP_obtain} and Fig. \ref{fig:day_operation}(a), where the HTE system and all its auxiliaries or parameters are assumed to have already stabilized to specific states or values.
In other words, model \eqref{eq:OP_PtG} is not capable of describing the transient process between different steady states, such as the transient process from MPP$_{\rm I}$ to MPP$_{\rm II}$, or from MPP$_{\rm II}$ to MPP$_{\rm III}$ in Fig. \ref{fig:MPP_obtain} and Fig. \ref{fig:day_operation}(a).
In fact, the operating points during the transient process need temporal coupling descriptions such as the change rate of temperature ${\rm d}\bar{T}/{\rm d}t$, and cannot be depicted by any points within the steady-state model \eqref{eq:OP_PtG} or the steady-state illustration plot Fig. \ref{fig:MPP_obtain}.
}

\added[id=R2C4]{
Nevertheless, the proposed operation model \eqref{eq:OP_PtG} can still be applicable to dynamic power input in some scenarios.
Usually the transient time is limited by the system's largest internal time constant, which is at the minute level in the case of Section \ref{ssec:dynamicPower}.
Therefore, it is reasonable to neglect the transient processes for the simulated numerical system under a power input changing hourly or more slowly, and to focus on the long-term system performance with the proposed steady-state operation model \eqref{eq:OP_PtG}.
This is a commonly used method to simulate the behaviors of a controllable system with large dispatch period \cite{Clegg_P2G_2015_els,Khani_SchedulingP2G_2017_els}.
In fact, economic load dispatch and peak regulation of the HTE system in smart grid is an obvious extending direction of current steady-state operation research, where renewable integration, time-of-use electricity price, secondary frequency regulation and other grid-side operation topics can all be studied and the HTE system is only one of the dispatchable resources.
}

\added[id=R2C4]{
If we want to describe HTE behavior precisely under fast fluctuations of power input, however, a refined dynamic model considering the multi-scale dynamics of temperature (response time of minutes), gas flow (response time of seconds) and electrolysis current (response time of milliseconds) is necessarily required \cite{Gasser_PEMFC_SystemModel_2006,Milewski_AuxiliaryDynamic_2014}.
It is also important to design a dynamic controller to coordinate the operating parameters all the time to ensure a rapid and smooth transition in a right direction, which brings about another potential research extension.
Apparently, to ensure both long-term and short-term performance, the optimal design of such a dynamic controller has to be developed from both a thorough study of steady-state operation as presented in this paper to locate the right target (such as MPP$_{\rm II}$ in Fig. \ref{fig:day_operation}(a)), and a smart coordination of multi-scale dynamics to achieve rapid response (such as the fast transition from MPP$_{\rm I}$ to MPP$_{\rm II}$ in Fig. \ref{fig:day_operation}(a)).
This is actually the extended research that we are currently working on.
}

\section{Conclusions}
In this paper, a lumped operation model of an \replaced[id=EC2]{high-temperature electrolysis}{HTE} system with auxiliaries was \added[id=EC3]{explicitly} developed from a comprehensive energy flow model and verified by a refined 3D-model in COMSOL.
\added[id=EC3]{The sub-maximum production point and maximum production point operating strategies were then proposed to coordinate energy consumptions of various auxiliaries and maximize hydrogen yield by optimizing the temperature and the feedstock flow rates.}

\added[id=EC3]{The proposed strategies have proved to be quite effective in increasing the energy conversion efficiency at a specific load as depicted in Fig. \ref{fig:Tpi_contour}, which is accompanied by lowered energy allocations to furnace, preheater and pump gruop, but elevated energy allocations to electrolyser and compressor.
The electrolysis stack is found operated near the thermo-neutral point at maximum production points as shown in Fig. \ref{fig:UP_energy}.
When applied to a fluctuant power input, these strategies can increase the overall hydrogen production by improving the system efficiency and enlarging the available load range simultaneously as shown in Fig. \ref{fig:MPPT_operation}(c) and Table \ref{tab:performance}, and thus increase the utilization ratio of existing assets and surplus power generations.
Both the grid operation and plant itself will be benefited due to increased hydrogen yield and enlarged load range.}

\added[id=EC3]{As the basis of local control of high-temperature electrolysis plants, this study is expected to provide a basis for further studies on dynamic coupling and economic dispatch or scheduling of high-temperature power-to-gas implementations within the power grid.}
\appendix
\section{\added[id=R2C1]{Expansion of Reversible Voltage $U_{\rm rev}(T,\tilde{\bm{p}})$}}\label{app:U_rev}
\added[id=R2C1]{The detailed expansion of $U_{\rm rev}$ can be derived from its definition in \eqref{eq:UrevUth}.
According to thermodynamic theories, the enthalpy change $\Delta_{\rm r} H_T$ is uniquely determined by temperature $T$:}
\begin{equation}\label{eq:form_DH}
  \Delta_{\rm r} H_T = \Delta_{\rm r}H_{T_{\rm H}} + (T-T_{\rm H}) \Delta_{\rm r}C_{T_{\rm H}}
\end{equation}
\added[id=R2C1]{where $C_{T_{\rm H}}$ represents the isobaric molar heat capacity of specific substance at temperature $T_{\rm H}$.
On the other hand, the entropy change $\Delta_{\rm r} S_{T,\tilde{\bm{p}}}$ is dependent on both temperature and pressure conditions \cite{Bruun_BG_SOFC_2009}.
Its dependence on partial pressures $\tilde{\bm{p}}$ can be described as:}
\begin{equation}\label{eq:form_DS_p}
  \Delta_{\rm r} S_{T,\tilde{\bm{p}}} = \Delta_{\rm r} S^{\ominus}_T - R{\rm ln}\left(\frac{p_{\rm H_2}p_{\rm O_2}^{0.5}}{p_{\rm H_2O}p^{\ominus \, 0.5}}\right)
\end{equation}
\added[id=R2C1]{where $S^{\ominus}_T$ denotes the entropy at standard state (gas pressure equals $p^{\ominus}$) and temperature $T$.
Then $S^{\ominus}_T$ can be formulated as the following function of temperature $T$:}
\begin{equation}\label{eq:form_DS_T}
  \Delta_{\rm r} S^{\ominus}_T = \Delta_{\rm r}S^{\ominus}_{T_{\rm H}} + (\Delta_{\rm r}C_{T_{\rm H}}) {\rm ln} \frac{T}{T_{\rm H}}.
\end{equation}
\added[id=R2C1]{Note that $H_{T_{\rm H}}$ and $S^{\ominus}_{T_{\rm H}}$ in \eqref{eq:form_DH} and \eqref{eq:form_DS_T} are reference enthalpy and entropy values of specific substance at a reference temperature $T_{\rm H}$.
Finally, by substituting \eqref{eq:form_DH}, \eqref{eq:form_DS_p} and \eqref{eq:form_DS_T} into \eqref{eq:UrevUth}, the following expansion form of $U_{\rm rev}(T,\tilde{\bm{p}})$ can be explicitly obtained:}
\begin{subequations}\label{eq:form_Urev_expansion}
\begin{align}
  U_{\rm rev}(T,\tilde{\bm{p}}) &\stackrel{\mbox{\tiny def}}{=} \frac{\Delta_{\rm r} G_{T,\tilde{\bm{p}}}}{n_{\rm e}F} \\
  &= \frac{\Delta_{\rm r} H_T - T\Delta_{\rm r} S_{T,\tilde{\bm{p}}} }{n_{\rm e}F} \\
  &= \frac{1}{n_{\rm e}F} \left[ \Delta_{\rm r} H_T - T \Delta_{\rm r} S^{\ominus}_T + RT{\rm ln}\left(\frac{p_{\rm H_2}p_{\rm O_2}^{0.5}}{p_{\rm H_2O}p^{\ominus \, 0.5}}\right) \right] \\
  &= \frac{\Delta_{\rm r}H_{T_{\rm H}} \!\!-\!\! T\Delta_{\rm r}S^{\ominus}_{T_{\rm H}} \!\!+\!\! (T\!\!-\!\!T_{\rm H}\!\!-\!\!T{\rm ln} \frac{T}{T_{\rm H}}) \Delta_{\rm r}C_{T_{\rm H}}}{n_{\rm e}F} \!+\! \frac{RT}{n_{\rm e}F}{\rm ln}\left(\!\frac{p_{\rm H_2}p_{\rm O_2}^{0.5}}{p_{\rm H_2O}p^{\ominus \, 0.5}}\!\right).
\end{align}
\end{subequations}
\added[id=R2C1]{In other words, $U_{\rm rev}^{\ominus}(T)$ in \eqref{eq:form_Urev} is actually}
\begin{equation}\label{eq:form_Urev0_expansion}
  U_{\rm rev}^{\ominus}(T) = \frac{\Delta_{\rm r}H_{T_{\rm H}} - T\Delta_{\rm r}S^{\ominus}_{T_{\rm H}} + (T\!\!-\!\!T_{\rm H} - T{\rm ln} \frac{T}{T_{\rm H}}) \Delta_{\rm r}C_{T_{\rm H}}}{n_{\rm e}F}.
\end{equation}
\added[id=R2C1]{The employed thermal properties in this paper are listed in Table \ref{tab:properties}, which is extracted from \cite{Bruun_BG_SOFC_2009} and \cite{EngineeringToolBox}.}
\begin{table}[!t]
  \renewcommand{\arraystretch}{1.3}
  \caption{\added[id=R2C1]{The Employed thermal properties of substances.}}\label{tab:properties}
  \centerline{
  \begin{threeparttable}
  \begin{tabular}{ccccc}\toprule
  Substance &$H_{T_{\rm H}}$($\rm J/mol$) &$S^{\ominus}_{T_{\rm H}}$($\rm J/(mol \cdot K)$) &$C_{T_{\rm H}}$($\rm J/(mol \cdot K)$) &$C_{T_{\rm L}}$($\rm J/(mol \cdot K)$) \\ \midrule
  $\rm H_2$ &35264 &171.79 &30.992 &28.92 \\
  $\rm H_2O$ &$-194511$ &240.485 &43.768 &34.02 \\
  $\rm H_2O$ (liquid) &- &- &- &75.95 \\
  $\rm O_2$ &38444 &250.01 &35.667 &29.89 \\
  $\rm N_2$ &36779 &234.226 &33.723 &29.18 \\ \bottomrule
  \end{tabular}
  \begin{tablenotes}
  \footnotesize
  \item{*} $T_{\rm H}=1200{\rm K}$, $T_{\rm L}=373.15{\rm K}$, $p^{\ominus}=1{\rm bar}$.
  \end{tablenotes}
  \end{threeparttable}
  }
\end{table}

\section{\added[id=R2C1]{Derivation of Lumped Pressure Values $\tilde{\bm{p'}}(\tilde{\bm{\pi}})$}}\label{app:p_pi}
According to \cite{Bird_Transport_2002}, the following mass balance equation \eqref{eq:massBalance} describes the transport continuity of $\rm H_2$ \added[id=R2C1]{along the flow direction $l$} at steady state:
\begin{equation}\label{eq:massBalance}
  -\frac{\partial}{\partial l}\left( \rho_{\rm H_2} v_{\rm ca} + j_{\rm H_2} \right) + \frac{i_{\rm el}M_{\rm H_2}}{n_{\rm e}F L_{z,{\rm cha}}} = 0
\end{equation}
\added[id=R2C1]{where $v_{\rm ca}$ is the flow velocity of cathode gas mixture, and $j_{\rm H_2}$ is the mass diffusion rate of $\rm H_2$ within that mixture.
Note that $\rho_{\rm H_2}$ and $M_{\rm H_2}$ are density and molar mass of $\rm H_2$, respectively, and $L_{z,{\rm cha}}$ is the channel height as depicted in Fig. \ref{fig:HTE_cell}.}
The definite integral of \eqref{eq:massBalance} along $l$ on $[l_{\rm s3},l_{\rm s4}]$ yields:
\begin{equation}\label{eq:massBalance_int}
  \frac{p_{\rm H_2,s3} M_{\rm H_2}}{R T_{\rm s3}}v_{\rm ca,s3} - \frac{p_{\rm H_2,s4} M_{\rm H_2}}{R T_{\rm s4}}v_{\rm ca,s4} + \frac{I_{\rm cha}M_{\rm H_2}}{n_{\rm e}F L_{z,{\rm cha}} L_{y,{\rm cha}}} = 0
\end{equation}
where $\rho_{\rm H_2}$ is substituted using the ideal gas law $p_{\rm H_2}\frac{M_{\rm H_2}}{\rho_{\rm H_2}}=R T$ (which is also employed \added[id=R2C1]{for HTE modeling} in \cite{Udagawa_SOECmodel_2007}, \cite{Udagawa_SOECoperation_2008}, and \cite{Ni_2D_SOEC_2012}), and $L_{y,{\rm cha}}$ is the channel width as depicted in Fig. \ref{fig:HTE_cell}.
Note that \added[id=R2C1]{the $j_{\rm H_2}$ term vanishes in \eqref{eq:massBalance_int} because} there is no mass diffusion at stack inlet $l_{\rm s3}$ or stack outlet $l_{\rm s4}$ where the concentration of $\rm H_2$ no longer varies.

Based on definition \eqref{eq:pis}, $\frac{I_{\rm cha}}{n_{\rm e}F}$ in the third term of \eqref{eq:massBalance_int} can be rewritten as:
\begin{subequations}
\begin{align}
\frac{I_{\rm cha}}{n_{\rm e}F}&{}={}\frac{\frac{w_{\rm ca,s3}}{N \rho_{\rm ca,s3}}p_{\rm H_2O,s3}}{\pi_{\rm ca}}\frac{\rho_{\rm ca,s3} \omega_{\rm H_2O,s3}}{M_{\rm H_2O}p_{\rm H_2O,s3}} \label{eq:InF_1} \\
&{}={}\frac{L_{z,{\rm cha}} L_{y,{\rm cha}} v_{\rm ca,s3} p_{\rm H_2O,s3}}{\pi_{\rm ca}} \frac{\rho_{\rm ca,s3} \omega_{\rm H_2O,s3}}{M_{\rm H_2O}p_{\rm H_2O,s3}} \label{eq:InF_2}\\
&{}={}\frac{L_{z,{\rm cha}} L_{y,{\rm cha}} v_{\rm ca,s3} p_{\rm H_2O,s3}}{\pi_{\rm ca}} \frac{1}{R T_{\rm s3}} .\label{eq:InF_3}
\end{align}
\end{subequations}
By substituting \eqref{eq:InF_3} into \eqref{eq:massBalance_int}, we obtain:
\begin{equation}\label{eq:p_H2_1}
  p_{\rm H_2,s4} = \frac{v_{\rm ca,s3}T_{\rm s4}}{v_{\rm ca,s4}T_{\rm s3}}\left(p_{\rm H_2.s3}+\frac{p_{\rm H_2O,s3}}{\pi_{\rm ca}}\right)
\end{equation}
With the same \added[id=R2C1]{derivation \eqref{eq:massBalance}-\eqref{eq:p_H2_1}} for $\rm H_2O$ transport, we have:
\begin{equation}\label{eq:p_H2O_1}
  p_{\rm H_2O,s4} = \frac{v_{\rm ca,s3}T_{\rm s4}}{v_{\rm ca,s4}T_{\rm s3}}\left(p_{\rm H_2O.s3}-\frac{p_{\rm H_2O,s3}}{\pi_{\rm ca}}\right)
\end{equation}
Thus, the outlet volume fractions of cathode pipe are obtained from \eqref{eq:p_H2_1} and \eqref{eq:p_H2O_1} together:
\begin{equation}\label{eq:p_H2}
\frac{p_{\rm H_2,s4}}{p_{\rm ca,s4}} = \frac{\pi_{\rm ca}p_{\rm H_2,s3}+p_{\rm H_2O,s3}}{\pi_{\rm ca}p_{\rm ca,s3}}
\end{equation}
\begin{equation}\label{eq:p_H2O}
\frac{p_{\rm H_2O,s4}}{p_{\rm ca,s4}} = \frac{(\pi_{\rm ca}-1)p_{\rm H_2O,s3}}{\pi_{\rm ca}p_{\rm ca,s3}}
\end{equation}
Analogously, the outlet volume fractions of anode pipe can be obtained as:
\begin{equation}\label{eq:p_O2}
\frac{p_{\rm O_2,s4}}{p_{\rm an,s4}} = \frac{(\pi_{\rm an}+1)p_{\rm O_2,s3}}{\pi_{\rm an}p_{\rm an,s3}+p_{\rm O_2,s3}}
\end{equation}
\begin{equation}\label{eq:p_N2}
\frac{p_{\rm N_2,s4}}{p_{\rm an,s4}} = \frac{\pi_{\rm an}p_{\rm N_2,s3}}{\pi_{\rm an}p_{\rm an,s3}+p_{\rm O_2,s3}}
\end{equation}
Here, we reach a useful conclusion according to \eqref{eq:p_H2}-\eqref{eq:p_N2}: with the inlet composition fixed, the outlet composition ratios of the HTE system are \emph{determined} by feed factors $\tilde{\bm{\pi}}$.
This is how $\tilde{\bm{\pi}}$ influences $\tilde{\bm{p}}$ and thus the electrolysis status and the system performance.

\added[id=R2C1]{Now that the outlet fractions \eqref{eq:p_H2}-\eqref{eq:p_N2} are obtained, we can calculate the lumped pressures $\tilde{\bm{p'}}$ required for cell voltage evaluation $U_{\rm cel}(\bar{i}_{\rm el},\bar{T},\tilde{\bm{p'}})$ in \eqref{eq:P_el_1}.}
For example, the lumped value of hydrogen partial pressure $p_{\rm H_2}'$, which is calculated in a geometric mean manner due to the logarithmic form in \eqref{eq:form_Urev}, can be formulated from \eqref{eq:p_H2} as:
\begin{subequations}
\begin{align}
p_{\rm H_2}'(\pi_{\rm ca})&{}={}p_{\rm H_2,s3}^{1-\lambda} \cdot p_{\rm H_2,s4}^{\lambda} \label{eq:p_p_H2_1} \\
&{}={}p_{\rm H_2,s3}^{1-\lambda}\left(p_{\rm ca,s4} \frac{\pi_{\rm ca}p_{\rm H_2,s3}+p_{\rm H_2O,s3}}{\pi_{\rm ca}p_{\rm ca,s3}}\right)^{\lambda} \label{eq:p_p_H2_2}\\
&{}={}\left(p_{\rm ca,s3}^{1-\lambda} p_{\rm ca,s4}^{\lambda}\right) \chi_{\rm H_2,s3}^{1-\lambda} \left(\frac{\pi_{\rm ca}\chi_{\rm H_2,s3}+\chi_{\rm H_2O,s3}}{\pi_{\rm ca}}\right)^{\lambda} \label{eq:p_p_H2_3}
\end{align}
\end{subequations}
Note that $\lambda \in (0,1)$ is a constant selected through a parametric study.
In fact, the pressure drop \added[id=R2C1]{$(p_{\rm ca,s3}-p_{\rm ca,s4})$} caused by gas flows in both pipes can usually be neglected compared with the value of absolute pressures \added[id=R2C1]{$p_{\rm ca,s3}$ and $p_{\rm ca,s4}$}.
In this situation, \added[id=R2C1]{$p_{\rm ca,s3} \simeq p_{\rm ca,s4}$, and \eqref{eq:p_p_H2_3} can be rewritten as}:
\begin{equation}\label{eq:p_p_H2_app}
p_{\rm H_2}'(\pi_{\rm ca}) \simeq p_{\rm ca,s4} \chi_{\rm H_2,s3}^{1-\lambda} \left(\frac{\pi_{\rm ca}\chi_{\rm H_2,s3}+\chi_{\rm H_2O,s3}}{\pi_{\rm ca}}\right)^{\lambda}
\end{equation}
\added[id=R2C1]{which is actually the aforementioned \eqref{eq:p_p_H2}.}
Analogously, $p_{\rm H_2O}'(\pi_{\rm ca})$, $p_{\rm O_2}'(\pi_{\rm an})$ and $p_{\rm N_2}'(\pi_{\rm an})$ can be formulated explicitly from \eqref{eq:p_H2O}, \eqref{eq:p_O2} and \eqref{eq:p_N2}, respectively, \added[id=R2C1]{as shown in \eqref{eq:p_p_H2O}-\eqref{eq:p_p_N2}.}
The equations \eqref{eq:p_p_H2}-\eqref{eq:p_p_N2} together present \added[id=R2C1]{the gas-flow submodel $\tilde{\bm{p'}}(\tilde{\bm{\pi}})$ as mentioned in Section \ref{ssec:gasFlow}.}

\section*{References}
\bibliography{myRef}
\end{document}